\documentclass[preprint,12pt,authoryear]{elsarticle}

\usepackage{amssymb}
\usepackage{amsmath}

\usepackage{amsfonts}
\usepackage{esint}
\usepackage{graphicx}
\usepackage{subfigure}
\usepackage{color}
\usepackage{epstopdf}
\usepackage{tikz}
\usepackage{pdflscape}
\usepackage{mathtools}

\graphicspath{{figures/}} 

\journal{International Journal of Engineering Science}

\begin{document}

\begin{frontmatter}



\title{Peculiarities of hydraulic fracture propagation in media with heterogeneous toughness: the energy balance, elastic battery and fluid backflow} 


\author[AU]{Daniel Peck}
\ead[url]{dtp@aber.ac.uk}

\author[RO]{Gaspare {Da Fies}}

\author[KY]{Ivan Virshylo}

\author[AU]{Gennady Mishuris}

\affiliation[AU]{
	organization={Department of Mathematics},
    addressline={Aberystwyth University},
	city={Aberystwyth},
	postcode={SY23 3BZ},
	state={Wales},
	country={UK}}

\affiliation[RO]{
	organization={Rockfield},
	addressline={Kings Road, Prince of Wales Dock},
	city={Swansea},
	postcode={SA1 8AS},
	state={Wales},
	country={UK}}

\affiliation[KY]{
	organization={Taras Shevchenko National University of Kyiv},
	addressline={Institute of Geology},
	city={Kyiv},
	country={Ukraine}}

\begin{abstract}
This paper investigates hydraulic fracture in a {medium} with periodic heterogeneous toughness. Results for the plane-strain (KGD) model are analysed. The energy distribution as the fracture propagates is examined, along with the evolution of the crack geometry. It is shown that the solid layer acts as an elastic battery, discharging to promote rapid propagation through weaker material layers. {The limiting case of an infinite-length crack is discussed.} The velocity of the fluid throughout the crack length is also considered. For fractures in high-toughness material it is shown that fluid backflow can occur, with its profile dependent on the toughness distribution. The implications of these findings are discussed.
\end{abstract}


\begin{highlights}
	\item Highly accurate (numerical) results for crack propagation in a material with heterogeneous fracture toughness are analysed. 
	\item The changing energy distribution during propagation between layers with differing material toughness is provided.
	\item It is demonstrated that the solid layer acts as an elastic battery, discharging to promote rapid propagation through weaker material layers. This explains the rapid oscillation of the crack tip velocity.
	\item Fluid backflow, where some fraction of the fluid volume insider the crack moves in the opposite direction to the crack tip motion, is observed. The form of the backflow is dependent on the toughness distribution.
\end{highlights}


\end{frontmatter}




\section{Introduction}

Hydraulic fracture (HF) is the process of a {fluid-driven} crack propagating in a solid material. 
On a planetary scale, hydraulic fracture is induced by the tectonics of the lithosphere and is closely related to the origin of earthquakes. Induced seismic risk is an important issue in geothermal energy projects and underground gas storage (carbon dioxide, natural gas, hydrogen, see e.g. \cite{LiEarth2024,Moein2023}), for which predicting and controlling  HF behavior is crucial (see e.g. \cite{s24175775,en17163865} and references therein). Understanding the processes of crack formation is also relevant in the assessment of natural hazards - the formation of landslides, earthquakes, and volcanic dykes, as well as soil flooding in engineering and construction applications.

Consequently, hydraulic fracture modelling and the study of HF-related effects have been an active field of research for decades. Amongst the earliest approaches are the 1D models, including the Kristianovich-Geertsma-de Klerk (KGD, plane strain) approximation \cite{Khristianovic1955,Geertsma1969}. Although since superseded in most applications, the fact that it is more easily modified than later models makes it useful for incorporating and testing previously unincorporated effects, and analysing their impact on the HF process.

One such important effect is the impact of heterogeneity of the fracture media, which has typically been neglected in HF models due to the complexity of the problem. In particular, while homogenisation strategies exist for the Young's modulus and other parameters, homogenisation of the fracture toughness $K_{Ic}$ cannot generally be done in a straightforward way (see e.g.\  \cite{KachanovSevostianov,Ponson2023}). Therefore, if heterogeneous toughness is to be incorporated via homogenisation, further investigation is needed to develop effective strategies. Heterogeneity of the toughness can also lead to secondary consequences of importance, such as crack deflection \cite{ZENG2017235}, and the impact on transport of dissolved CO$_2$ within fractures related to carbon-storage (and therefore the potential for CO$_2$ leak-off) \cite{ChenCO22024}. Homogenization of elastic properties for more complicated media such as highly anisotropic rocks, the effects of inelasticity or nonlinear elastic behavior, are beyond the scope of this study.

Concerning homogenization of the fracture toughness in the context of HF, a general discussion was given in the introductions of \cite{Gaspare2022,OYEDOKUN2017351}. There is also the recent work of \cite{Sarmadi2024}, who proposed an approach to dealing with fracture in heterogeneous media where there was no clear distinction between rock layers utilizing a stochastic-based method, as well as the work of \cite{ARMA2021,Gaspare2022,DaFies2023} where a temporal-averaging based approach was proposed, claiming that it provided the optimal `average' toughness. 

Testing these homogenisation strategies, and the perculiarities of HF propagating in heterogeneous media, requires numerical simulations. FEM techniques (see for example \cite{LI2020167} and references therein) constitute an effective approach; we mention, in particular, the mesh-transition technique for use within an ABAQUS environment by \cite{RuedaARMA2023}, the phase-field approach for porous media was developed by \cite{Sarmadi2024}, and the sandwich-layered structure model of multi-layered rockbeds by \cite{ZHONG2024110115}.

An alternative method for modelling HF propagating in media with heterogeneous toughness is based on the `universal algorithm' approach. This was developed by \cite{Wrobel2015} in the context of the 1D fracture models (PKN, KGD, radial), before later being extended to the case of heterogeneous media as outlined in \cite{Gaspare2020}. We can mention methods of modelling  more general planar fractures in layered rocks, such as an approach using the 3D-displacement discontinuity method by \cite{LI2024121}. An alternative approach, based on incorporating conditions for initiation of crack propagation between distinct rock layers  \cite{Carlos2021,Peruzzo2024b} within the level-set method was given by \cite{PeruzzoThesis}. This has been applied to analyses of the effect of rock heterogeneity on crack arrest in the case of buoyant fractures \cite{Mori2024}. 
It should be noted that \cite{WANG2024104000} suggested the system asymptotics, which are utilized by these approaches when obtaining solutions\footnote{Both due to their high correspondence with the full solution and to evaluate the singular elasticity equation \eqref{inverse_KGD_0}, see \cite{Savitski2002}.} need to be updated, since the solutions may lose accuracy in cases of cracks propagating between material layers. They have suggested that an additional 'intermediate' asymptotic solution  needs to be utilized, in addition to the crack-tip and far-field asymptotics.

Experimental investigations have also been conducted. For example, HF propagating along weak interfaces was examined by \cite{Stanchits2015}, where the fracture path was found to be neither symmetrical nor planar. {Investigations by \cite{SAPAY2025110856} into fracture in concrete demonstrated that moisture content plays a key role in determining material toughness, with wetting of the medium delaying crack initiation (due to viscous forces between the crack faces) but leading to higher velocity fracture growth once propagation initiates. They also found that the measured (effective) material toughness increases with the fracture length, which they interpreted as demonstrating increasing resistance of the crack to further growth.} Other experimental results for materials with heterogeneous toughness  \cite{SHI2023211673} confirmed that the {presence} of heterogeneities promotes more tortuous fracture paths. Also, a difference between near-wellbore and far-field behaviour was observed, particularly regarding fracture arrest and diversion (around inhomogeneities), hypothesising that they were related to the elastic energy available over the fracture length. 

We note that analyses of the fracture energy can be completed utilizing existing formulations for the energy balance during HF propagation. This was derived for a planar hydraulic fracture with fluid lag by \cite{LECAMPION20074863}, and extended to the case of multiple hydraulic fractures by \cite{BUNGER20131538}. Very recently, an investigation of the energy balance in a hydraulic fracture at depth was provided in \cite{Peruzzo2024a}. However, these works assumed homogeneous fracture toughness; the effect of its heterogeneity has yet to be examined.

The present work carries out this investigation of the energy balance for HF in heterogeneous media, while also extending the results from \cite{Gaspare2022}. That paper focused on strategies for evaluating the average toughness, and as such neglected discussion of processes not involved in the averaging procedure. Here, we provide an overview of peculiar features of hydraulic fracture propagation in heterogeneous materials, concentrating on those that are extremely relevant for
HF modelling. Results obtained for the KGD model using the ‘universal-solver’ algorithm \cite{Gaspare2020} are analysed, with particular focus on the crack geometry, energy balance, crack/ fluid velocities, and the limiting behaviour of the propagation regime.

The paper is organised as follows. The problem formulation is introduced in Sect.~\ref{Sect:ProbOverview}. This includes the governing equations, distribution of the periodic toughness heterogeneity, and approaches to characterising the fracture regime local to the crack tip. Limiting behaviour of this regime parameter are investigated. In Sect.~\ref{Sect:Energy}, the energy distribution during fracture of heterogeneous media is considered in detail. In Sect.~\ref{Sect:Velo} the velocity of the crack tip is investigated, followed by an examination of the fluid behaviour within the fracture. Concluding remarks are given in Sect.~\ref{Sect:Conc}. Some of the limiting analysis and figures are relegated to the Appendices for the sake of readability.

\section{The KGD problem with a periodic fracture toughness distribution}\label{Sect:ProbOverview}

\subsection{Problem formulation with periodic toughness}

\begin{figure}[t!]
	\centering
	\begin{tikzpicture}[scale = 1.2]
		\draw[black] (-4,2.5) .. controls (1,1.35) and (2,0.9) .. (2.85,0); 
		\draw [black,thick,dotted,->] (-4,0) -- (-4,3); 
		\node at (-4,3.25) {$z$}; 
		\node at (-4.35,1.3) {$Q_0$}; 
		\draw [black,->] (-4,0) -- (3.75,0); 
		\node at (4,0) {$x$}; 
		\draw[black,->] (-1,1.1) -- (-1,1.6); 
		\draw[black,->] (-0.8,1.05) -- (-0.8,1.55); 
		\draw[black,->] (-0.6,1) -- (-0.6,1.5); 
		\node at (-0.8,0.8) {$p$};  
		\draw[black,<->] (-2.2,0.05) -- (-2.2,2); 
		\node at (-2,1) {$w$}; 
		\draw[black,<->] (-3.9,-0.1) -- (2.85,-0.1); 
		\node at (-0.5,-0.4) {$l(t)$}; 
		\draw[blue,thick,->] (-3.99,2.1) -- (-3.45,2.1); 
		\draw[blue,thick,->] (-3.99,1.6) -- (-3.45,1.6); 
		\draw[blue,thick,->] (-3.99,1.1) -- (-3.45,1.1); 
		\draw[blue,thick,->] (-3.99,0.6) -- (-3.45,0.6); 
		\draw[blue,thick,->] (-3.99,0.1) -- (-3.45,0.1); 
		\node at (-3.5,1.3) {{\tiny Fluid flow}}; 
		\draw[black,thick,dotted] (-1.5,3) -- (-1.5,1.9); 
		\node at (-2.75,2.75) {{\tiny Material 1}};
		\draw[black,thick,dotted] (1,3) -- (1,1.15); 
		\node at (-0.25,2.75) {{\tiny Material 2}};
		\draw[black,thick,dotted] (3.5,3) -- (3.5,0); 
		\node at (2.25,2.75) {{\tiny Material 1}};
	\end{tikzpicture}
	\caption{Cross-section of the KGD crack, propagating through periodically distributed toughness layers.}
	\label{Fig:KGD}
\end{figure}


The Kristianovich-Geertsma-de Klerk (KGD, plane strain) model \cite{Khristianovic1955,Geertsma1969} assumes of crack situated in the plane $x\in [-l(t), l(t)]$, that is symmetrical about the midpoint $x=0$ such that only the length $x\geq 0$ needs to be considered. {The crack length} $2l(t)$ and aperture $w(t,x)$ change in time as the fracture advances (see Fig.~\ref{Fig:KGD}). The crack growth is driven by a Newtonian fluid that is injected at the point $x=0$, {with the pumping rate inducing a flux per unit height of $Q_0 (t)$.} A periodic toughness distribution is assumed, with spacial period $X$, the material toughness at the crack tip being a function of the crack length $K_{Ic}(l(t))$. {Throughout this paper we take $X=1$ for simplicity, however the results can be generalised to other spacial periods}. Further details on the toughness distribution are provided in Sect.~\ref{Sect:MaterialToughness}. Fluid leak-off into the surrounding domain, alongside any fluid-lag at the crack tip, {is} neglected.

\subsubsection{Distribution of the material toughness}\label{Sect:MaterialToughness}

{For ease of analysis, a periodic toughness distribution is assumed. Simulations performed by the authors for randomly layered toughness distributions showed the same effects and trends as reported in the rest of this paper.}

The material toughness distribution is assumed to vary between two values, $K_{Ic}\equiv K_{Ic}^{max}$ and $K_{Ic} \equiv K_{Ic}^{min}$, over {some fixed spacial period $X$. Throughout this paper we take $X=1$ for the sake of simplicity. The results are however generalisable to other spacial periods via the rescaling of the crack length as $L(t)=l(t)/X$.} 

\begin{figure}[t]
	\centering
	\includegraphics[width=0.40\textwidth]{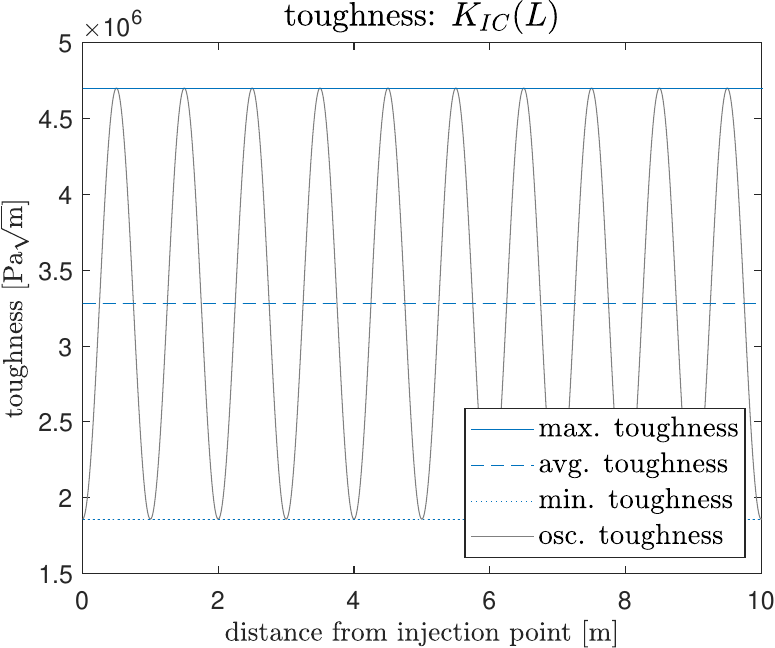}
	\put(-175,140) {{\bf (a)}}
	\put(-105,140) {{\bf Sinusoidal}}
	\hspace{12mm}
	\includegraphics[width=0.40\textwidth]{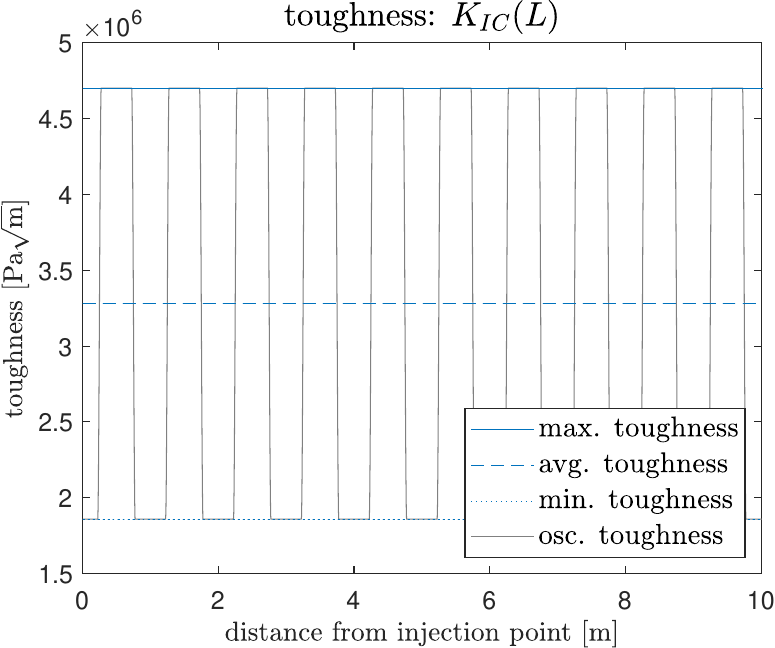}
	\put(-175,140) {{\bf (b)}}
	\put(-105,140) {{\bf Step-wise}}
	\caption{The {\bf (a)} sinusoidal and {\bf (b)} step-wise toughness distribution, shown over the spatial length $x$ of the material (with $\delta_{max}=10$, $\delta_{min}=1$, see Sect.~\ref{Sect:Delta1}).}
	\label{Fig:ToughDis1}
\end{figure}

The values of $K_{Ic}^{max}$ and $K_{Ic}^{min}$ are chosen to investigate the different fracturing regimes as discussed in Sect.~\ref{Sect:Delta}.  
We consider both a step-wise and a sinusoidal periodic distribution of the toughness (see Fig.~\ref{Fig:ToughDis1}). They represent a simple two-layered material and the case of a many-layered material (in the limit as the layer width tends to zero).  In all figures {that} follow, the sinusoidal distribution {is} shown on the left-hand side of the page, and step-wise distribution on the right-hand side. {A previous investigation for the radial model demonstrated that uneven layering of the toughness had minimal effect on results for heterogeneous hydraulic fracture \cite{DaFies2023}. Consequently, we will assume even layering in this work. }

\subsubsection{Governing equations}\label{Sect:GovEqn}

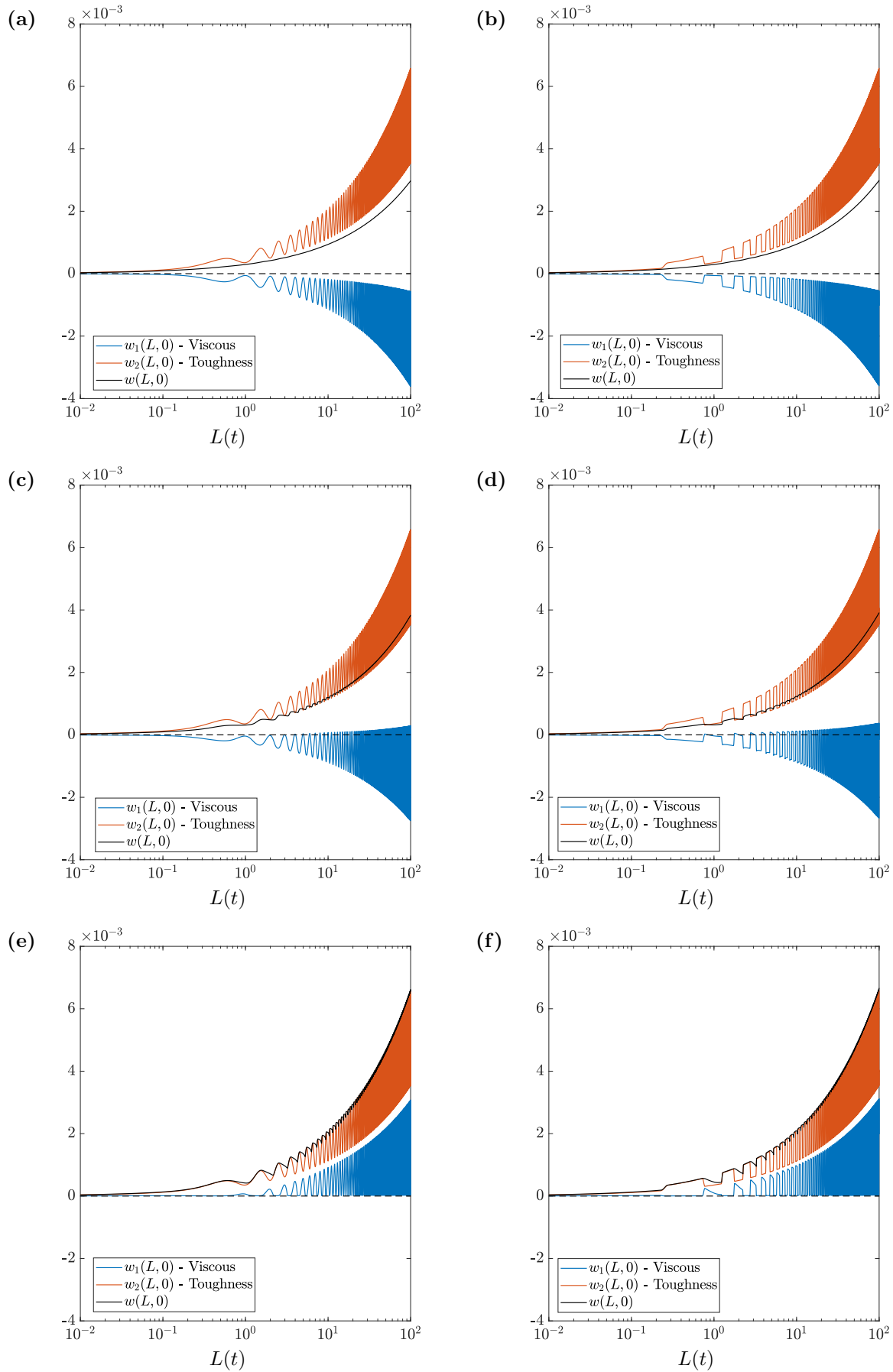
\begin{figure}[hp!]
	\centering
	\includegraphics[width=0.45\textwidth]{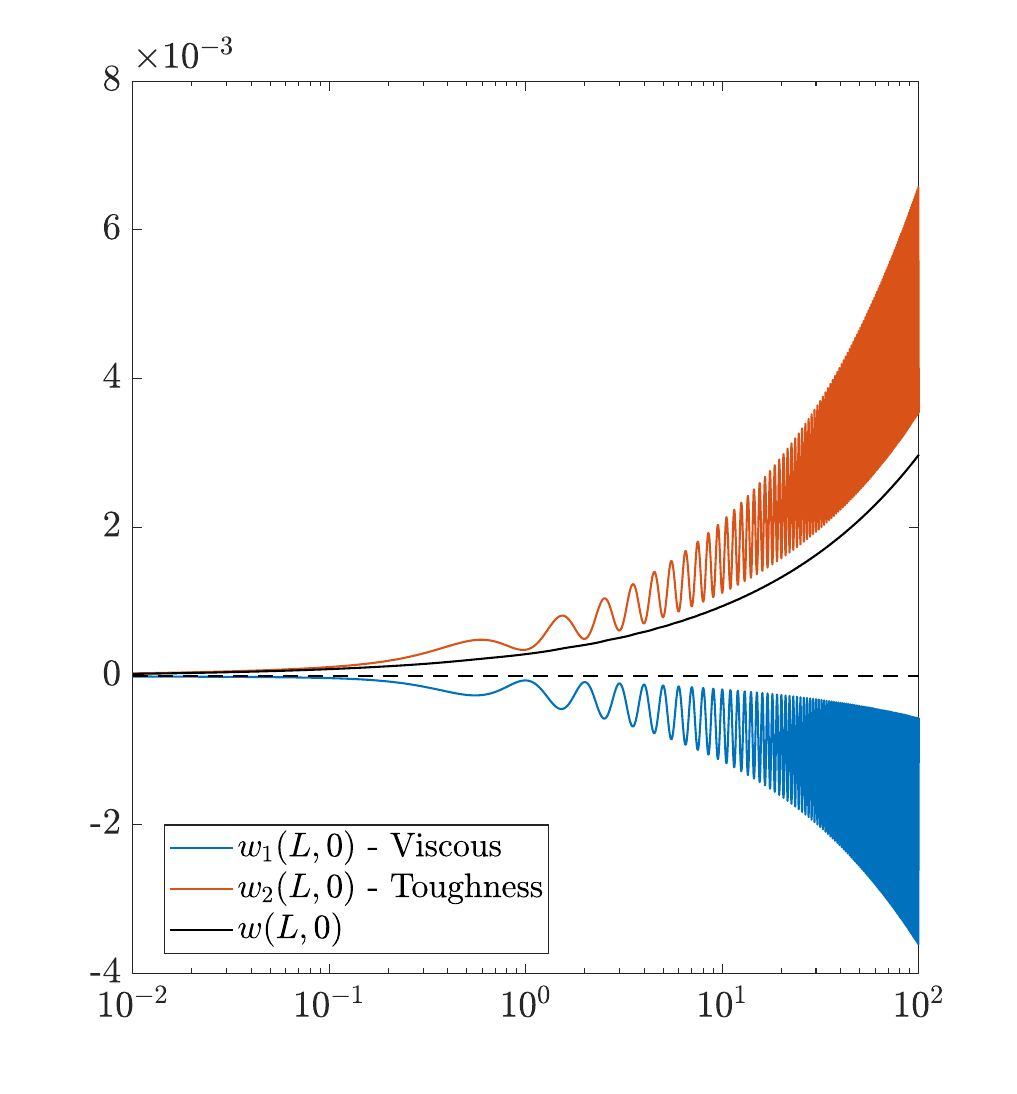}
	\put(-90,0) {$l(t)$}
	\put(-180,180) {{\bf (a)}}
	\hspace{6mm}
	\includegraphics[width=0.45\textwidth]{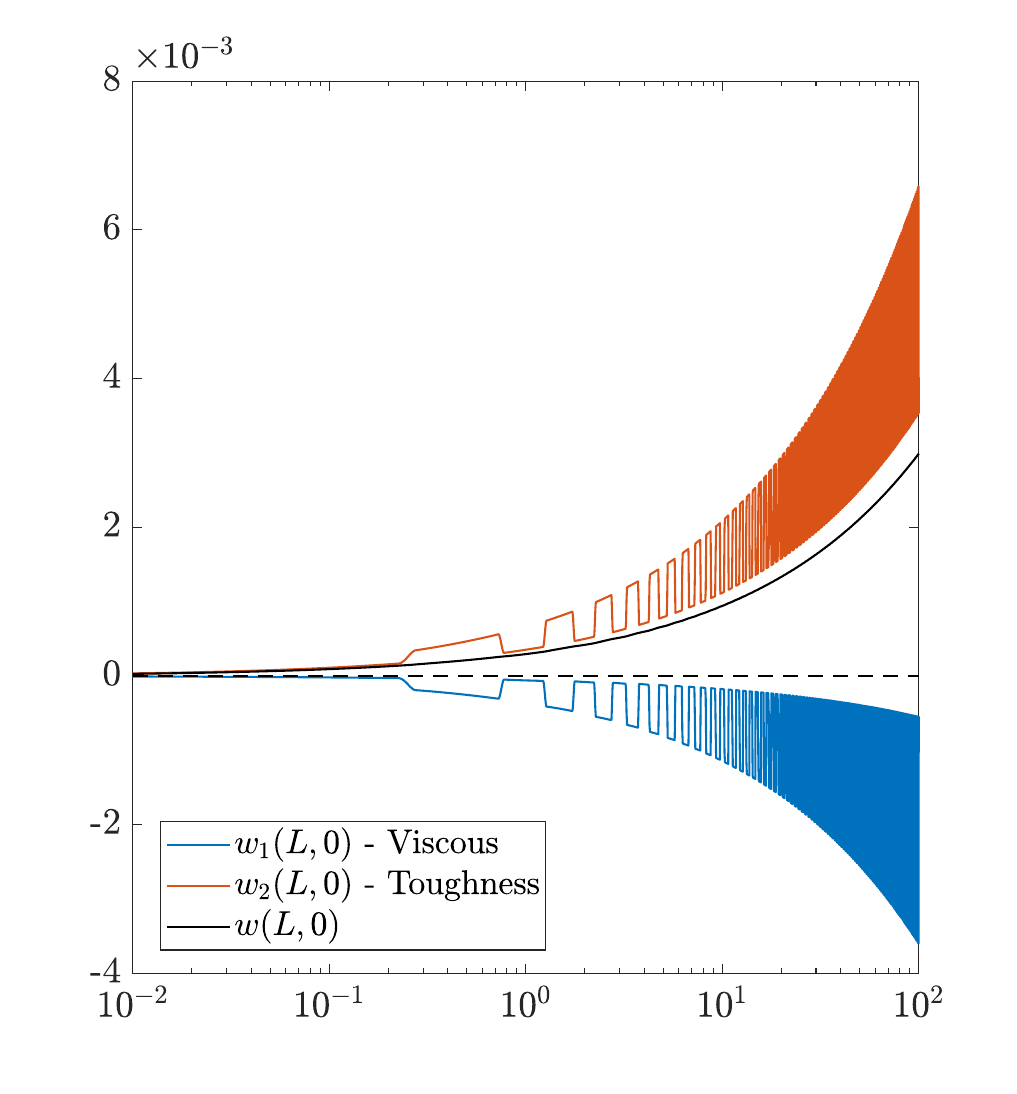}
	\put(-90,0) {$l(t)$}
	\put(-180,180) {{\bf (b)}}

	\includegraphics[width=0.45\textwidth]{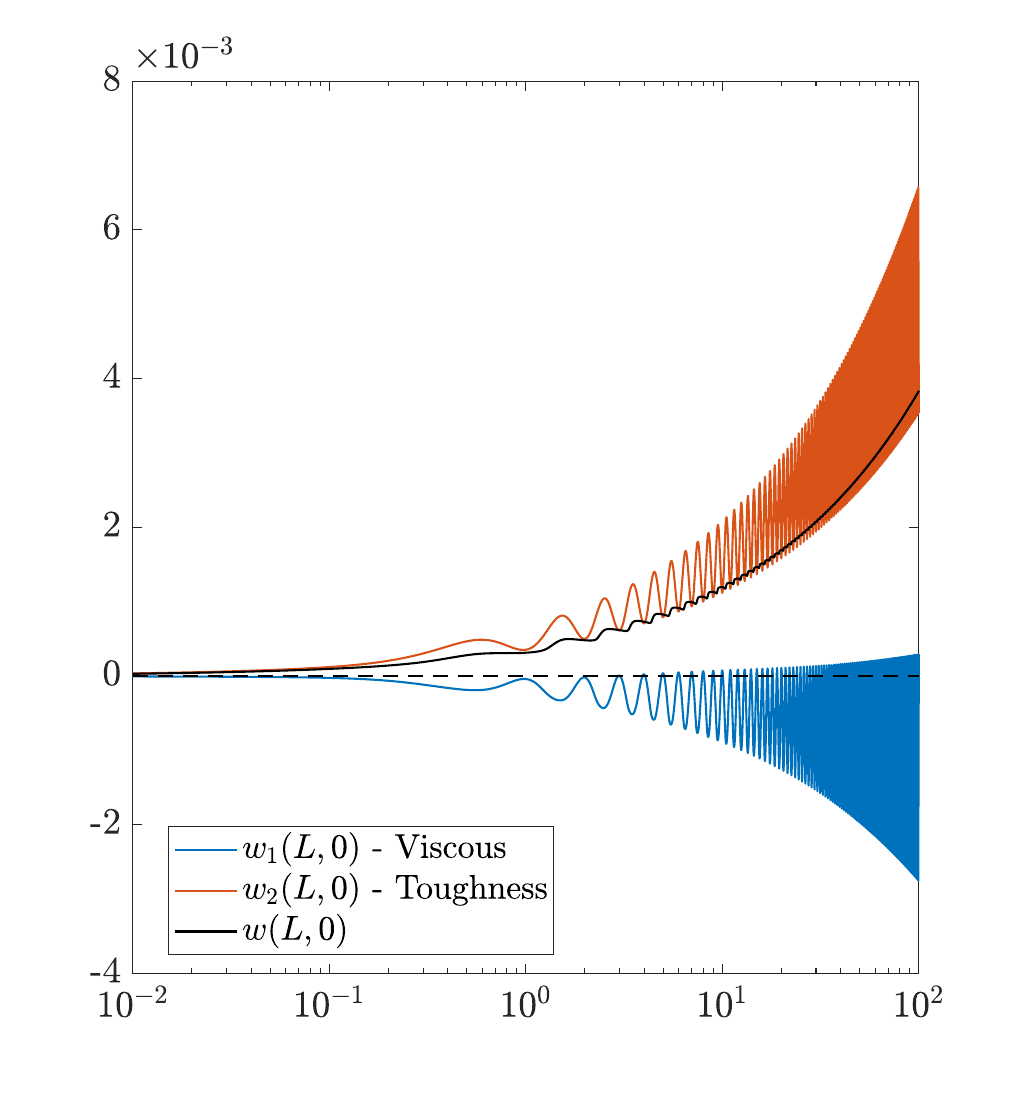}
	\put(-90,0) {$l(t)$}
	\put(-180,180) {{\bf (c)}}
	\hspace{6mm}
	\includegraphics[width=0.45\textwidth]{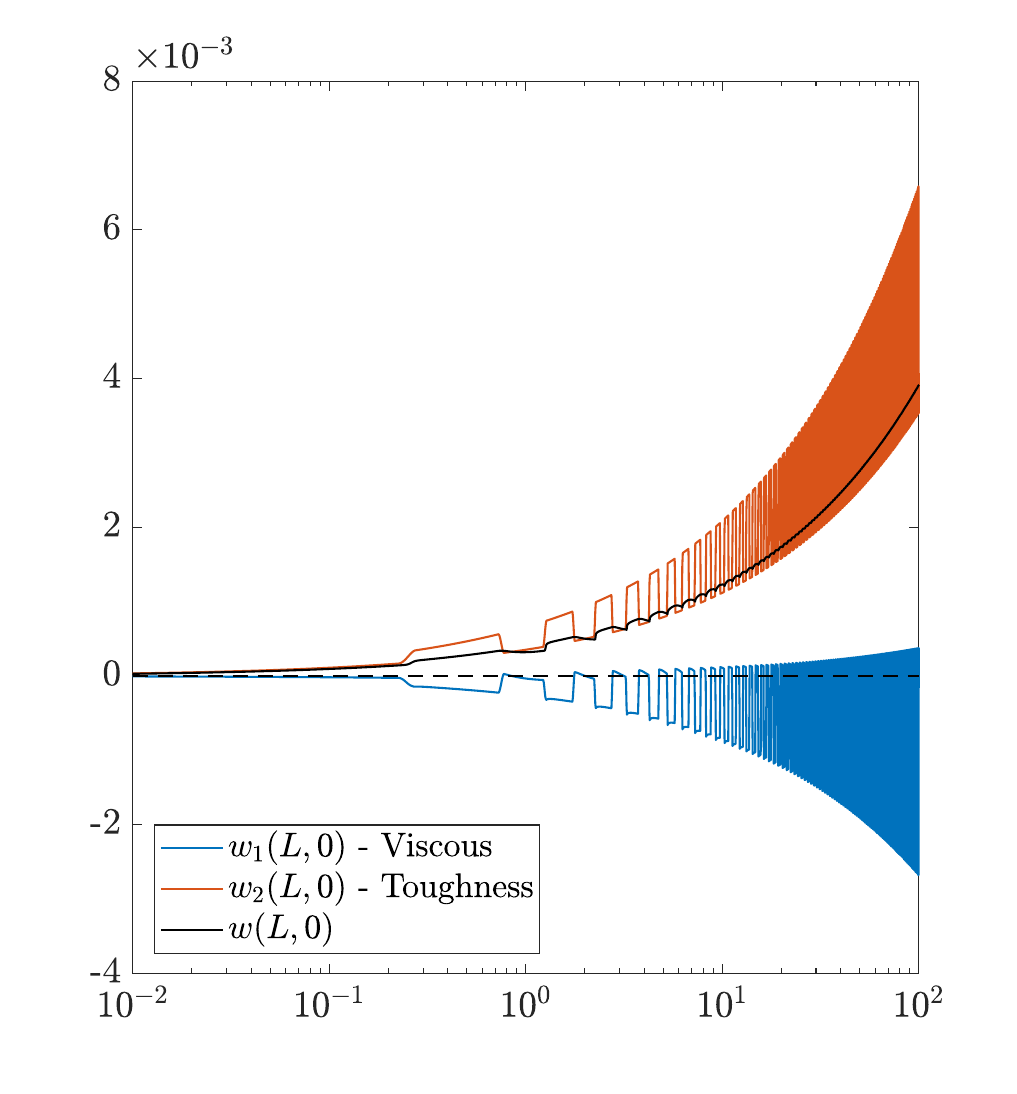}
	\put(-90,0) {$l(t)$}
	\put(-180,180) {{\bf (d)}}

	\includegraphics[width=0.45\textwidth]{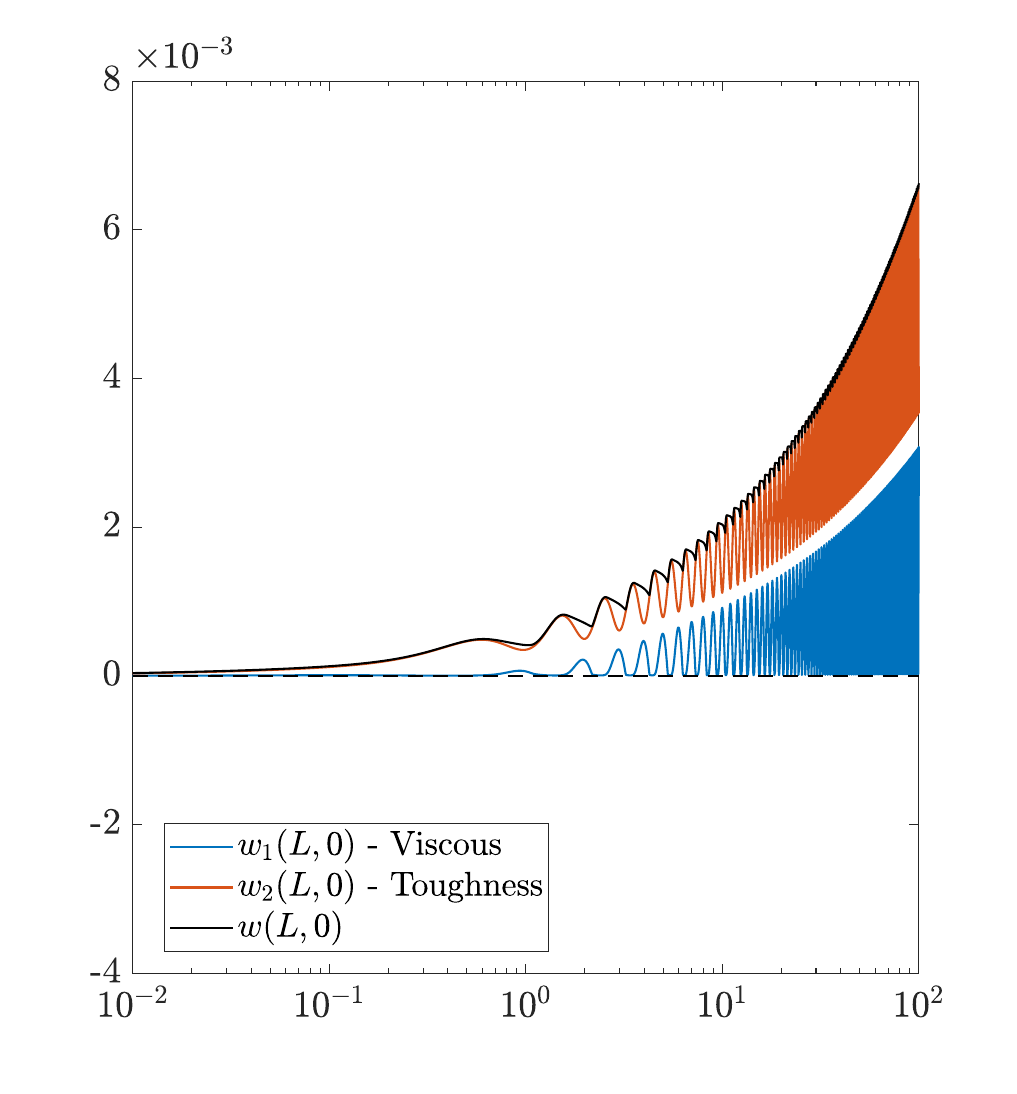}
	\put(-90,0) {$l(t)$}
	\put(-180,180) {{\bf (e)}}
	\hspace{6mm}
	\includegraphics[width=0.45\textwidth]{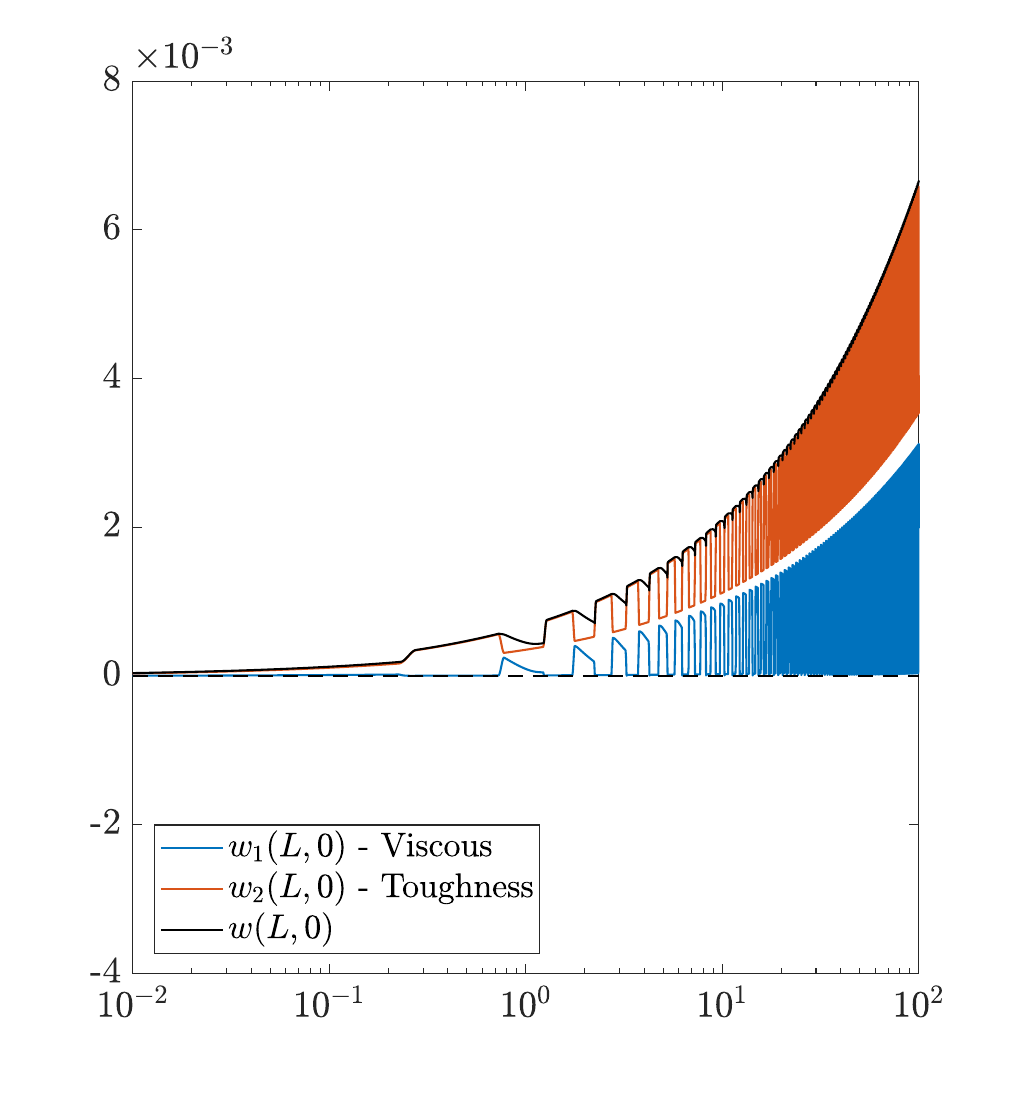}
	\put(-90,0) {$l(t)$}
	\put(-180,180) {{\bf (f)}}
	\caption{Toughness and viscosity components of the aperture \eqref{inverse_KGD_0} at $x=0$, (a), (b) $\delta_{max}=1$, $\delta_{min}=0.1$, (c), (d) $\delta_{max}=10$, $\delta_{min}=1$, (e), (f) $\delta_{max}=100$, $\delta_{min}=10$. }
	\label{Fig_w1_w2}
\end{figure}

The 1D formulations of KGD/ radial {hydraulic} fracture with varying material toughness are provided in \cite{Gaspare2022,DaFies2023} (see also references therein). Here we will only provide a summary of the main equations for the KGD model with variable toughness.\\
\noindent{\bf Continuity equation (mass conservation):}
\begin{equation}\label{continuity}
	\frac{\partial w}{\partial t}+\frac{\partial q}{\partial x}=0,\quad t > t_0,\quad 0< x < l(t),
\end{equation}
where $w(t,x)$ is the crack aperture (opening), and $q(t,x)$ is the fluid flow rate.\\
\noindent{\bf The global fluid balance equation:}\\
Obtained by integrating \eqref{continuity} over space and time:
\begin{equation} \label{fluid_balance}
	{\int_0^{l(t)} w(t,x) - w(0,x) \, dx =  \frac{1}{2}\int_0^t Q_0 (\tau ) \, d\tau ,}
\end{equation}
where $Q_0$ is the {flux per unit height}, and we have utilized \eqref{bc_0}.\\
\noindent{\bf Poiseuille equation (fluid flow):}
\begin{equation}\label{poiseuille}
	q = - \frac{1}{M'} w^{3} \frac{\partial p}{\partial x},
\end{equation}
where $p=p(t,x)$ ($p=p_{f}-\sigma_0$, with $\sigma_0$ the confining stress {and $p_f$ the fluid pressure}) is the net fluid pressure, while constant $M'$ is the modified dynamic viscosity $M'= 12 \mu$. \\
\noindent{\bf Inverse elasticity equation (solid-fluid coupling):}
\begin{equation}
	\label{inverse_KGD_0}
	w(t,x)=\underbrace{-\frac{4}{\pi {E^\prime}}\int_0^{l(t)} \frac{\partial p(t,\eta)}{\partial \eta} K\left(\frac{\eta}{l(t)}, \frac{x}{l(t)}\right)d\eta}_{w_1(t,x)}+
	\underbrace{\frac{4 K_{Ic}(l(t))}{\sqrt{\pi l(t)} {E^\prime}}\sqrt{l^2(t)-x^2}}_{w_2(t,x)}    ,
\end{equation}
where $E^\prime = E / (1 - \nu^2)$, with $E$ the Young's modulus and $\nu$ the Poisson's {ratio, while $K_{IC}(l(t))$ is the material toughness (see Sect.~\ref{Sect:MaterialToughness}) local to the crack tip. Meanwhile, the kernel $K(\eta,x)$ is given by (see \cite{Wrobel2015})
\[
K\left(\eta, x\right) = \left(\eta - x\right) \ln\left| \frac{\sqrt{1-x^2} + \sqrt{1-\eta^2}}{\sqrt{1-x^2} - \sqrt{1-\eta^2}} \right| + x \ln\left(\frac{1+\eta x+\sqrt{1-x^2}\sqrt{1-\eta^2}}{1+\eta x-\sqrt{1-x^2}\sqrt{1-\eta^2}} \right).
\]
Note that in \eqref{inverse_KGD_0} the} term $w_1(t,x)$ describes the effect of the fluid viscosity, while $w_2(t,x)$ describes the impact of the material toughness, with the decomposition shown in Fig.~\ref{Fig_w1_w2}.\\

\noindent{\bf Stress intensity factor:}
\begin{equation}
	\label{K_I}
	K_{I}(t)= 2\sqrt{\frac{l(t)}{\pi}} \int_0^{l(t)} \frac{p(t,s)ds}{\sqrt{l^2(t)-s^2}},
\end{equation}
where $K_I (t)$ is the stress intensity factor.\\
\noindent{\bf Irwin criterion (fracture criterion from Linear Elastic Fracture Mechanics):}
\begin{equation}
	\label{K_I_criterion}
	K_{I}(t)=K_{IC}(l(t)),\quad t\ge0,
\end{equation}
\noindent{\bf Crack opening boundary conditions:}\\
\noindent {Due to symmetry, the flux per unit height $Q_0$ in one direction is given by}
\begin{equation}\label{bc_0} 
	q(t,0)= \frac{1}{2}Q_0(t).
\end{equation}
In addition, the form of operator \eqref{inverse_KGD_0} implies that:
\begin{equation}
	\label{bc_p} \frac{\partial w}{\partial x}(t,0)=0.
\end{equation}
\noindent{\bf Crack tip boundary conditions:}
\begin{equation}\label{bc_1}
	w(t,l(t))=0, \quad q(t,l(t))=0.
\end{equation}
\noindent{\bf Fluid velocity:}\\
In addition to these equations we introduce the fluid velocity, $v$, as a new parameter. It describes the average speed of fluid flow through the fracture cross section. This follows from \eqref{poiseuille} as:
\begin{equation}\label{particle_velocity}
	v(t,x)=\frac{q(t,x)}{w(t,x)}=-\frac{1}{M'} w^{2} \frac{\partial p}{\partial x}.
\end{equation}
Provided the fluid front coincides with the crack tip (i.e.\ no fluid lag and finite fluid leakoff at the tip), the {\it speed equation} holds \cite{Linkov2011}
\begin{equation}\label{SE}
	\frac{dl}{dt}=v(t,l(t))=-\frac{1}{M'} w^{2} \frac{\partial p}{\partial x}\bigg|_{x=l(t)}.
\end{equation}
The advantages of applying this in computations alongside the system asymptotics are well documented, see e.g. \cite{Detournay2014,Linkov_3,Wrobel2015}. Crucially for the present analysis, this allows for highly accurate tracing of the fracture front while also ensuring that the fracture velocity is computed as a component of the solution.\\

\subsubsection{Numerical solver for hydraulic fracture of a heterogeneous material}\label{Sect:Solver}

This system of equations is solved utilizing a time-space solver developed by the authors, which is of ``universal algorithm'' type, see e.g. \cite{Wrobel2015,Perkowska2015,Peck2018a}. The particular algorithm utilized in this paper is the same as that utilized in \cite{Gaspare2022,DaFies2023}, while a detailed overview of its construction can be found in \cite{Gaspare2020}. The algorithm has been demonstrated to achieve an exceptionally high level of solution accuracy, with the algorithm being computed to a specified level of global solution error. Throughout this paper, the key computational parameters $w$, $v$, {$\frac{\partial p}{\partial x}$}, have a global error below $10^{-4}$ (see e.g.\ the energy balance relative error presented in Fig.~\ref{Fig_Energy_Balance_Local}). Note however that the local error may be higher, particularly for high toughness ratios and the step-wise distribution. 
More accurate computations were performed where necessary. Checks have been performed to ensure that datasets are computed to a sufficient level of accuracy to verify the stated results. For full details see \cite{Gaspare2020}.

\subsubsection{Material parameters used in computations}

The material parameters used in simulations correspond to those typically encountered in hydraulic fracturing of rock (see \cite{Gaspare2022} and references therein). They are provided in Table.~\ref{Table:parameters}.

\begin{table}[t]
	\centering
	\begin{tabular}{c|c|c|c}
		$E$ & $\nu$ & $\mu$ & $Q_0 $ \\
		\hline \hline
		&&&\\
		$2.81 \times 10^{10}$ \, [Pa] & $0.25$ & $1 \times 10^{-3}$ \, [Pa s] &  $4.41 \times 10^{-3}$ \, [m$^2$ / s] \\
		&&&\\
		\hline \hline
	\end{tabular}
	\caption{Problem parameters used in simulations. {Material constants are taken inline with those expected for hydraulic fracturing operations in shale rock.} Note that the {flux per unit height} $Q_0 $ is taken to be constant.}
	\label{Table:parameters}
\end{table}

\subsection{Monitoring the local fracture regime}\label{Sect:Delta}

The regime of the fracture determines the extent and character of crack propagation (for more details see e.g. \cite{Garagash2011,Lecampion2017} and references therein). In its simplest description (and in absence of fluid leak-off), hydraulic fracture propagation is either fluid viscosity or material toughness dominated, with a transient regime in between. Unlike in conventional models, those incorporating heterogeneous toughness may experience ``rock-layer''-dependent propagation regimes, which vary rapidly as the crack advances, see e.g. \cite{Gaspare2022,DaFies2023}. 

Examining the behaviour of such fractures therefore requires outlining a measure with which to estimate the regime of fracture propagation at any given moment. We utilize the measure, denoted $\delta(t)$, first introduced in \cite{Gaspare2022} to characterise the regime local to the crack tip. As it was investigated in more details there, below we focus only on re-introducing it and briefly discussing its behaviour. 

\subsubsection{Parameter $\delta$}\label{Sect:Delta1}

\begin{figure}[hp!]
	\centering
	\includegraphics[width=0.45\textwidth]{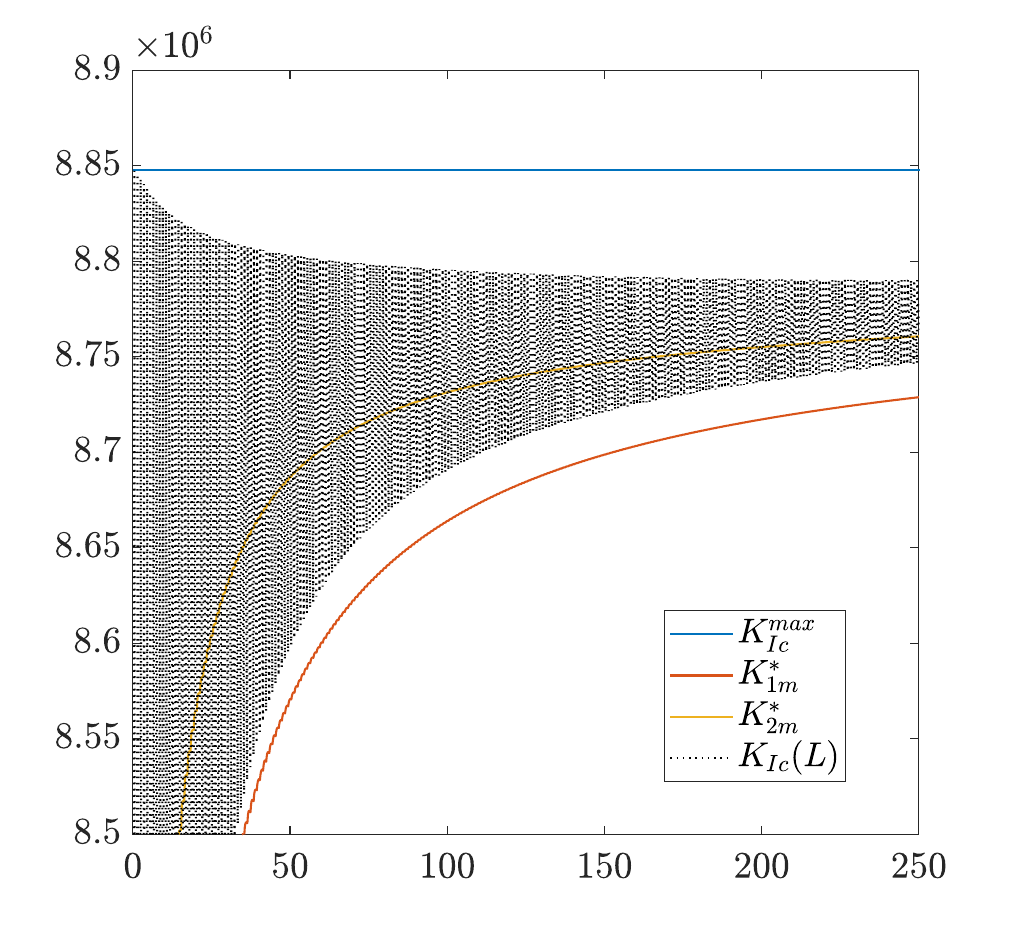}
	\put(-92,-3) {$l(t)$}
	\put(-195,85) {$C(l)$}
	\put(-180,160) {{\bf (a)}}
	\hspace{6mm}
	\includegraphics[width=0.45\textwidth]{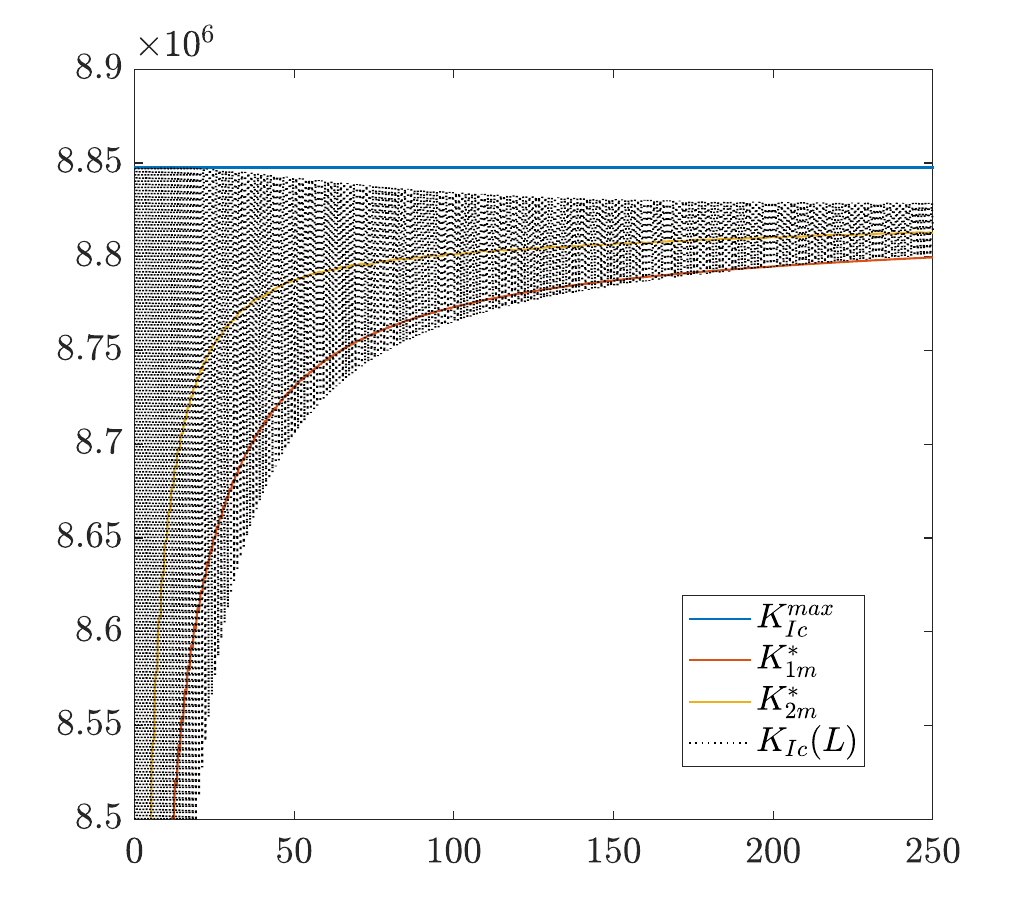}
	\put(-92,-3) {$l(t)$}
	\put(-195,85) {$C(l)$}
	\put(-180,160) {{\bf (b)}}
	
	\includegraphics[width=0.45\textwidth]{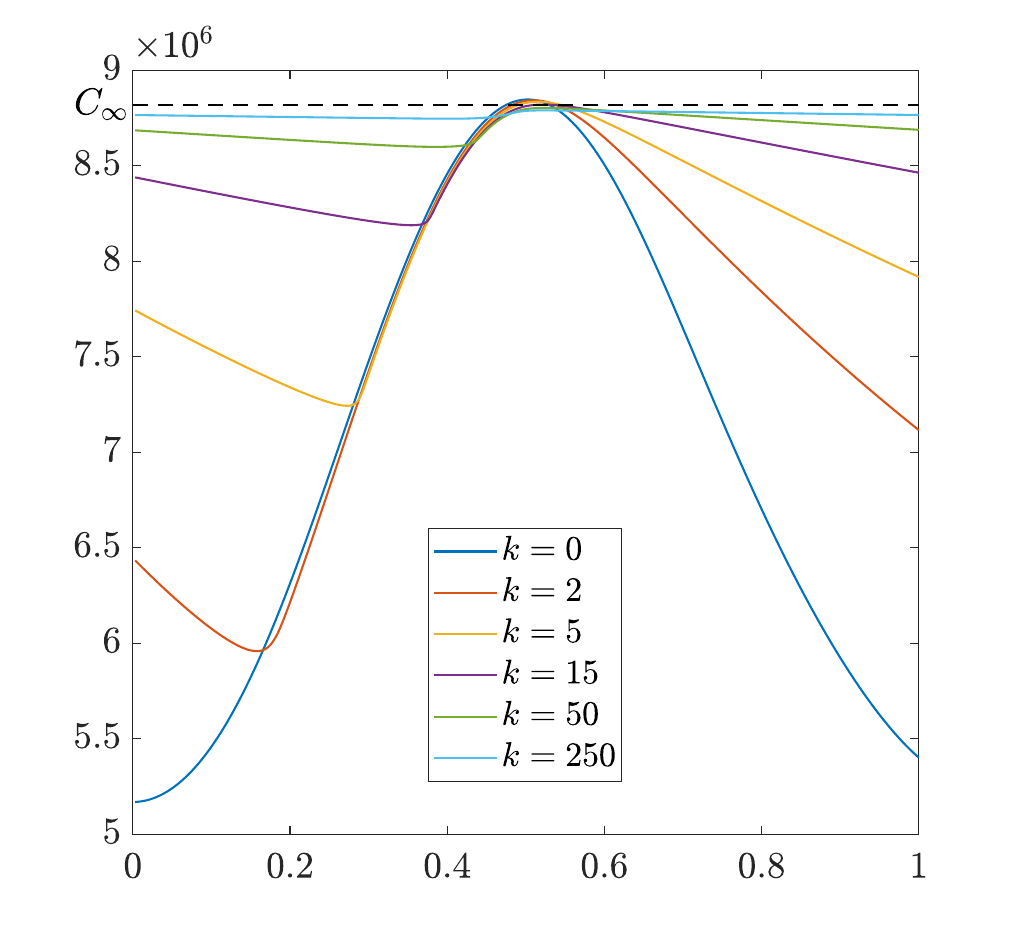}
	\put(-95,-3) {$l - k$}
	\put(-195,85) {$C(l)$}
	\put(-180,160) {{\bf (c)}}
	\hspace{6mm}
	\includegraphics[width=0.45\textwidth]{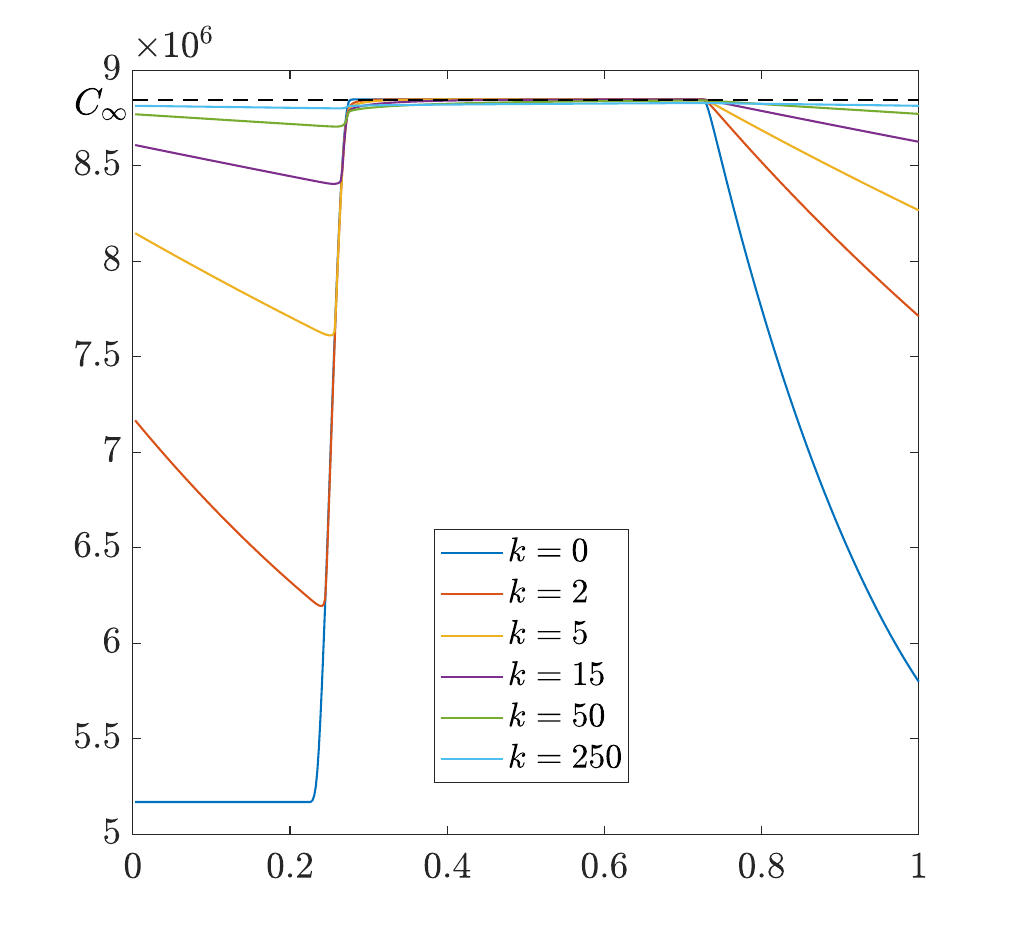}
	\put(-95,-3) {$l - k$}
	\put(-195,85) {$C(l)$}
	\put(-180,160) {{\bf (d)}}

	\includegraphics[width=0.45\textwidth]{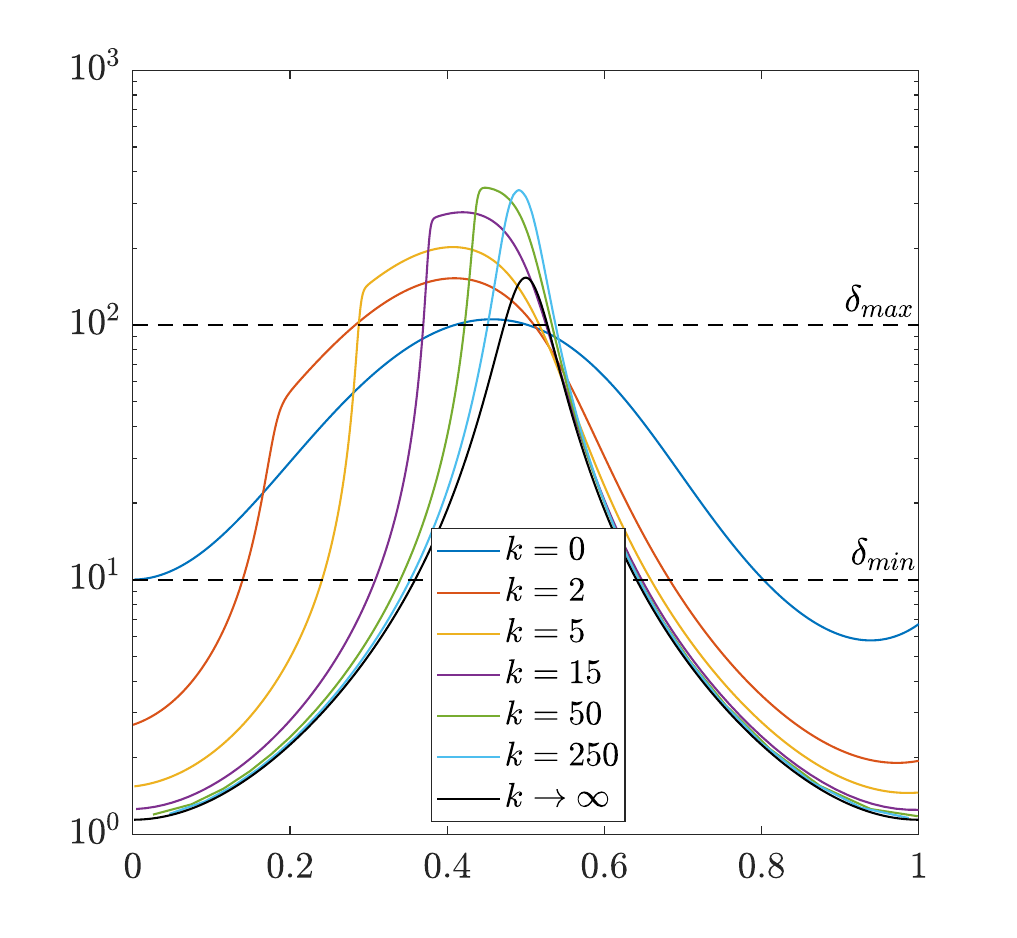}
	\put(-95,-3) {$l - k$}
	\put(-195,85) {$\delta(l)$}
	\put(-180,160) {{\bf (e)}}
	\hspace{6mm}
	\includegraphics[width=0.45\textwidth]{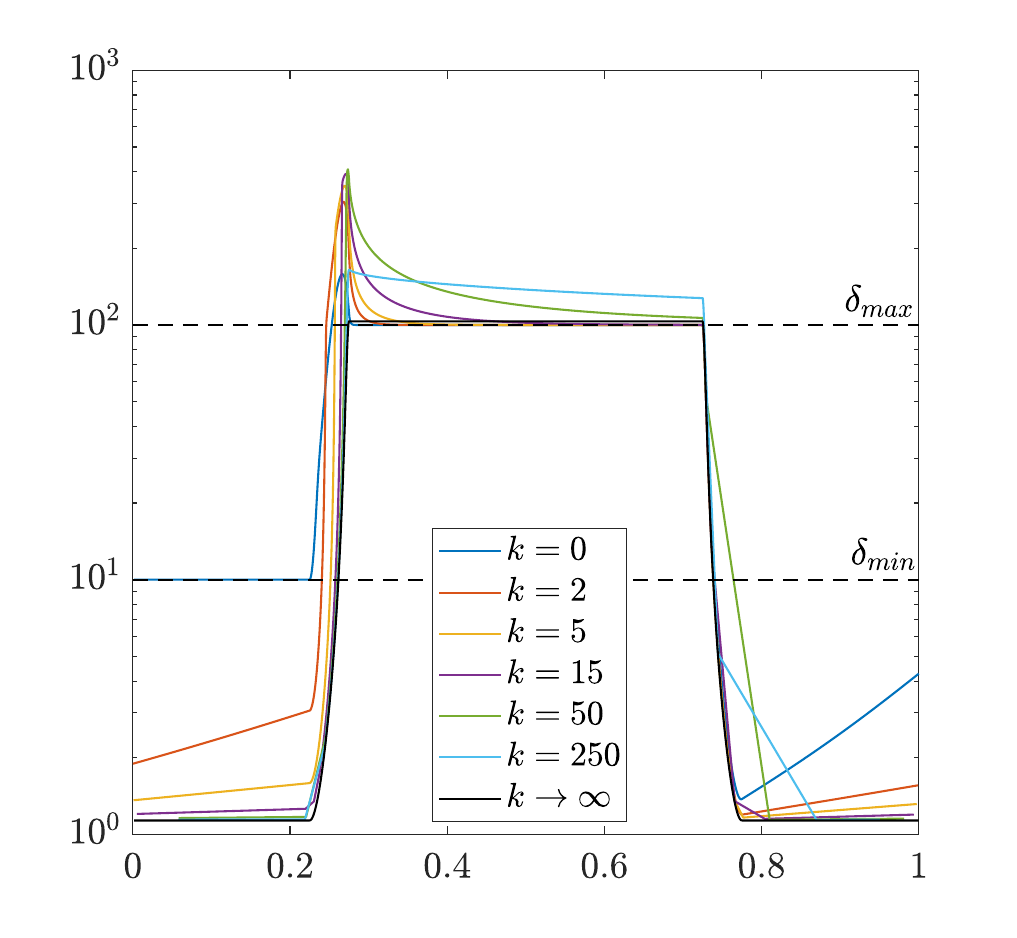}
	\put(-95,-3) {$l - k$}
	\put(-195,85) {$\delta(l)$}
	\put(-180,160) {{\bf (f)}}
	\caption{Behaviour of (a), (b) $C(l)$ over $l(t)$ (see \eqref{local_indicator}), and (c), (d) $C(l)$ and (e), (f) $\delta (l)$  against the value of $l(t)$ within the $k^{\text{th}}$ toughness period {of length $X=1$} for the (a), (c), (e) sinusoidal, (b), (d), (f) piece-wise toughness distribution. Data is taken for the toughness-toughness case ($\delta_{max}=100$, $\delta_{min}=10$). The (assumed constant) limiting value $C\to C_\infty$ {as} $l\to\infty$ is obtained using least-squares, with the limiting behaviour of $\delta$ following from \eqref{local_indicator_infty}. The parameters $K_{1m}^*$, $K_{2m}^*$ are temporal-averaged values of the toughness, obtained using the formulae \eqref{temporal_averages} from \cite{Gaspare2022}.}
	\label{fig:CL_infty}
\end{figure}
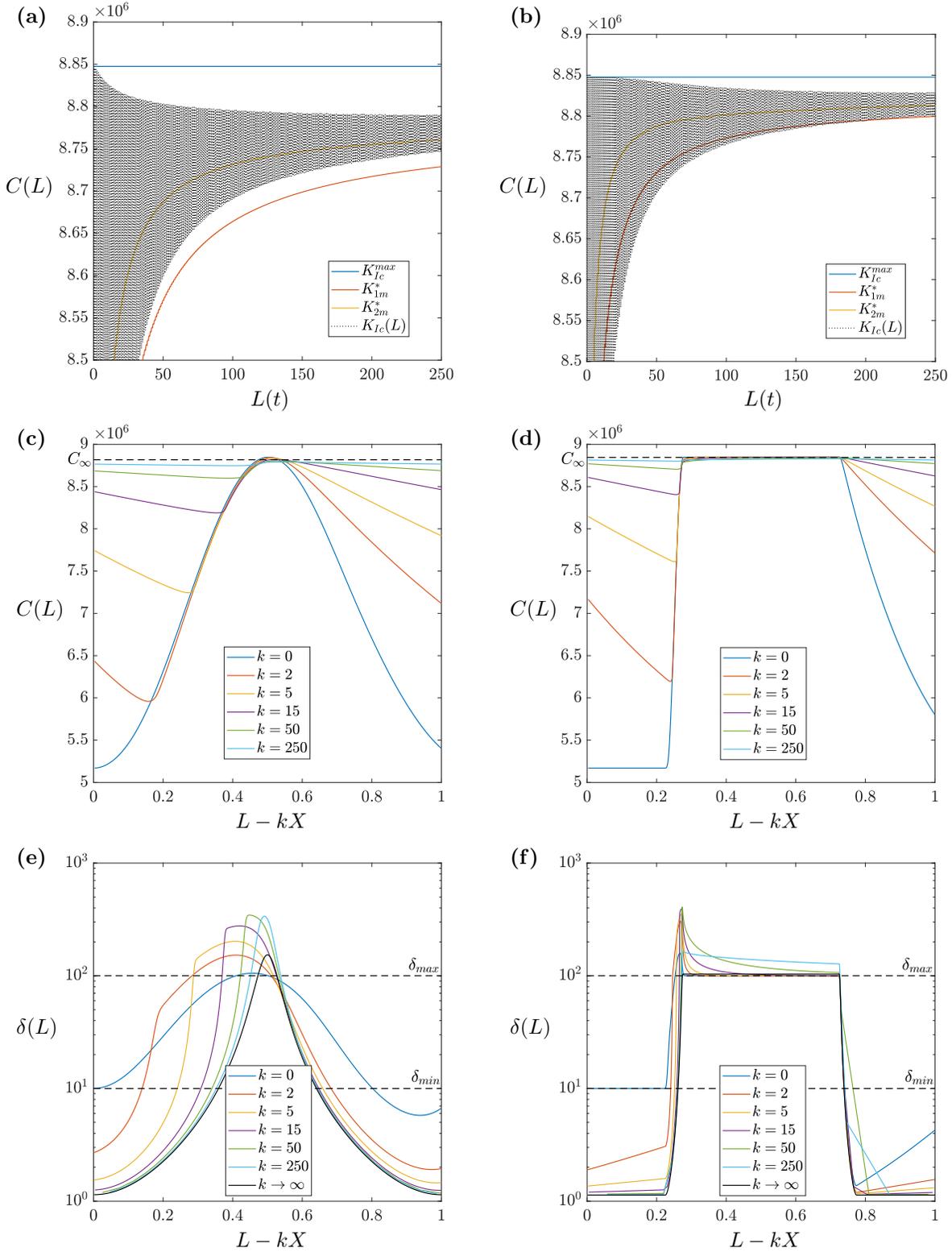

As noted in Sect.~\ref{Sect:GovEqn}, the elasticity equation \eqref{inverse_KGD_0} can be expressed as
\begin{equation}
	\label{inverse_KGD_1}
	w(t,x)=w_1(t,x)+
	w_2(t,x),
\end{equation}
where term $w_1(t,x)$ describes the effect of the fluid viscosity, while $w_2(t,x)$ describes the impact of the material toughness on the fracture profile (shown graphically in Fig.~\ref{Fig_w1_w2}). Let us examine the fracture volume, described as
\begin{equation}
	\label{volume_KGD}
	{V(t)= \int_0^{l(t)}w(t,x)dx,\quad V_j(t)=\int_0^{l(t)}w_j(t,x)dx,\quad j=1,2,}
\end{equation}
$$
V(t) = V_1 (t) + V_2 (t).
$$
In the case of a constant injection rate ($Q_0=const$), and where the crack starts 
from zero length, two of these terms can be computed explicitly (utilizing \eqref{fluid_balance})
\begin{equation}
	\label{volume_KGD}
	{V(t)=Q_0t,\quad 
	V_2(t)=\frac{\sqrt{\pi l(t)}}{E^\prime} l(t) K_{Ic}(l(t)).}
\end{equation}
This allows us to introduce the following measure,  first outlined in \cite{Gaspare2022}, which is given by the ratio of the fluid and toughness associated volumes
\begin{equation}
	\label{local_indicator}
	\delta(t)=\frac{V_2(t)}{V_1(t)}=\frac{V_2(t)}{V(t)-V_2(t)}=\frac{K_{Ic}(l(t))}{C(t)-K_{Ic}(l(t))},\quad C(t)=
	{\frac{Q_0 E^\prime}{\sqrt{\pi}}}\frac{t}{l^{3/2}(t)},
\end{equation}
{which} describes the local propagation regime at the crack tip: $\delta\ll 1$ corresponds to the viscosity dominated (small toughness) regime, while $\delta\gg 1$ indicates toughness dominated propagation. Note however that this is only intended as an approximate measure. The new parameter $C(t)$, shown in Fig.~\ref{fig:CL_infty}a,b, is useful for modelling and investigating limiting behaviour as $t\to\infty$, as will be discussed in the next section. The values of $\delta(l(t)))$, $C(l(t))$, as the crack propagates through various toughness periods are given in Fig.~\ref{fig:CL_infty}.

To account for the fracturing regime in our investigations, we take the material toughness' maximum, $K_{Ic}^{max}$, and minimum, $K_{Ic}^{min}$, values to correspond to specific values of $\delta$ (computed for a homogeneous material). These values are given in Table.~\ref{Table:delta}. For example, taking $K_{Ic}^{max}$ corresponding to $\delta = 10$, and $K_{Ic}^{min}$ corresponding to $\delta = 0.1$, we can investigate a material whose layers alternate between the Toughness and Viscosity dominated regimes respectively\footnote{Note that while $\delta=10$ may produce a purely toughness-dominate fracture for homogeneous media, in the heterogeneous case viscosity-dominated effects may occur when propagating through that layer. Similarly, the layer with $\delta=0.1$ may exhibit toughness-dominated behaviour. See e.g. the behaviour of $\delta(l)$ in Fig.~\ref{fig:CL_infty}c,d.}. Throughout the paper, we denote the toughness pairing using $\delta_{max}$, $\delta_{min}$, for the sake of simplicity. 

\begin{table}[t]
	\centering
	\begin{tabular}{|l||c|c|c|c|}
		\hline \hline
		${\delta} $ & $0.1$ & $1$ & $10$ & $100$ \\
		\hline
		$K_{Ic}$ [Pa $\sqrt{\text{m}}$] & $3.260 \times 10^{5}$ & $1.860 \times 10^{6}$ & $4.700 \times 10^{6}$ & $8.760\times 10^{6}$  \\
		\hline \hline
	\end{tabular}
	\put(-372,25) {{\bf (a)}}
	
	\vspace{4mm}
	
	\begin{tabular}{| l |c | c | c|}
		\hline \hline
		{\bf Regime} & Intermediate - Viscosity & Toughness - Intermediate & Toughness - Toughness \\
		\hline
		& $\delta_{max} = 1$, $\delta_{min} = 0.1$ & $\delta_{max} = 10$, $\delta_{min} = 1$ & $\delta_{max} = 100$, $\delta_{min} = 10$ \\
		\hline \hline
	\end{tabular}
	\put(-425, 25) {{\bf (b)}}
	\caption{{\bf (a)} Values of the material toughness $K_{Ic}$ (to $3$ {significant figures}) corresponding to the given value of the ratio $\delta$ \eqref{local_indicator}, for a rock whose other material parameters are as stated in Table.~\ref{Table:parameters}. {\bf (b)} Values of $\delta_{max}$ and $\delta_{min}$ (corresponding to $K_{Ic}^{max}$ and $K_{Ic}^{min}$) used in computations to model different regime combinations.}
	\label{Table:delta}
\end{table}

\subsubsection{Behaviour as $l\to\infty$}

{The limiting case of an infinite-length crack offers valuable insights regarding appropriate homogenisation strategies for the material toughness. Most notably: (i) it provides the limiting value that any homogenisation procedure should tend towards as $l\to\infty$, (ii) it highlights which heterogeneity-induced effects are only prevalent over short length-scales, which allows for the construction of simplified 'short-length' or 'long-length' approximations. For this reason, a brief investigation of this limiting case is provided below.}

{It was previously} demonstrated that the solution for heterogeneous toughness will tend to the maximum toughness solution as the crack length tends to infinity \cite{DONTSOV2021108144,Gaspare2022}. Combining this with results for the long-time asymptotics (see e.g. \cite{Garagash2011} and references therein), as $t\to\infty$:
\begin{equation}
	\label{mafia_length}
	l(t)=l_\infty(Q_0, \mu, E^\prime, K_{IC}^{\max})t^{2/3},
\end{equation}
for some $l_\infty$, which depends upon the process parameters.

Assuming {that} the function $C(t)$ tends to some constant limiting value ${C(t)}\to C_{\infty}$ as $t\to\infty$ (for discussion of this assumption, see Appendix.~A), then combining this representation with the definition of $\delta$ \eqref{local_indicator} we obtain the limiting behaviour as $t\to\infty$
\begin{equation}
	\label{local_indicator_infty}
	\delta_\infty(t)=\frac{K_{Ic}\left(l(t)\right)}{C_\infty-K_{Ic}\left(l(t)\right)}. 
\end{equation} 
It is clear that we must have $C_\infty > K_{Ic}^{max}$, while in the infinite toughness limit $C_\infty \to K_{Ic}^{max}$ as $K_{Ic}^{max}\to\infty$. We also have that, for two periodic toughness distributions $K^{(1)}_{IC}(x)$ and $K^{(2)}_{IC}(x)$ such that $K^{(1)max}_{IC}=K^{(2)max}_{IC}$, and 
$K^{(1)}_{IC}(x)\ge K^{(2)}_{IC}(x)$, then $\delta_\infty^{(1)max}=\delta_\infty^{(2)max}$ and $\delta_\infty^{(1)}(x)\ge\delta_\infty^{(2)}(x)$.

The behaviour of $C(t)$ and $\delta(t)$, alongside the limiting behaviour as $t\to\infty$, are provided in Fig.~\ref{fig:CL_infty}. The limiting value $C_\infty$ is also shown, with the values for various toughness distributions provided in Table.~\ref{Table_Cinfty}, alongside those for the heterogeneous case with $K_{Ic}(l(t))\equiv K_{Ic}^{max}$. {The values of $C_\infty$ are obtained using a least-squares approximation, the details of which are provided in Appendix.~A, with the behaviour of $\delta_\infty$ following from \eqref{local_indicator_infty}. We also include the values predicted by the temporal-averaged toughness $K_{Ic}(l(t))\equiv K_{1,2m}^*$ from \cite{Gaspare2022}, which are given by
\begin{equation}  \label{temporal_averages}
\begin{split}
K_{1m}^* (l, dl) = \left( \int_l^{l+dl} \frac{d\xi}{v_0 (\xi)} \right)^{-1} \int_l^{l+dl} K_{Ic}(\xi) \frac{d\xi}{v_0 (\xi)} , \\[4mm]
K_{2m}^* (l,dl) = \sqrt{ \left( \int_l^{l+dl} \frac{d\xi}{v_0 (\xi)} \right)^{-1} \int_l^{l+dl} K_{Ic}^2(\xi) \frac{d\xi}{v_0 (\xi)} },
\end{split}
\end{equation}
where $v_0 (l(t))$ is the crack-tip velocity, and we take $dl=X=1$ in all cases. The limiting value is obtained using the same least-squares approach as for $C_\infty$.}

\begin{landscape}
	\begin{table}[tb!]
		\centering
		\begin{tabular}{||l || c | c | c|c| c||}
			\hline \hline
			$\delta_{max} - \delta_{min}$ & $1-0.1$ & $10-1$ & $100-0.1$ & $100-1$ & $100 - 10$ \\
			\hline \hline
			$C_{\infty}$ [Pa $\sqrt{\text{m}}$] & $3.690\times 10^{6}$ & $5.131\times 10^{6}$  & $8.806\times 10^{6}$&  $8.813\times 10^6$ & $8.816\times 10^{6}$  \\ 
			\hline
			$C_\infty$ for $K_{Ic}\equiv K_{1m}^{*}$ [Pa $\sqrt{\text{m}}$] & $3.690 \times 10^{6}$ & $5.118 \times 10^{6}$ & $8.790\times 10^{6}$ &  $8.798\times 10^{6}$ & $8.799\times 10^{6}$ \\
			\hline
			$C_\infty$ for $K_{Ic}\equiv K_{2m}^{*}$ [Pa $\sqrt{\text{m}}$] & $3.699 \times 10^{6}$ & $5.130 \times 10^{6}$ &$8.792\times 10^{6}$&   $8.805\times 10^{6}$ & $8.806\times 10^{6}$ \\
			\hline
			$C_\infty$ for $K_{Ic}\equiv K_{Ic}^{max}$ [Pa $\sqrt{\text{m}}$] & $3.723 \times 10^{6}$ & $5.170 \times 10^{6}$ &   $8.848\times 10^{6}$  &  $8.848\times 10^{6}$ & $8.848\times 10^{6}$ \\
			\hline
			$K_{Ic}^{max}$ [Pa $\sqrt{\text{m}}$] & $1.860 \times 10^{6}$ & $4.700 \times 10^{6}$ &$8.760\times 10^{6}$&   $8.760\times 10^{6}$ & $8.760\times 10^{6}$ \\
			\hline\hline
		\end{tabular}
		\put(-540,65) {{\bf (a)}}
		\put(-270,65) {{\bf Sinusoidal}}
		
		\vspace{6mm}
		
		\begin{tabular}{||l || c | c |c| c| c||}
			\hline \hline
			$\delta_{max} - \delta_{min}$ & $1-0.1$ & $10-1$ & $100-0.1$ & $100-1$ & $100 - 10$ \\
			\hline \hline
			$C_{\infty}$ [Pa $\sqrt{\text{m}}$] & $3.711\times 10^{6}$  & $5.158\times 10^{6}$  & $8.842\times 10^{6}$ & $8.843\times 10^6$ & $8.843\times 10^{6}$  \\
			\hline
			$C_\infty$ for $K_{Ic}\equiv K_{1m}^{*}$ [Pa $\sqrt{\text{m}}$] & $3.689 \times 10^{6}$ & $5.151 \times 10^{6}$ &  $8.828\times 10^{6}$ & $8.830\times 10^{6}$ & $8.832\times 10^{6}$ \\
			\hline
			$C_\infty$ for $K_{Ic}\equiv K_{2m}^{*}$ [Pa $\sqrt{\text{m}}$] & $3.700 \times 10^{6}$ & $5.158 \times 10^{6}$ & $8.830\times 10^{6}$ & $8.836\times 10^{6}$ & $8.842\times 10^{6}$ \\
			\hline
			$C_\infty$ for $K_{Ic}\equiv K_{Ic}^{max}$ [Pa $\sqrt{\text{m}}$] & $3.723 \times 10^{6}$ & $5.170 \times 10^{6}$ & $8.848\times 10^{6}$ & $8.848\times 10^{6}$ & $8.848\times 10^{6}$ \\
			\hline
			$K_{Ic}^{max}$ [Pa $\sqrt{\text{m}}$] & $1.860 \times 10^{6}$ & $4.700 \times 10^{6}$ & $8.760\times 10^{6}$ & $8.760\times 10^{6}$ & $8.760\times 10^{6}$ \\
			\hline\hline
		\end{tabular}
		\put(-540,65) {{\bf (b)}}
		\put(-270,65) {{\bf Step-wise}}
		\caption{Values of $C_\infty$ \eqref{local_indicator_infty} obtained for material toughness regimes $K_{Ic}(l(t))$ with the (a) sinusoidal, (b) step-wise, distribution. Here $K_{1m}^*$, $K_{2m}^*$, are the values of the 'average' material toughness obtained using the temporal-averaging strategy \eqref{temporal_averages} outlined in \cite{Gaspare2022}.}
		\label{Table_Cinfty}
	\end{table}
\end{landscape}

From Table.~\ref{Table_Cinfty}, we can see that the rock heterogeneity does play a role in the limiting behaviour. The values of $C(l)$ for the heterogeneous toughness distribution do not converge to those of the maximum toughness $K_{Ic}\equiv K_{Ic}^{max}$ as $l\to\infty$, with $C_\infty (K_{Ic}(l(t))) < C_\infty (K_{Ic}^{max})$ in all cases. The temporal-averaging based approaches from \cite{Gaspare2022} are effective at predicting the limiting value $C_\infty$ for all considered distributions, with the energy-based average $K_{2m}^*$ being a better predictor than $K_{1m}^*$. Meanwhile, the particular toughness distribution also influences the result, with the value of $C_\infty$ for the square distribution always being larger than that for the sinusoidal distribution. Finally, comparing the results when $\delta_{max} - \delta_{min}$ are $100 - 1$ and $100-10$, we can see that the minimum toughness only plays a minor role in the value of $C_\infty$ compared to that of the maximum toughness. 

\section{Energy distribution during heterogeneous fracture}\label{Sect:Energy}

As a hydraulic fracture propagates within a heterogeneous material the extent of `viscosity' and `toughness'-dominated behaviour is amplified. This can be seen from the behaviour of $\delta$ in Fig.~\ref{fig:CL_infty}c,d, with the maximum of $\delta$ noticeably exceeding $\delta_{max}$, and the minimum of $\delta$ being almost an order of magnitude below $\delta_{min}$. 

In the preceding work \cite{Gaspare2022} (and by others, see that paper), an explanation of this was given in terms of the fracture pressurisation. To obtain a fuller picture however, we need to examine the system behaviour from an energy perspective. {Towards this end, we first outline the different energy terms influencing fracture development in Sect.~\ref{Sect_Energy_Balance}. Following this, the behaviour of the energy terms as the fracture propagates through a heterogeneous material is provided and analysed in Sect.~\ref{Sect:Battery}. Finally, a brief investigation of the limiting behaviour is given in Sect.~\ref{Sect:EnergyLim}.}

\subsection{Overview of the energy distribution}\label{Sect_Energy_Balance}

\begin{figure}[b!]
	\centering
	\includegraphics[width=0.45\textwidth]{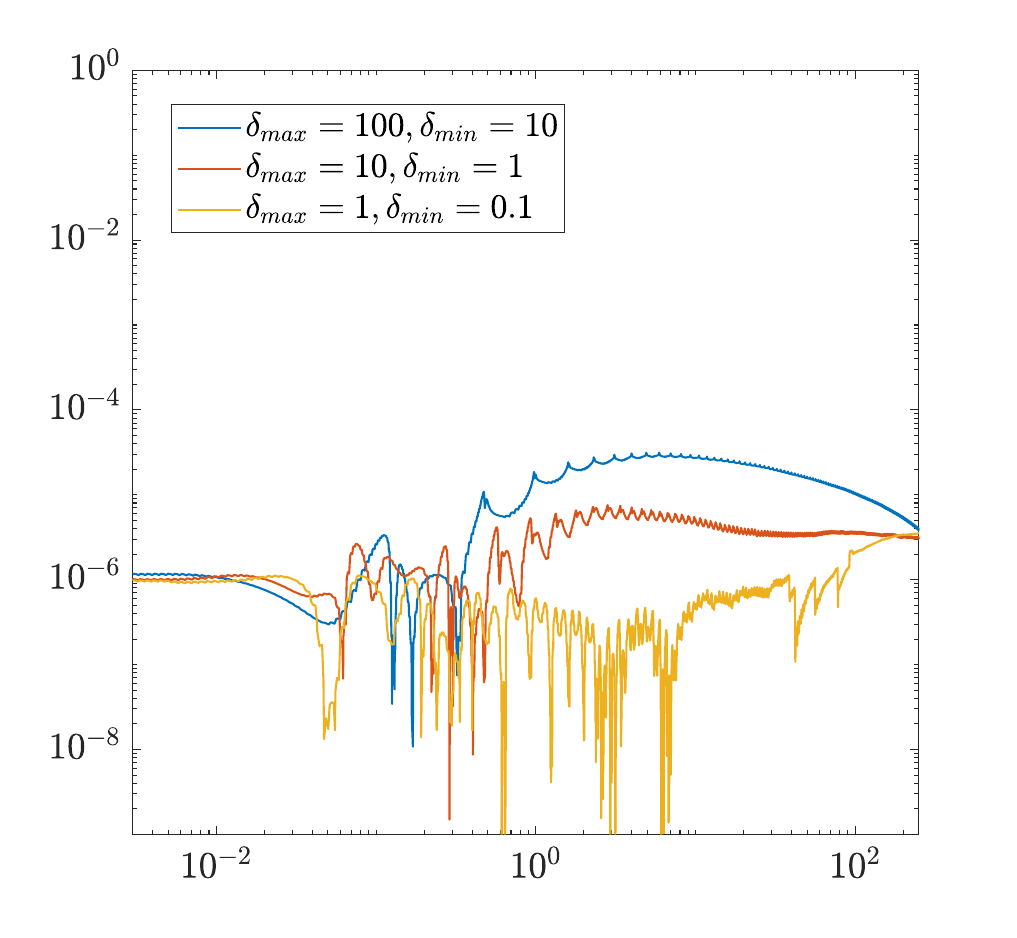}
	\put(-85,0) {$t$}
	\put(-230,90) {$\int_0^t \Delta P(\xi) {\rm d}\xi$}
	\put(-190,160) {{\bf (a)}}
	\hspace{6mm}
	\includegraphics[width=0.45\textwidth]{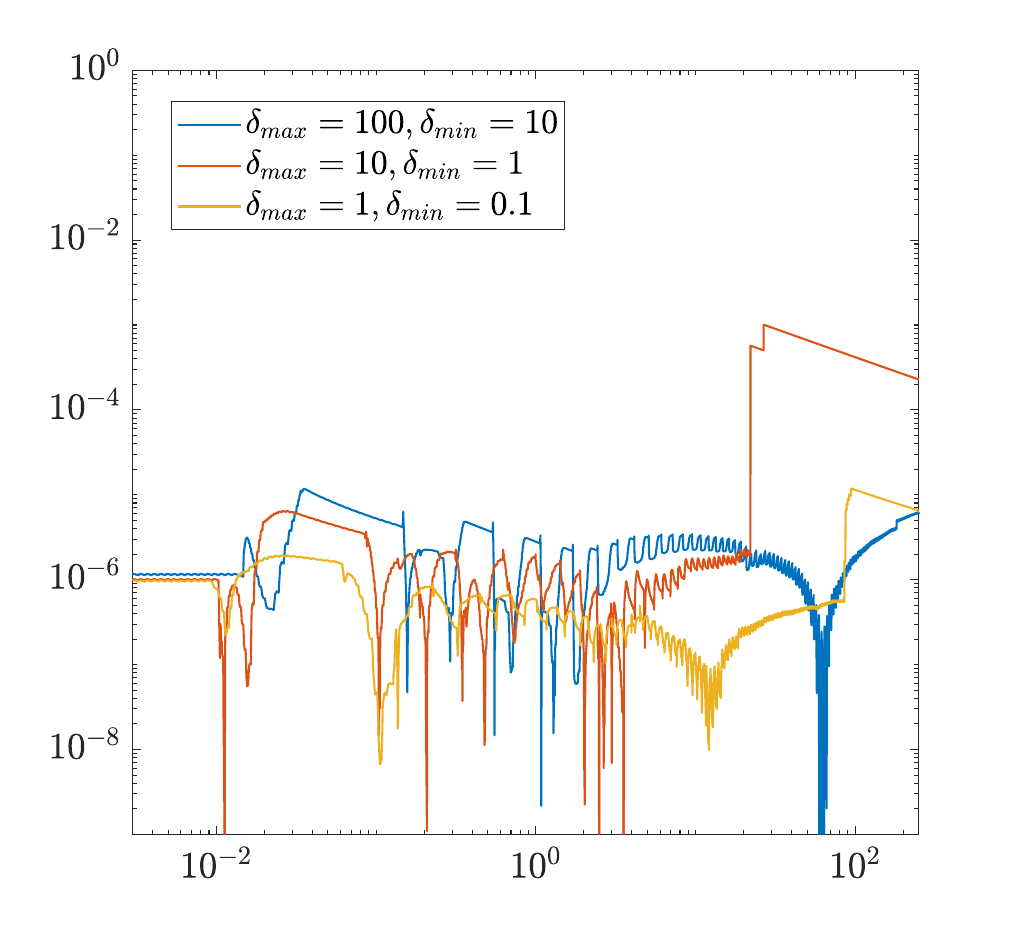}
	\put(-85,0) {$t$}
	\put(-190,160) {{\bf (b)}}
	\caption{The global (cumulative) relative error of numerical computations for the energy components against (propagation) time, computed from \eqref{energy_balance_1}, for the (a) sinusoidal and (b) step-wise toughness distributions.}
	\label{Fig_Energy_Balance_Local}
\end{figure}
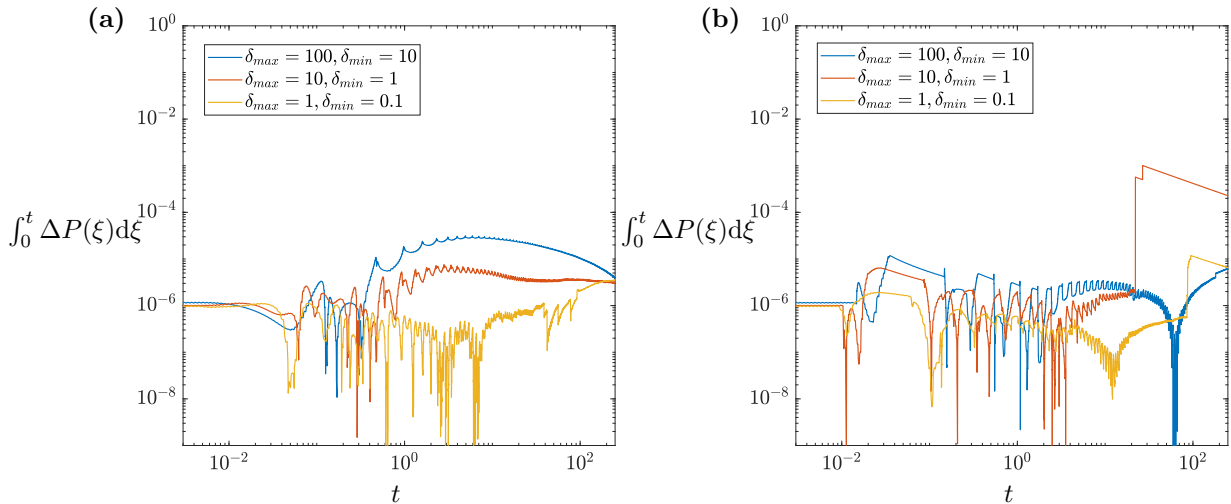

In order to investigate the physical processes occurring within the hydraulic fracture, we consider the energy balance within the fracture over time. In the case of a single hydraulic fracture propagating without fluid lag, the energy within {a fracture of length $l(t)$ is described in terms of}  four primary terms:
\begin{itemize}
	\item The power, $P(l(t))$, representing the energy added to the system due to fluid injection.
	\item The toughness energy, $S(l(t))$, representing the energy spent fracturing the solid material.
	\item The viscous energy, $D(l(t))$, {expended by the fluid due to shearing effects on the crack faces}.
	\item The elastic energy, $U(l(t))$, stored within the solid material.
\end{itemize}
For details on the derivation of these terms within a hydraulic fracture see for example \cite{LECAMPION20074863,BUNGER20131538} and references therein. The formulae given below are adapted from those provided in  \cite{BUNGER20131538}.

These four energy terms are related {at each moment in time} as 
\begin{equation} \label{energy_balance_1}
	P(l(t)) = D(l(t)) + S(l(t)) + U(l(t)).
\end{equation}
where each energy term is given by:
\begin{equation} \label{energy}
	P(l) = 2 p(0,t) q (0,t) , \quad  D(l) = \frac{1}{M^\prime} \int_0^{l(t)} w \left(\frac{\partial p}{\partial x}\right)^2 \, dx =  M^\prime \int_0^{l(t)} \frac{v^2 (t,x)}{w(t,x)} \, dx ,
\end{equation}
\[
S(l) =\frac{2}{E^\prime} v_0 (t) K_{Ic}^2\left(l(t)\right)  , \quad U(l) = \int_0^{l(t)} p\frac{\partial w}{\partial t} + w \frac{\partial p}{\partial t} \, dx =  \int_0^{l(t)}\frac{\partial}{\partial t} \left( p(t,x)w(t,x)  \right)  \, dx. 
\]

In the numerical solver, each of these terms is computed separately during post-processing from the system parameters. Consequently, the accuracy of the energy computations can be inferred using \eqref{energy_balance_1}, with the global (cumulative) relative error of this expression provided in Fig.~\ref{Fig_Energy_Balance_Local}. The solver is run to a prescribed tolerance, with the default setting ensuring that the error in the crack tip velocity remains below $10^{-4}$ at each time-step. As can be seen in Fig.~\ref{Fig_Energy_Balance_Local}, the relative error of the total energy
is of the same order throughout the simulations. {However, this} can still mean sharp spikes in the local error, particularly when the difference between the maximum and minimum toughness is large (e.g. $\delta_{max}=100, \delta_{min}=10$), and over long simulated time-scales. On the other hand, taking a stricter tolerance for the crack tip velocity improves the local energy balance, while the relative difference in results for the physical parameters (crack length, velocities, pressure) remains below the original threshold.

This however remains sufficiently accurate to verify the claims made in this paper. With this we can examine how the energy balance impacts the propagation of fracture within a heterogeneous material.

\subsection{Energy distribution and the elastic battery} \label{Sect:Battery}

{In the following analysis, we are interested in the dependence of the fracture energy on the toughness of the material being propagated through at each specific moment $t$. As the material toughness at each moment depends on the crack tip position, and thus the length $K_{Ic}(l(t))$ (i.e.\ the material currently being propagated through), in all figures we will take $l(t)$ on the $x$-axis.} {In the main text, we only include those figures needed to highlight the behaviour. Additional figures showing the energy distribution for all cases are in Appendix.~B (Figs.~B.1-B.3). }

\begin{figure}[hp!]
	\centering
	\includegraphics[width=0.45\textwidth]{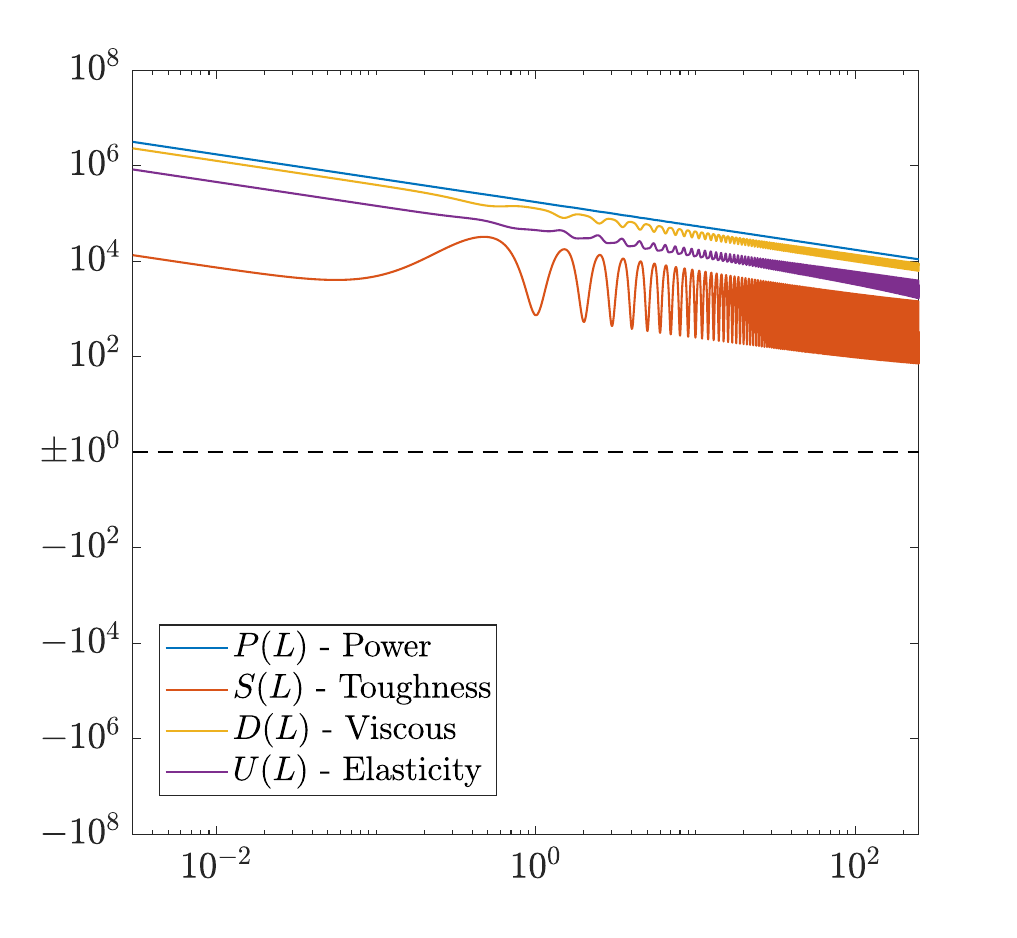}
	\put(-95,-3) {$l(t)$}
	\put(-205,85) {Power}
	\put(-195,155) {{\bf (a)}}
	\hspace{6mm}
	\includegraphics[width=0.45\textwidth]{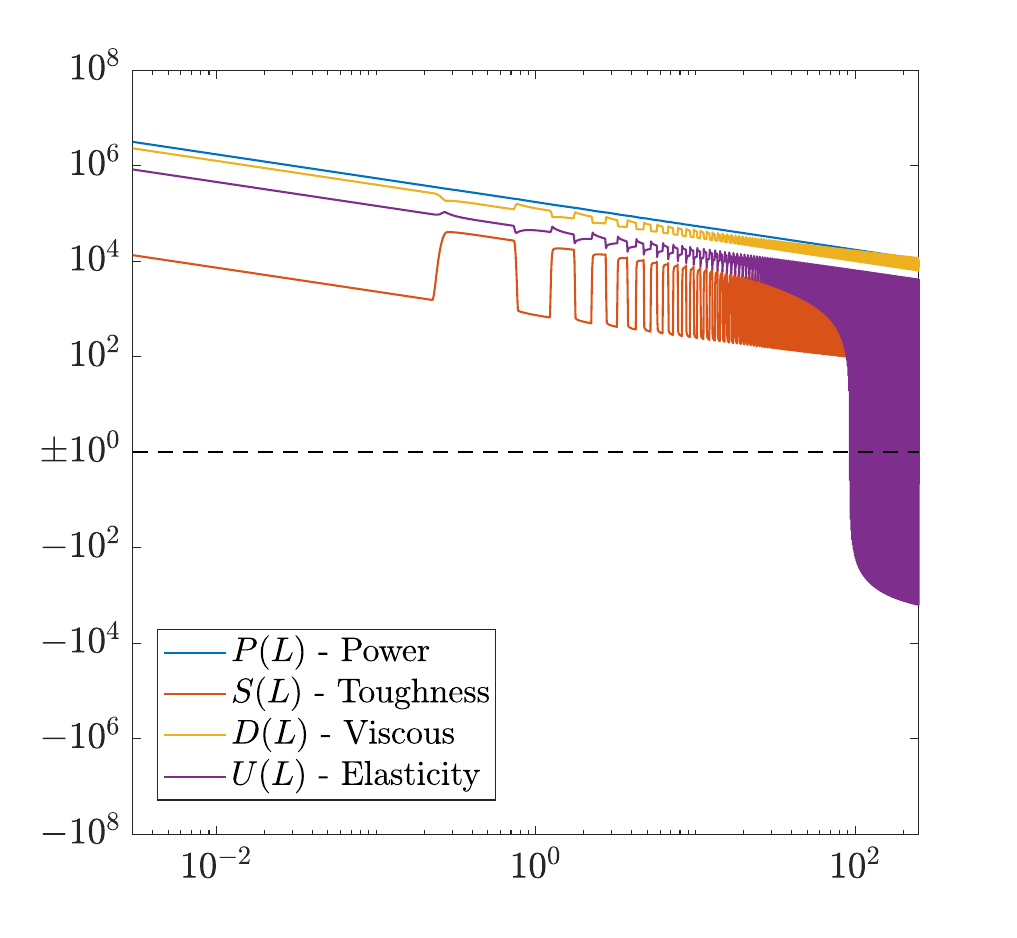}
	\put(-95,-3) {$l(t)$}
	\put(-205,85) {Power}
	\put(-195,155) {{\bf (b)}}
	
	\includegraphics[width=0.45\textwidth]{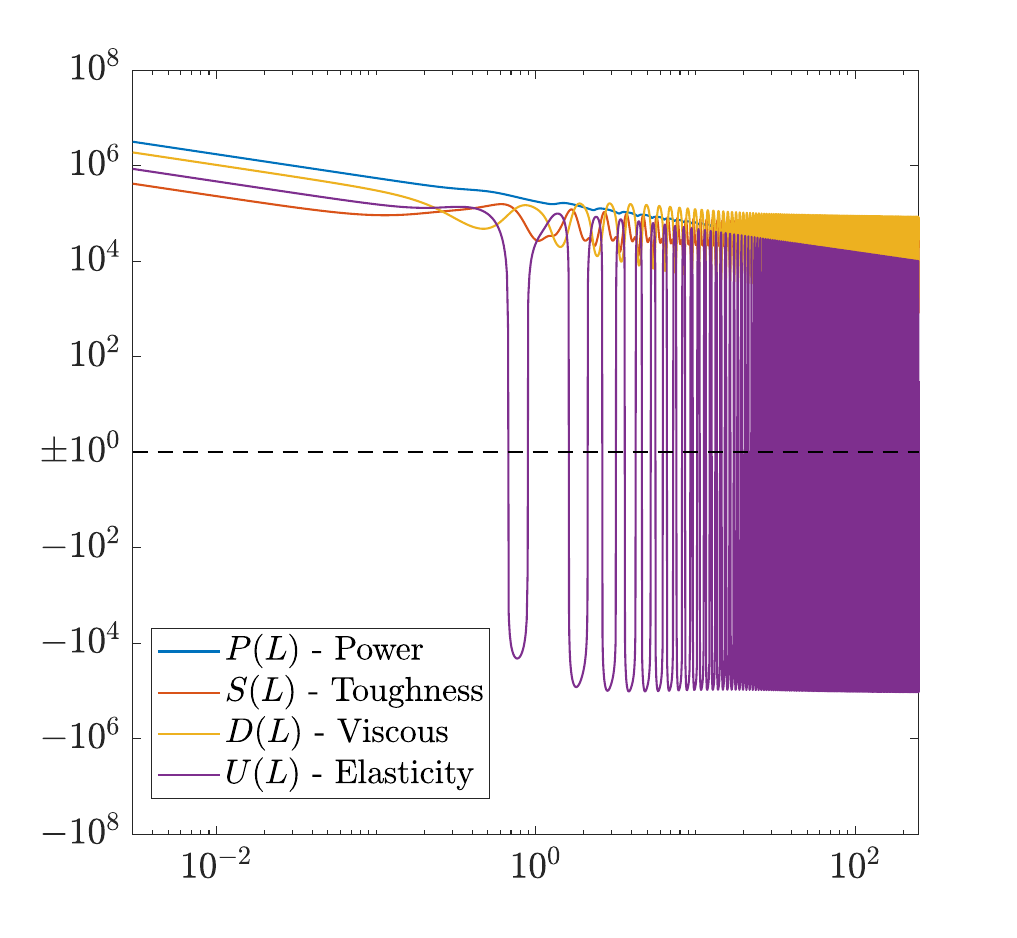}
	\put(-95,-3) {$l(t)$}
	\put(-205,85) {Power}
	\put(-195,155) {{\bf (c)}}
	\hspace{6mm}
	\includegraphics[width=0.45\textwidth]{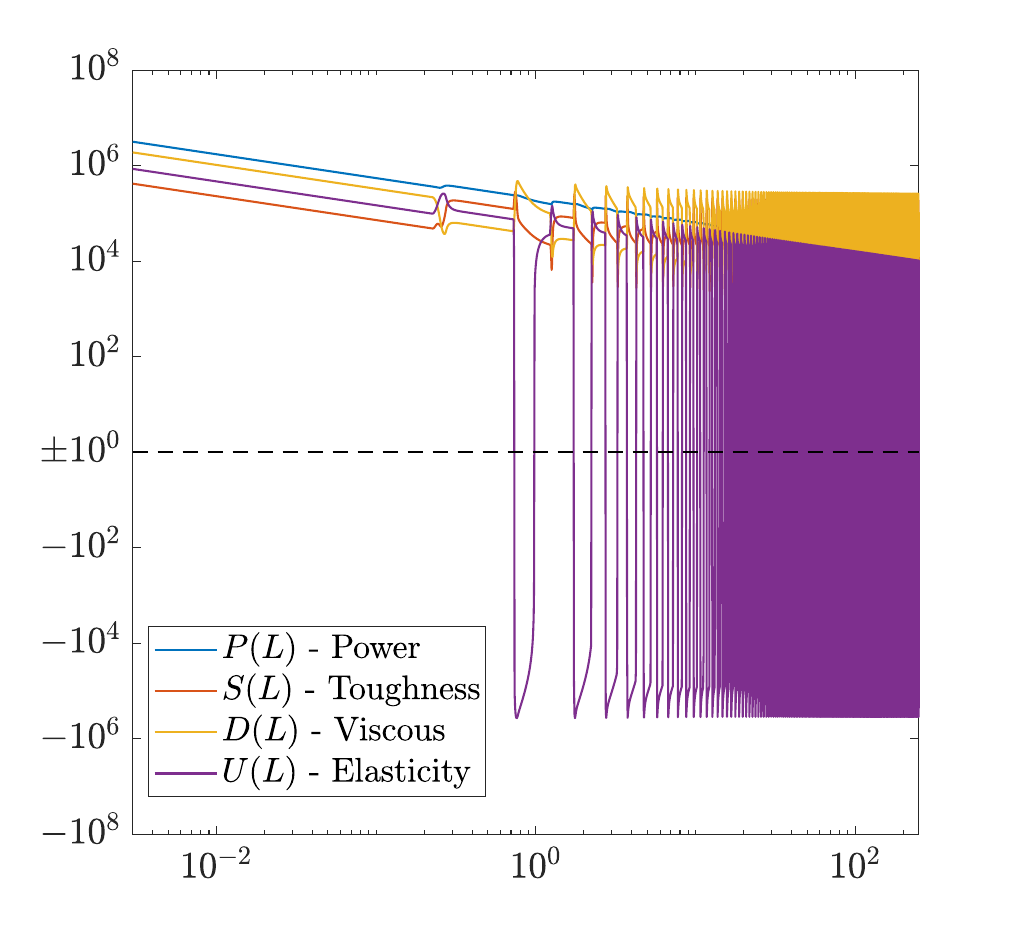}
	\put(-95,-3) {$l(t)$}
	\put(-205,85) {Power}
	\put(-195,155) {{\bf (d)}}
	
	\includegraphics[width=0.45\textwidth]{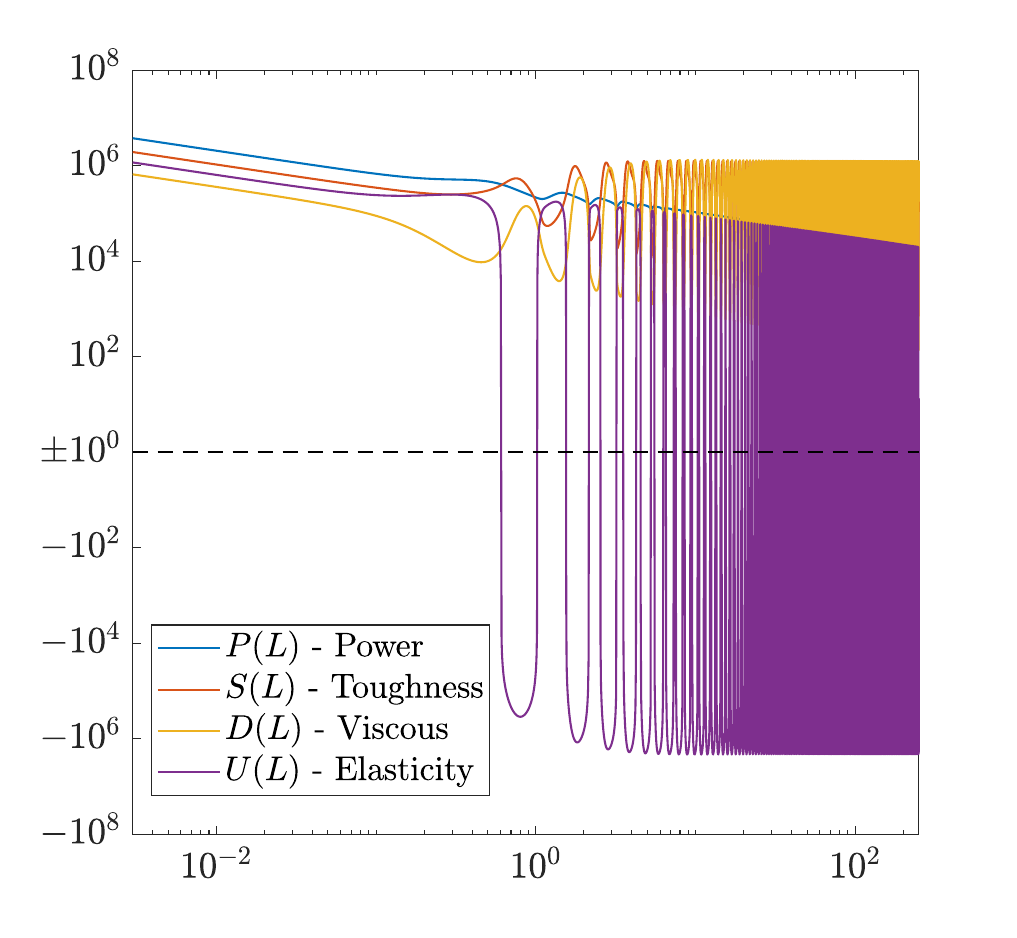}
	\put(-95,-3) {$l(t)$}
	\put(-205,85) {Power}
	\put(-195,155) {{\bf (e)}}
	\hspace{6mm}
	\includegraphics[width=0.45\textwidth]{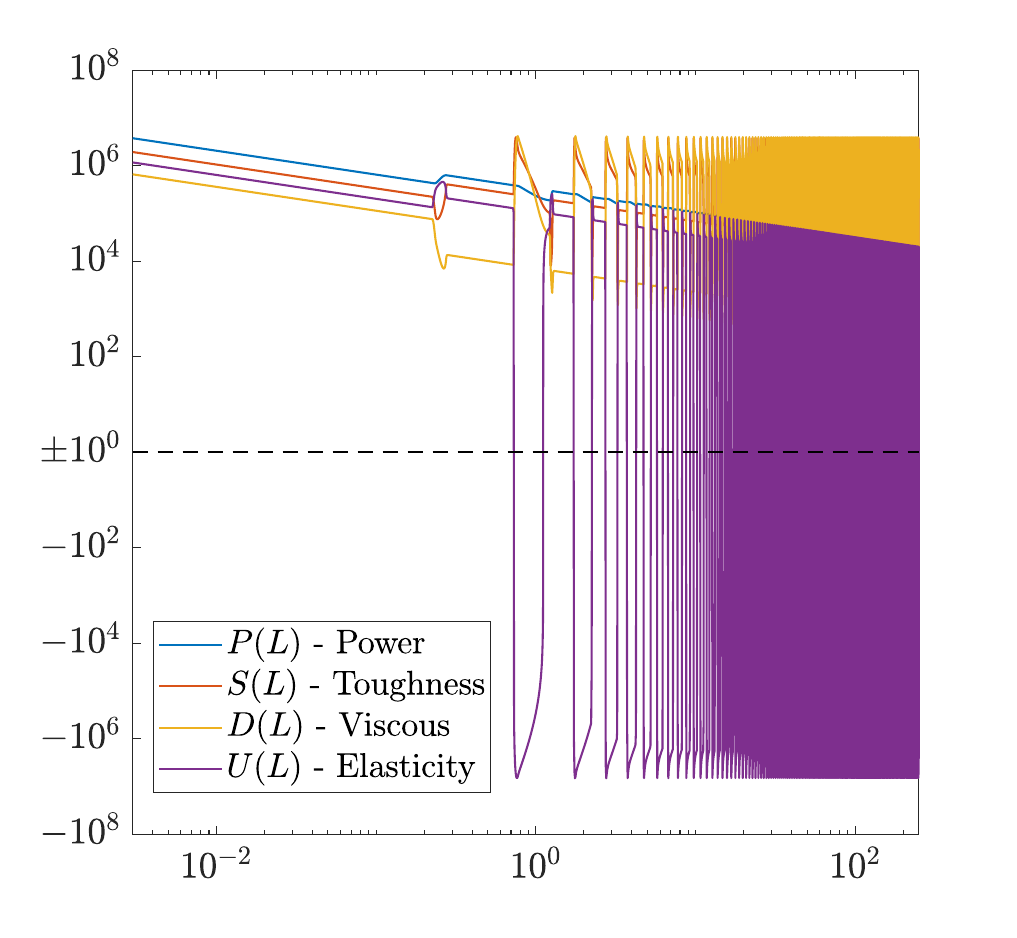}
	\put(-95,-3) {$l(t)$}
	\put(-205,85) {Power}
	\put(-195,155) {{\bf (f)}}
	\caption{Distribution of the power [W] \eqref{energy} within the hydraulic fracture as it propagates through the domain. This is for the (a), (b) intermediate-viscosity, (c), (d) toughness-intermediate, (e), (f) toughness-toughness, case with the (a), (c), (e) sinusoidal, (b), (d), (f) step-wise, toughness distribution. }
	\label{Energy_OverL_All}
\end{figure}

Figures showing the energy distribution \eqref{energy} as the fracture propagates through the heterogeneous material {over the entire fracture length} are provided in Fig.~\ref{Energy_OverL_All}.  {These results clearly show that the energy terms strongly depend on the toughness of the material being propagated through, as best evidenced by the energy terms. This effect is present in all cases, however is more pronounced when the maximum toughness layer is in the toughness dominated regime (Fig.~\ref{Energy_OverL_All}c-f).}

\begin{figure}[t!]
	\centering
	\includegraphics[width=0.45\textwidth]{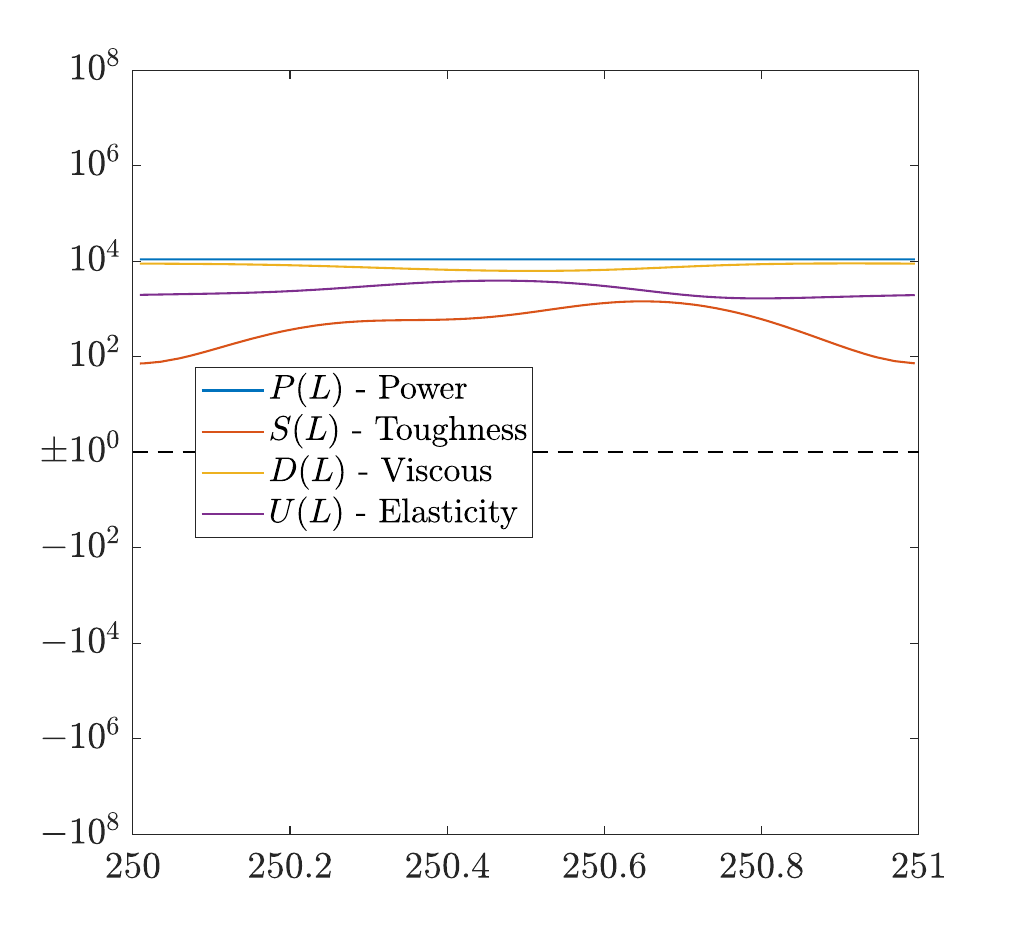}
	\put(-95,-3) {$l(t)$}
	\put(-205,80) {Power}
	\put(-195,155) {{\bf (a)}}
	\put(-105,155) {$k=250$}
	\hspace{6mm}
	\includegraphics[width=0.45\textwidth]{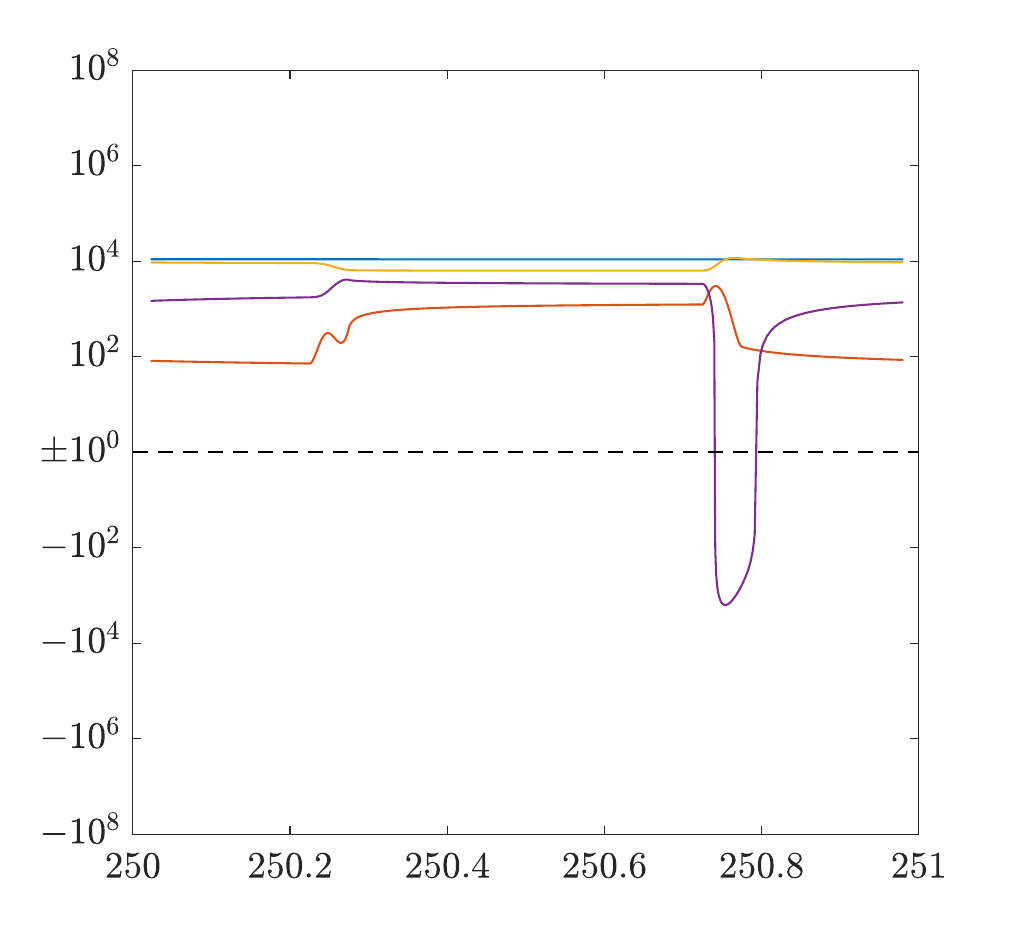}
	\put(-95,-3) {$l(t)$}
	\put(-205,80) {Power}
	\put(-195,155) {{\bf (b)}}
	\put(-105,155) {$k=250$}
	\caption{Distribution of the power [W] \eqref{energy} within the hydraulic fracture as it propagates through the $250^{\text{th}}$ toughness period. This is for the intermediate-viscosity case ($\delta_{max}=1$, $\delta_{min}=0.1$) with the (a), sinusoidal, (b) step-wise, toughness distribution. Recall that for the stepwise distribution, $0.25<l-k<0.75$ corresponds to the maximum toughness layer, with the remainder of the domain corresponding to the minimum toughness layer.}
	\label{Energy_Periods_1_01_main}
\end{figure}
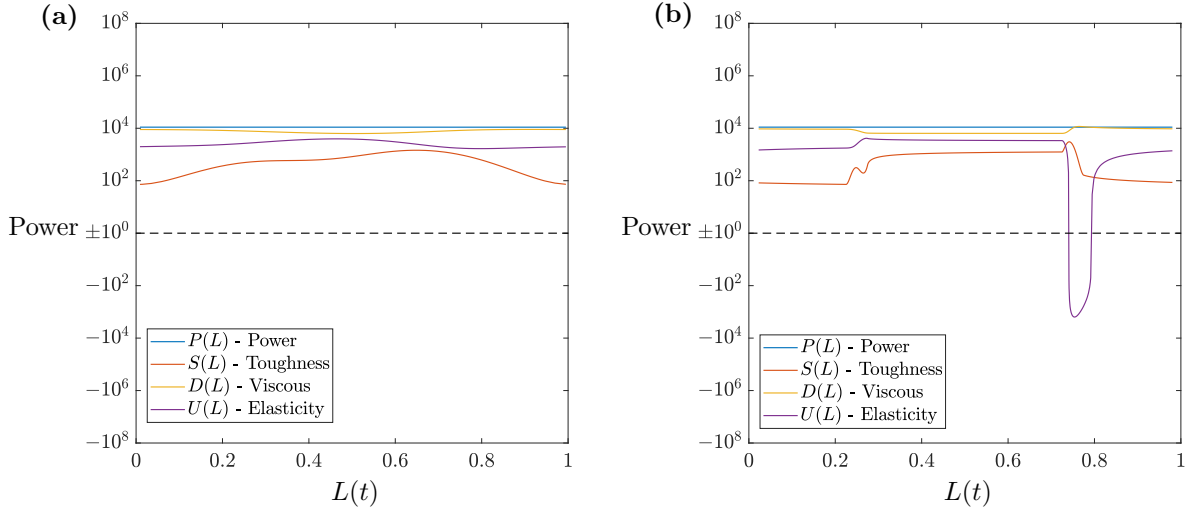

\begin{figure}[tp!]
	\centering
	\includegraphics[width=0.45\textwidth]{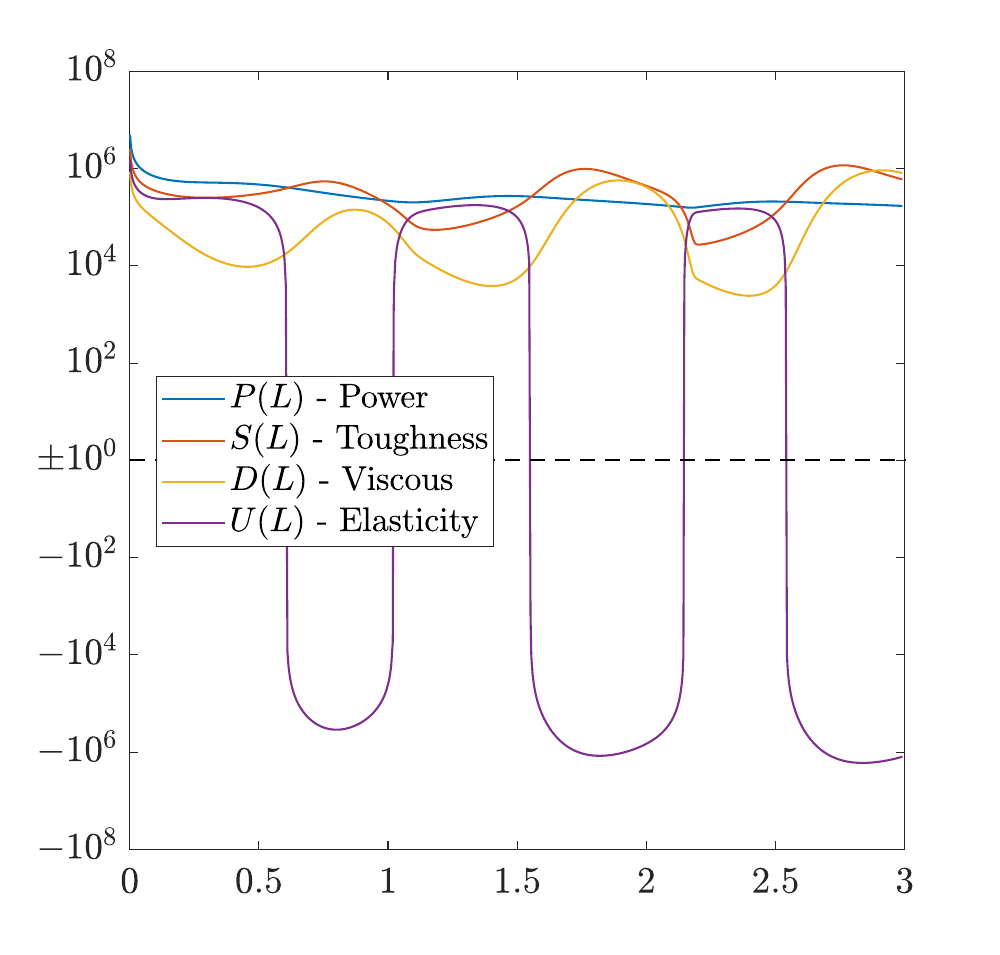}
	\put(-95,-3) {$l(t)$}
	\put(-205,85) {Power}
	\put(-195,155) {{\bf (a)}}
	\put(-115,160) {$k=1,2,3$}
	\hspace{6mm}
	\includegraphics[width=0.45\textwidth]{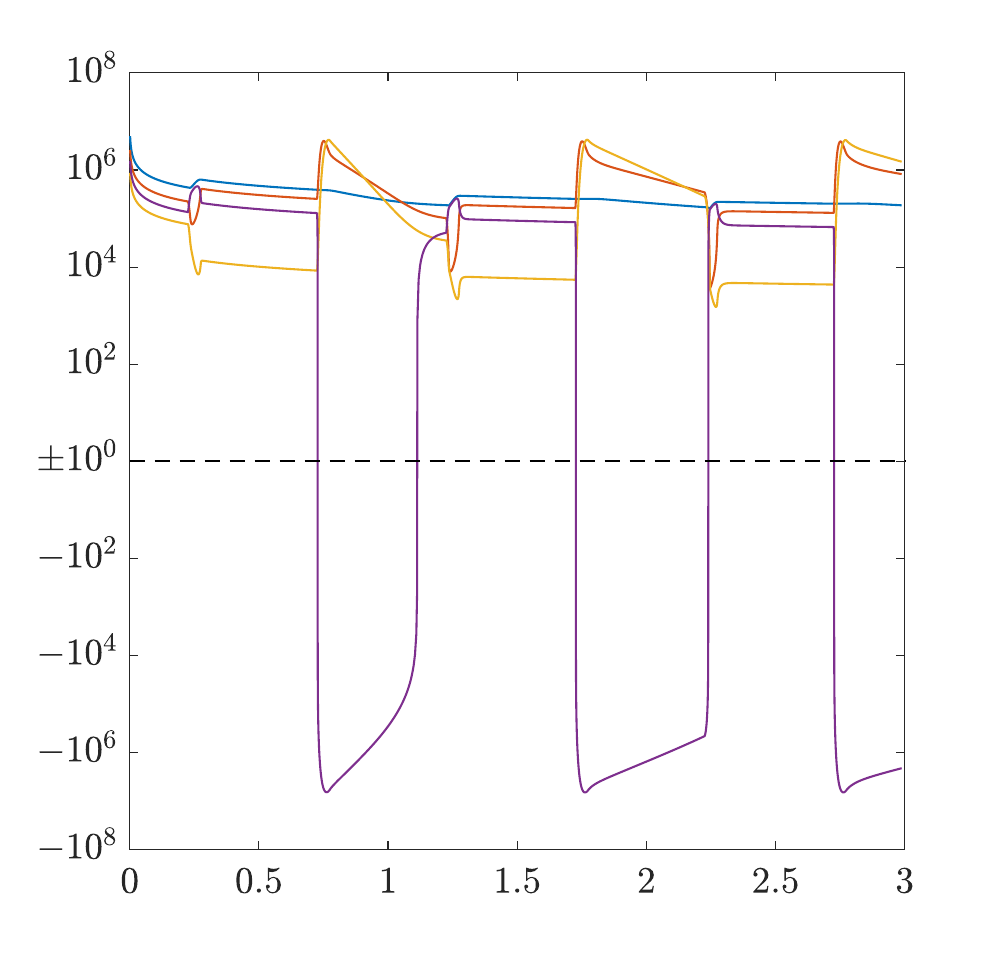}
	\put(-95,-3) {$l(t)$}
	\put(-205,85) {Power}
	\put(-195,155) {{\bf (b)}}
	\put(-115,160) {$k=1,2,3$}
	
	\vspace{2mm}
	
	\includegraphics[width=0.45\textwidth]{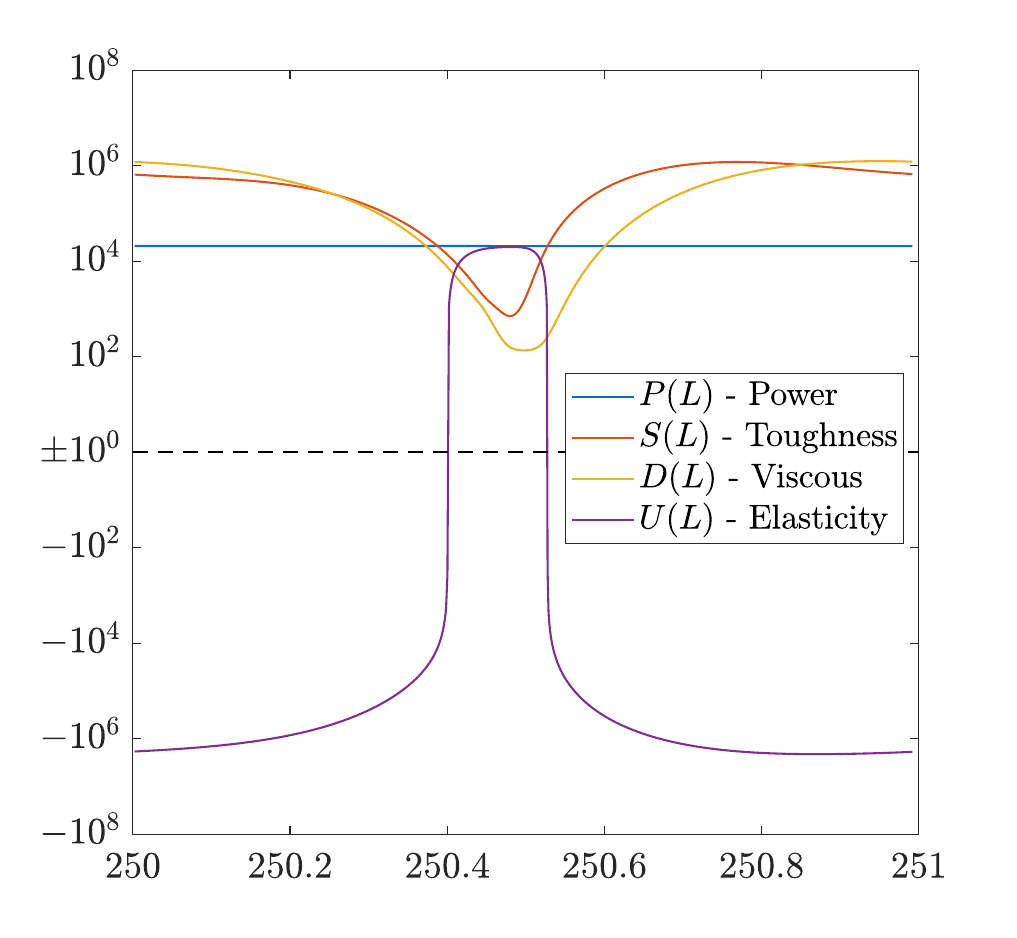}
	\put(-95,-3) {$l(t)$}
	\put(-205,85) {Power}
	\put(-195,155) {{\bf (c)}}
	\put(-105,155) {$k=250$}
	\hspace{6mm}
	\includegraphics[width=0.45\textwidth]{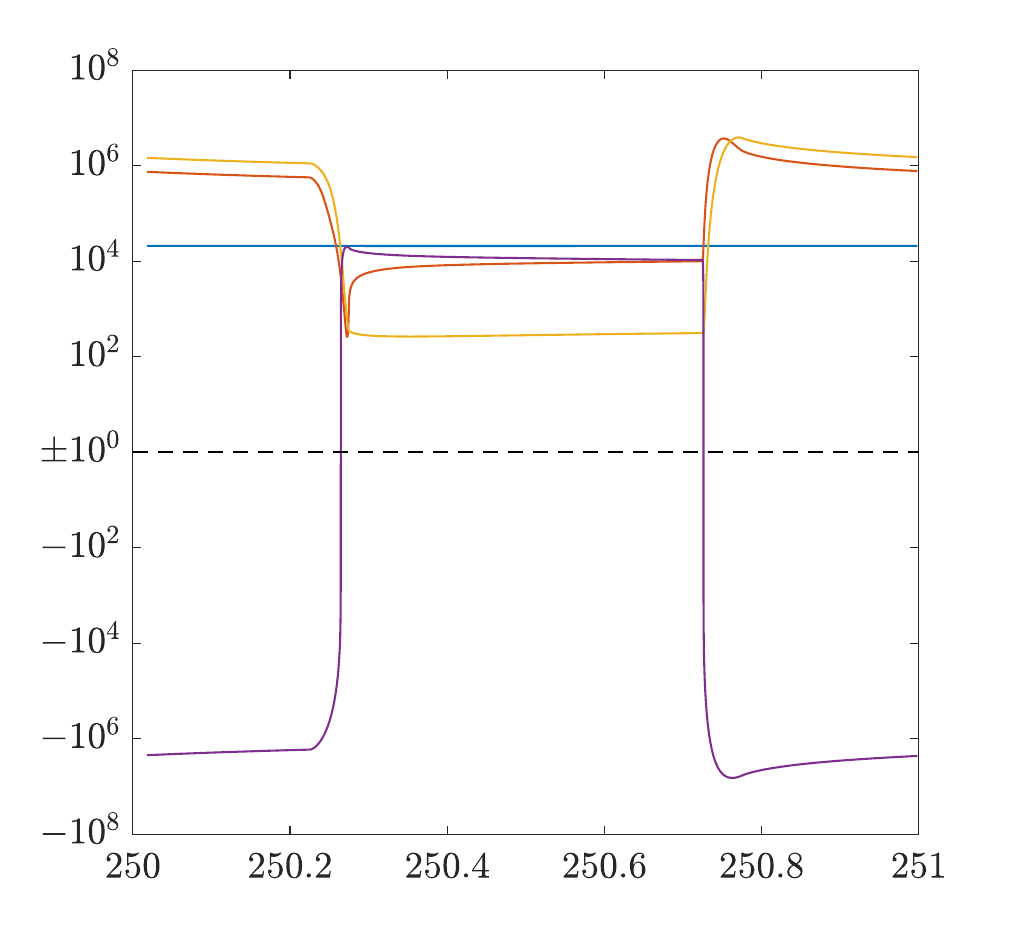}
	\put(-95,-3) {$l(t)$}
	\put(-205,85) {Power}
	\put(-195,155) {{\bf (d)}}
	\put(-105,155) {$k=250$}
	\caption{Distribution of the power [W] \eqref{energy} within the hydraulic fracture as it propagates through (a), (b) the first three toughness periods, and the $k^{\text{th}}$ toughness period with (c), (d) $k=250$. This is for the toughness-toughness case ($\delta_{max}=100$, $\delta_{min}=10$) with the (a), (c) sinusoidal, (b), (d) step-wise, toughness distribution. Recall that for the stepwise distribution, $0.25<l-k<0.75$ corresponds to the maximum toughness layer, with the remainder of the domain corresponding to the minimum toughness layer.}
	\label{Energy_Periods_100_10_main}
\end{figure}

{This motivates investigating the behaviour of the energy terms within a single toughness period, to highlight the differing behaviour in high and low toughness rock layers. This comparison of energy terms within a single period is provided in Figs.~\ref{Energy_Periods_1_01_main}-\ref{Energy_Periods_100_10_main}. It can be seen that, after the initial two periods, the energy distribution settles into a periodic oscillation to match the toughness, with only small changes over time. The subsequent behaviour depends strongly on whether one of the rock layers is in the toughness dominated regime, and we therefore analyse these two cases separately.}  

\medskip

\noindent {\bf With at least one layer in the toughness dominated regime (Fig.~\ref{Energy_Periods_100_10_main})}
\begin{itemize}
	\item As can be seen clearly in Fig.~\ref{Energy_Periods_100_10_main}, while propagating in the maximum toughness layer, the `power' being injected into the fracture is stored within the elastic material - with far less being used to propagate the fracture, resulting in a slower propagation rate. 
	\item Once the fracture reaches the minimum toughness layer, this elastic energy is rapidly transferred to both the fracture propagation and the fluid. 
\end{itemize}

This behaviour is observed in both the toughness-toughness and toughness-intermediate cases (see Fig.~\ref{Energy_OverL_All}, also Figs.~B.1-B.3), indicating that the presence of a 'toughness-dominated' layer is the key factor for initiating this behaviour.\\

\noindent {\bf With all layers in the viscosity dominated/ intermediate regime (Fig.~\ref{Energy_Periods_1_01_main})}
\begin{itemize}
	\item If both layers are in the viscosity dominated regime then the energy distribution is largely independent of the material toughness. Storage within the elastic material does however play a significant role over long length-scales, due to the abundance of elastic material, however this is dependent on the toughness distribution (compare Fig.~\ref{Energy_Periods_1_01_main}a,b).
\end{itemize}

\medskip

\begin{figure}[hp!]
	\centering
	\includegraphics[width=0.45\textwidth]{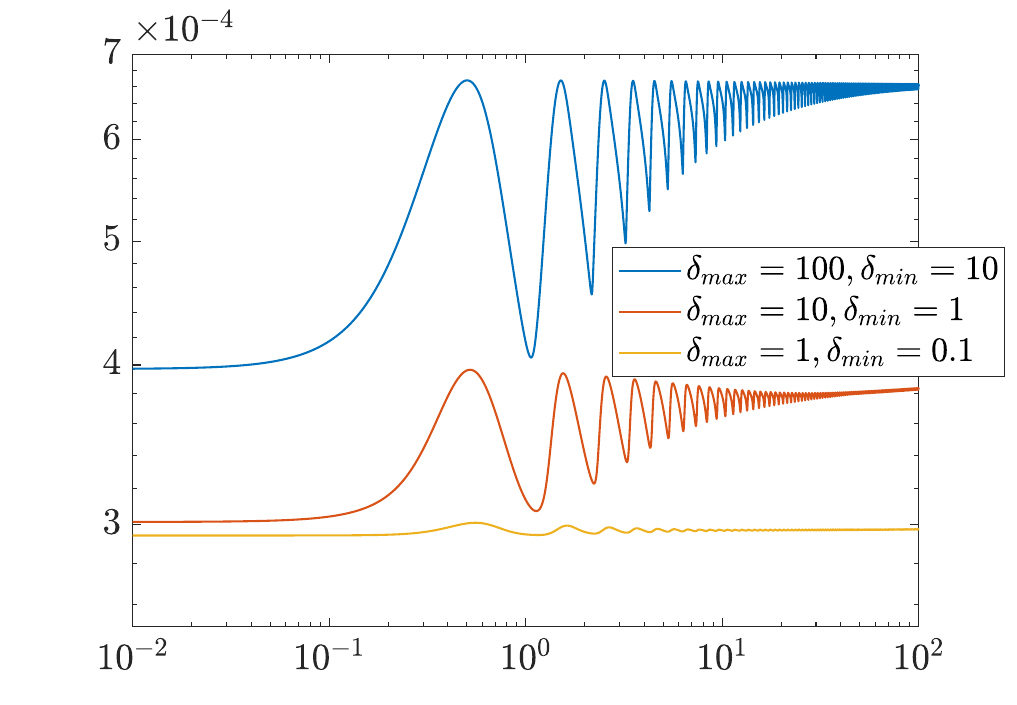}
	\put(-95,-5) {$l(t)$}
	\put(-195,60) {$\frac{w(l,0)}{\sqrt{l}}$}
	\put(-185,120) {{\bf (a)}}
	\hspace{6mm}
	\includegraphics[width=0.45\textwidth]{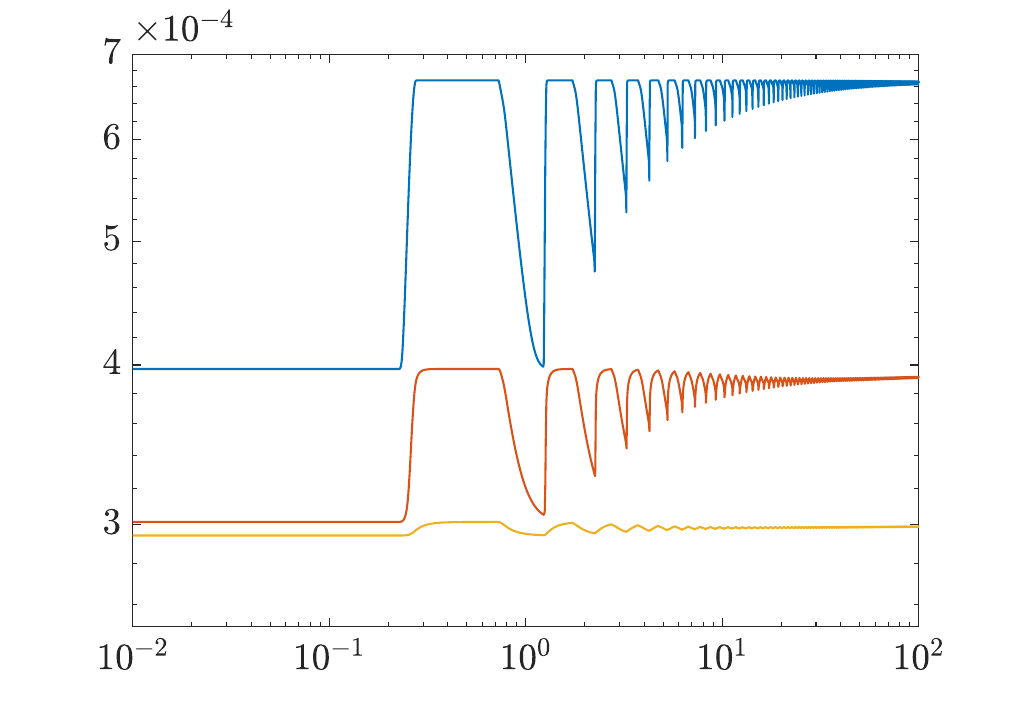}
	\put(-95,-5) {$l(t)$}
	\put(-185,120) {{\bf (b)}}
	
	\includegraphics[width=0.45\textwidth]{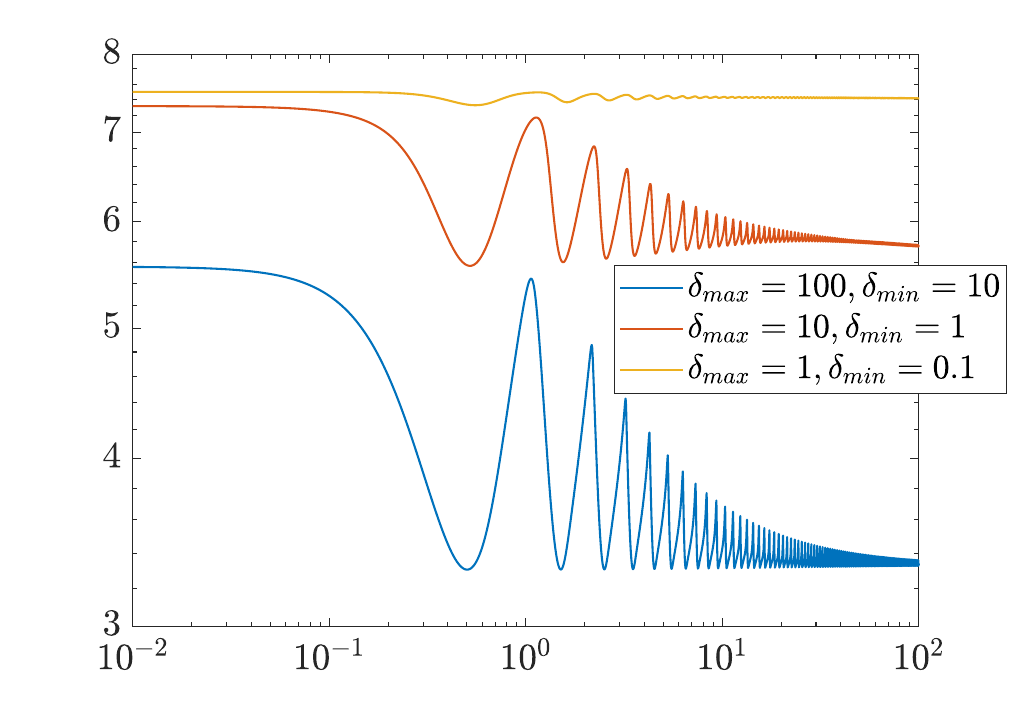}
	\put(-95,-5) {$l(t)$}
	\put(-215,60) {$\sqrt{l}v(l,0)$}
	\put(-185,120) {{\bf (c)}}
	\hspace{6mm}
	\includegraphics[width=0.45\textwidth]{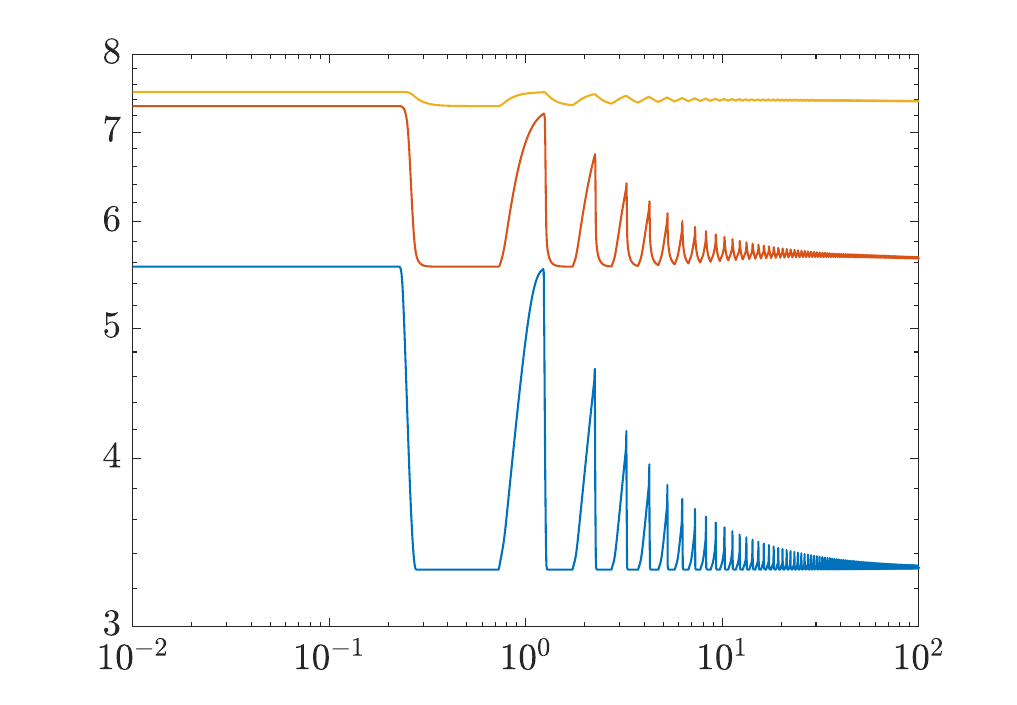}
	\put(-95,-5) {$l(t)$}
	\put(-185,120) {{\bf (d)}}
	
	\includegraphics[width=0.45\textwidth]{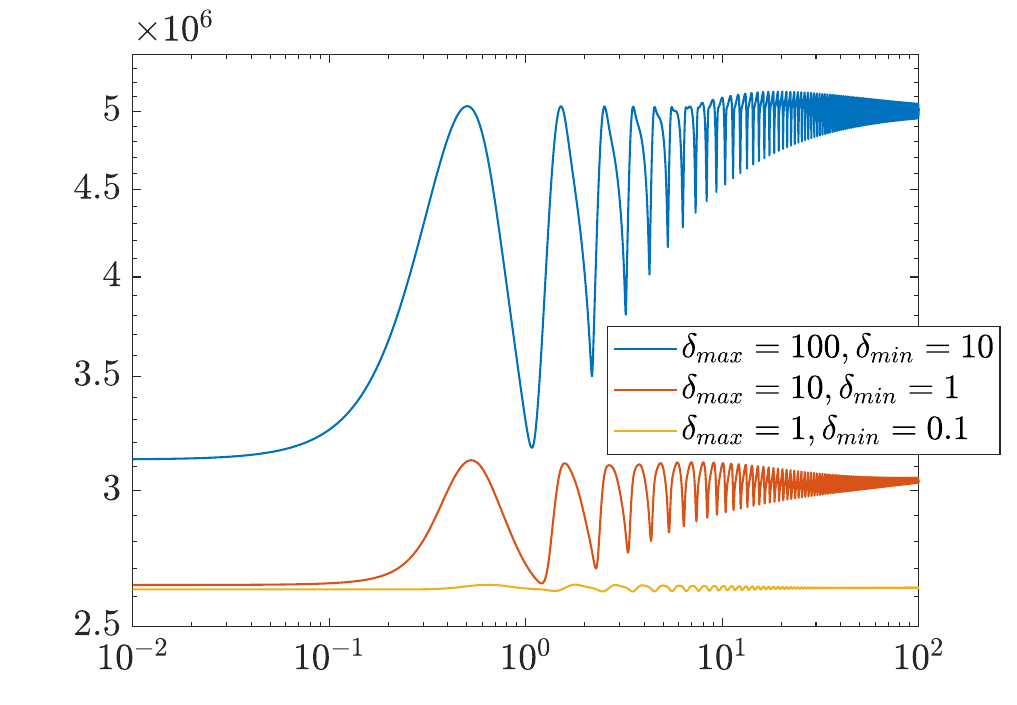}
	\put(-95,-5) {$l(t)$}
	\put(-215,60) {$\sqrt{l}p(l,0)$}
	\put(-185,120) {{\bf (e)}}
	\hspace{6mm}
	\includegraphics[width=0.45\textwidth]{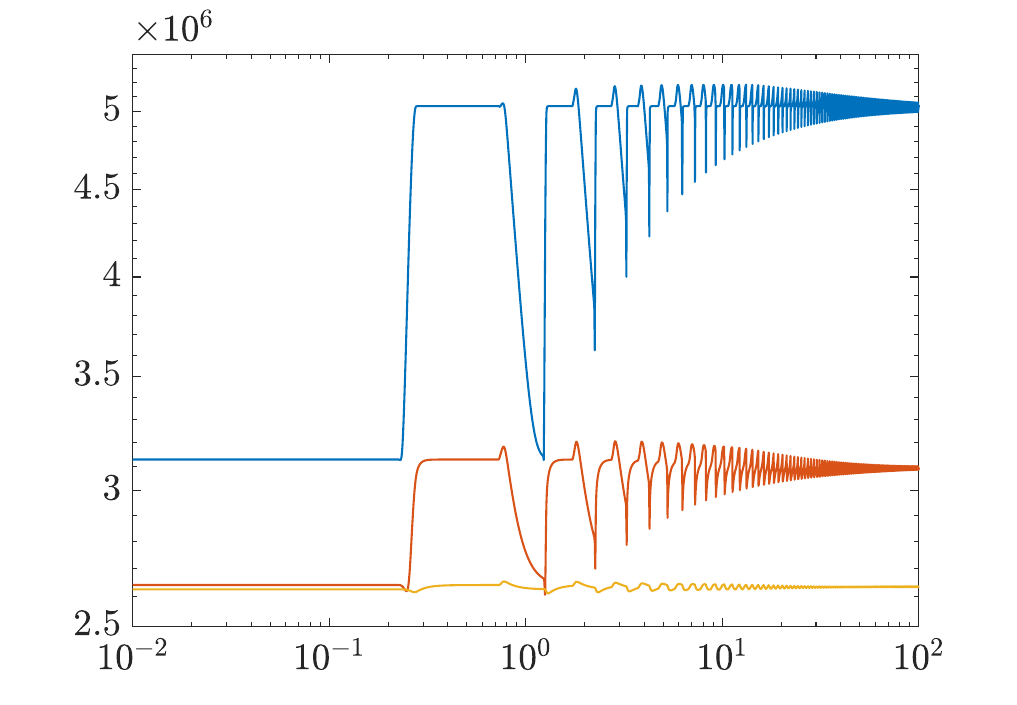}
	\put(-95,-5) {$l(t)$}
	\put(-185,120) {{\bf (f)}}
	
	\includegraphics[width=0.45\textwidth]{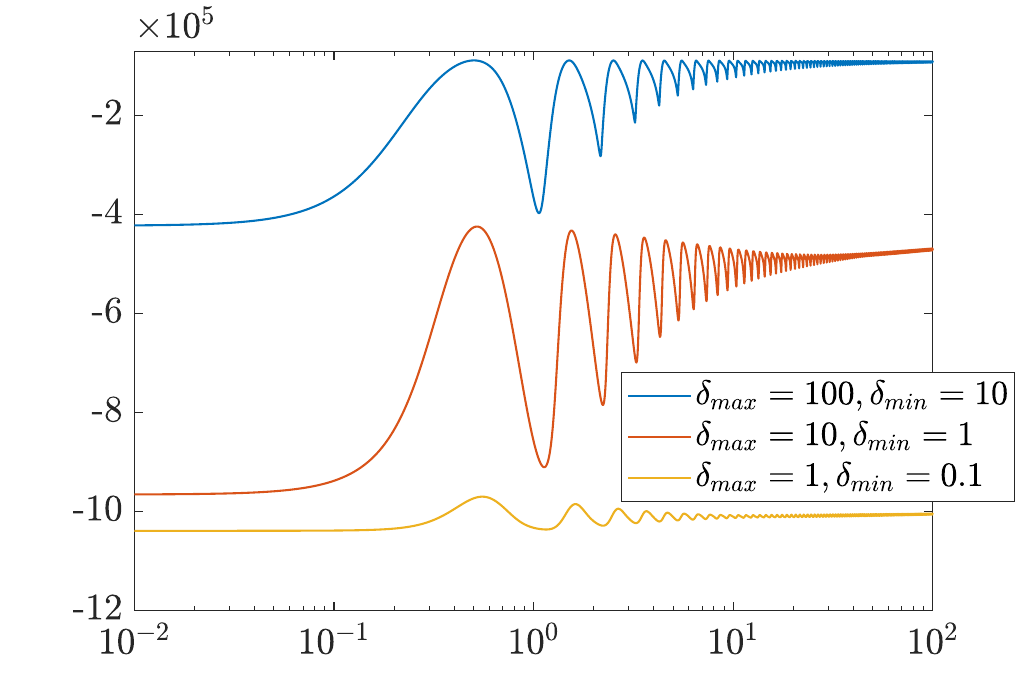}
	\put(-95,-5) {$l(t)$}
	\put(-215,60) {$l^{\frac{3}{2}}\frac{\partial p}{\partial x}(l,0)$}
	\put(-185,120) {{\bf (g)}}
	\hspace{6mm}
	\includegraphics[width=0.45\textwidth]{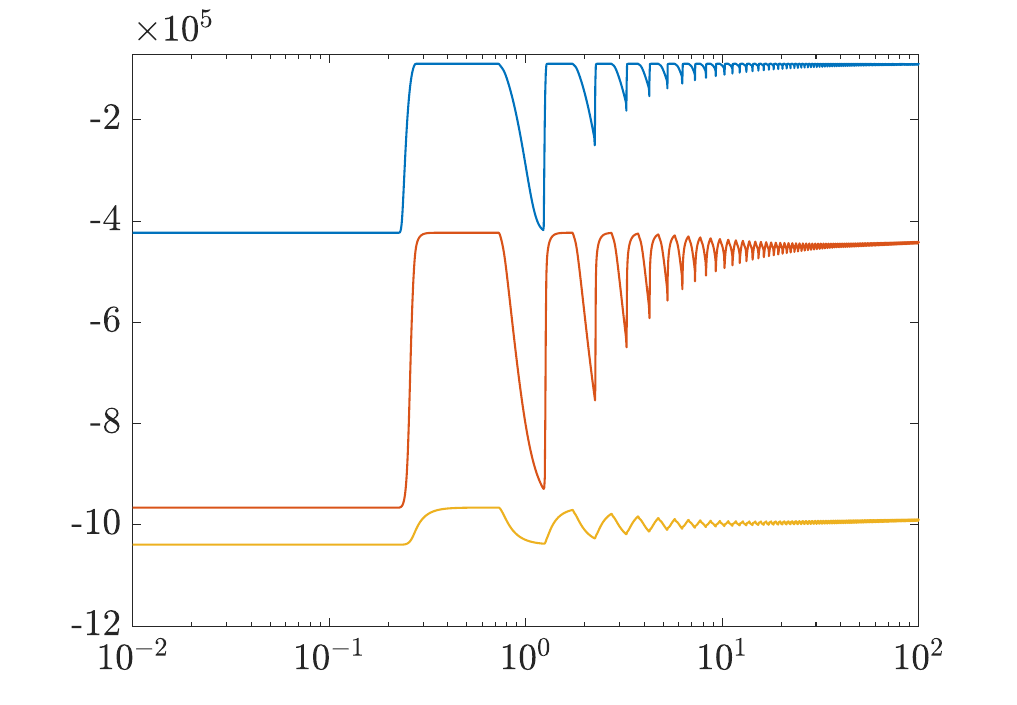}
	\put(-95,-5) {$l(t)$}
	\put(-185,120) {{\bf (h)}}
	\caption{The {rescaled:} (a), (b)  aperture $w(l(t),x) / \sqrt{l(t)}$, (c), (d) fluid velocity $\sqrt{l(t)} v(l(t),x)$, {(e), (f) fluid pressure $\sqrt{l(t)}p(l(t),x)$, (g), (h) fluid pressure derivative $l^{3/2}(t) \frac{\partial p}{\partial x} (l(t),x)$}, at the fracture opening $x=0$ as the crack propagates through the heterogeneous material whose periodic toughness has a (a), (c), (e), (g) sinusoidal, (b), (d), (f), (h) step-wise, distribution.}
	\label{Energy_Pressure}
\end{figure}

These behaviours explain the rapid fracture propagation through the weaker material. However, this also induces more fundamental changes in the fracture geometry. {To highlight this, the value of the fracture aperture at the point of injection $x=0$ is provided in  Fig.~\ref{Energy_Pressure}a,b. It can be seen that the crack geometry itself oscillates} as the crack passes between high and low toughness {layers, with the distance between the crack faces rapidly decreasing once propagation through a lower-toughness layer begins.} In this way, the elastic medium behaves as a battery, storing energy while propagating through tougher material, and then releasing it during propagation through the weaker layers. {In Figs.~\ref{Energy_Pressure}c-h, it can be seen that there are corresponding layer-dependent oscillations in the fluid velocity, pressure and pressure derivative. The impact of these rapid oscillations on the velocity of fluid throughout the crack length will be the subject of Sect.~\ref{Sect:Backflow}.}

\subsection{Limiting behaviour of the energy distribution}\label{Sect:EnergyLim}

\begin{figure}[hp!]
	\centering
	\includegraphics[width=0.45\textwidth]{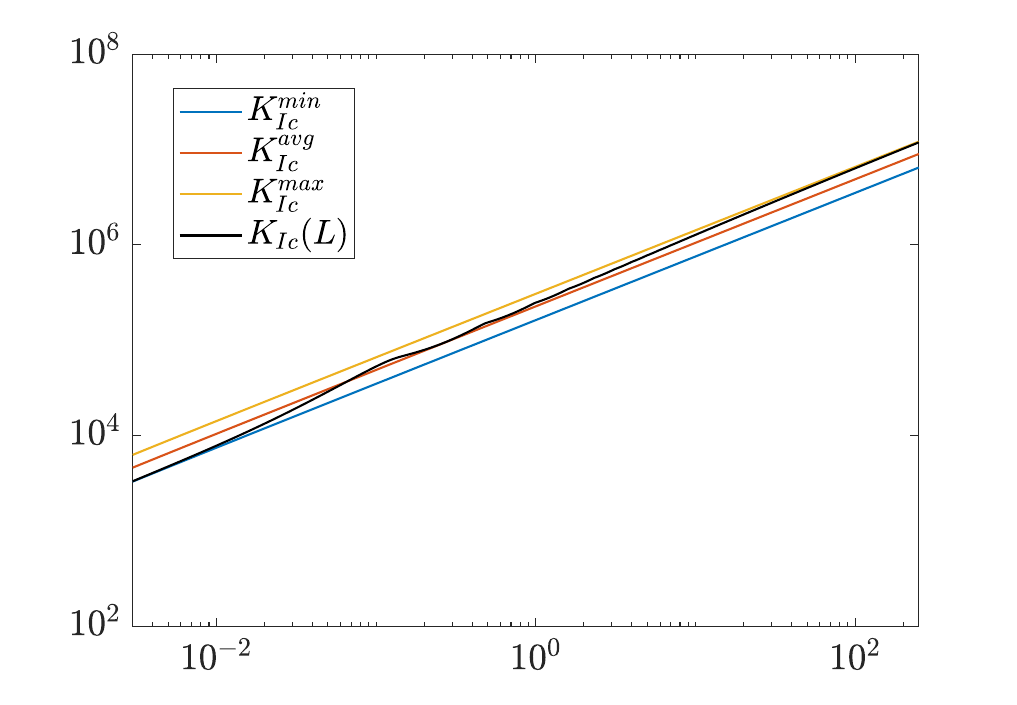}
	\put(-95,-5) {$l(t)$}
	\put(-220,60) {$\int_0^t P(\xi){\rm d}\xi$}
	\put(-185,120) {{\bf (a)}}
	\hspace{6mm}
	\includegraphics[width=0.45\textwidth]{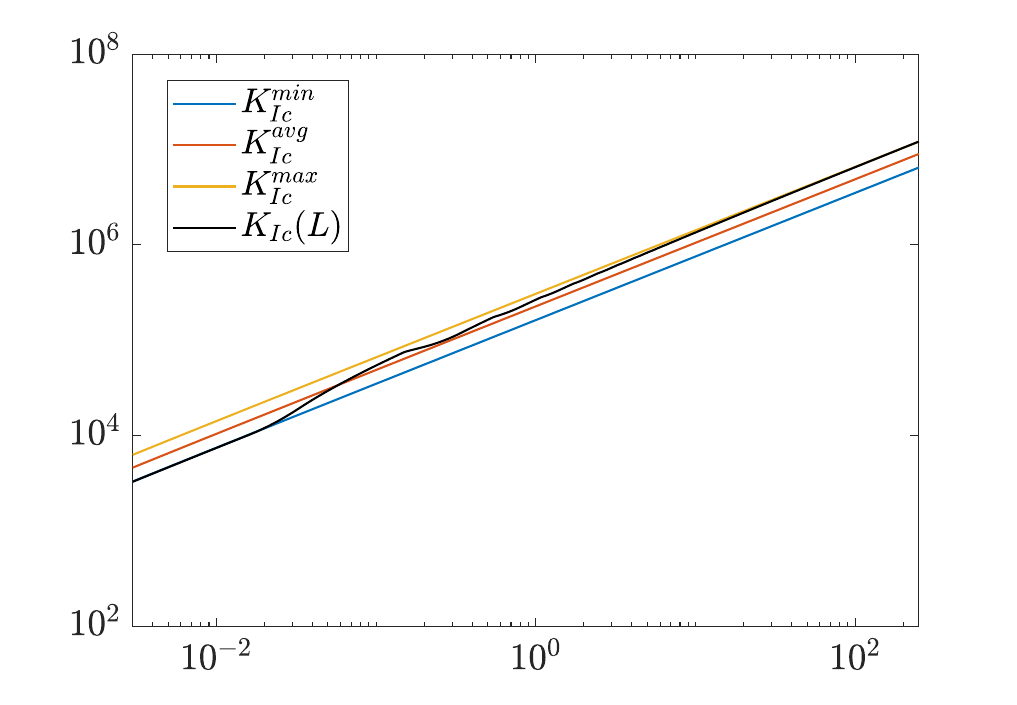}
	\put(-95,-5) {$l(t)$}
	\put(-185,120) {{\bf (b)}}
	
	\includegraphics[width=0.45\textwidth]{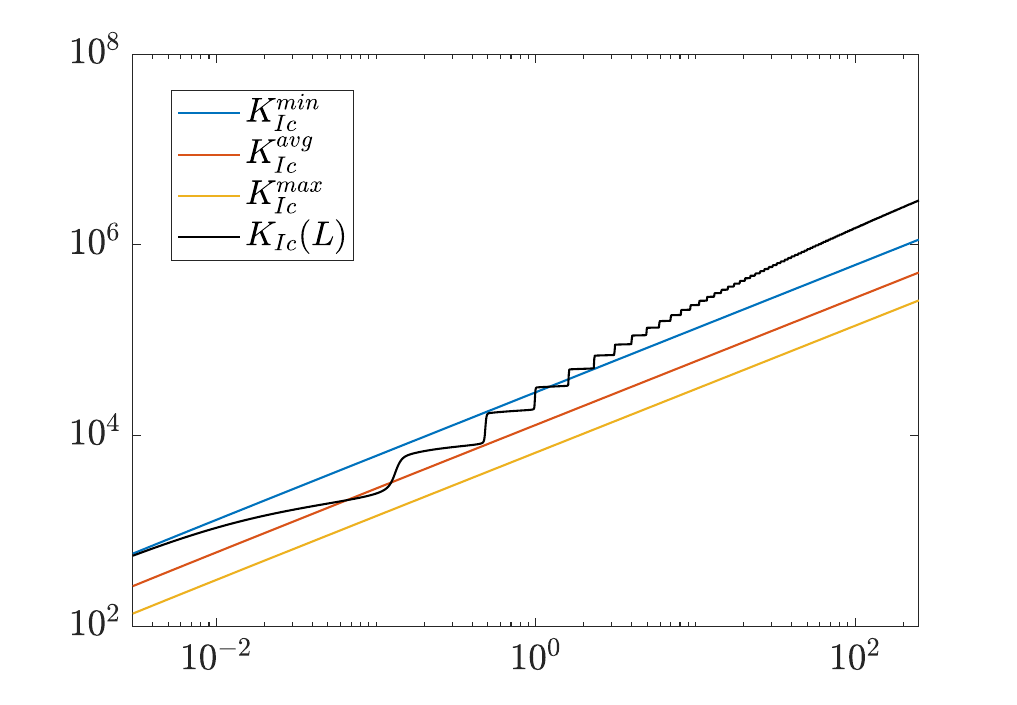}
	\put(-95,-5) {$l(t)$}
	\put(-220,60) {$\int_0^t D(\xi){\rm d}\xi$}
	\put(-185,120) {{\bf (c)}}
	\hspace{6mm}
	\includegraphics[width=0.45\textwidth]{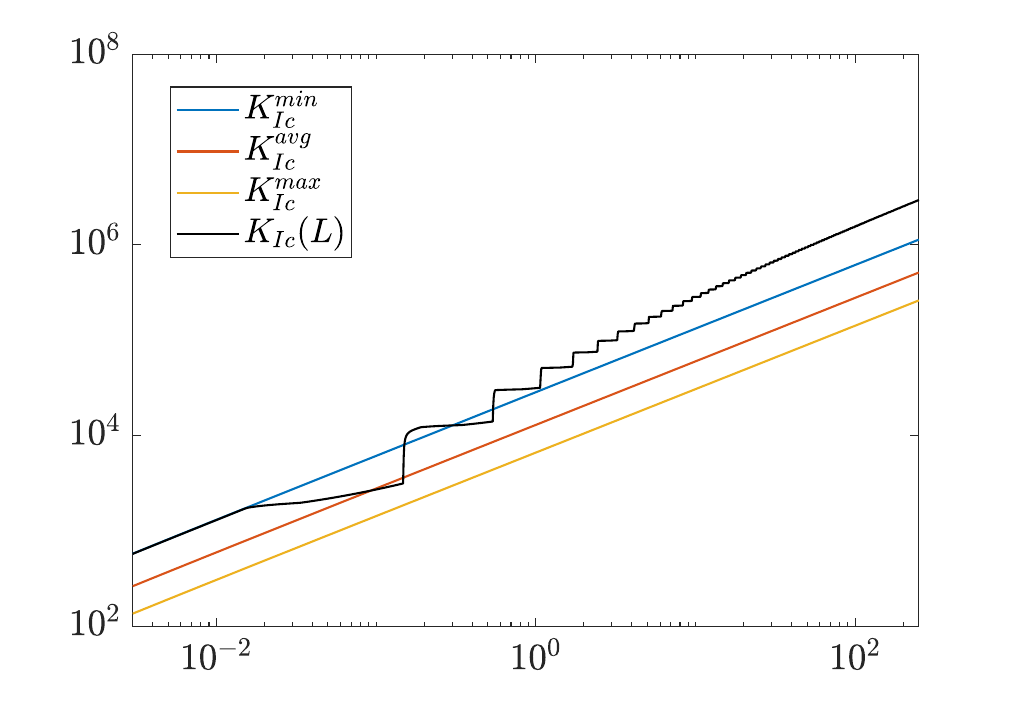}
	\put(-95,-5) {$l(t)$}
	\put(-185,120) {{\bf (d)}}
	
	\includegraphics[width=0.45\textwidth]{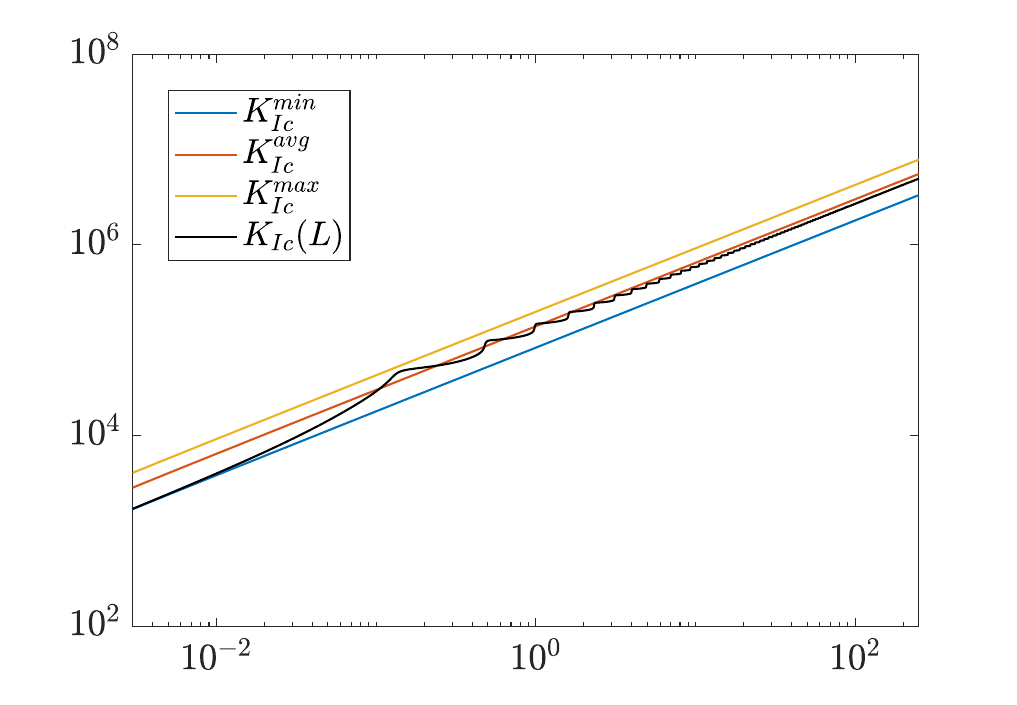}
	\put(-95,-5) {$l(t)$}
	\put(-220,60) {$\int_0^t S(\xi){\rm d}\xi$}
	\put(-185,120) {{\bf (e)}}
	\hspace{6mm}
	\includegraphics[width=0.45\textwidth]{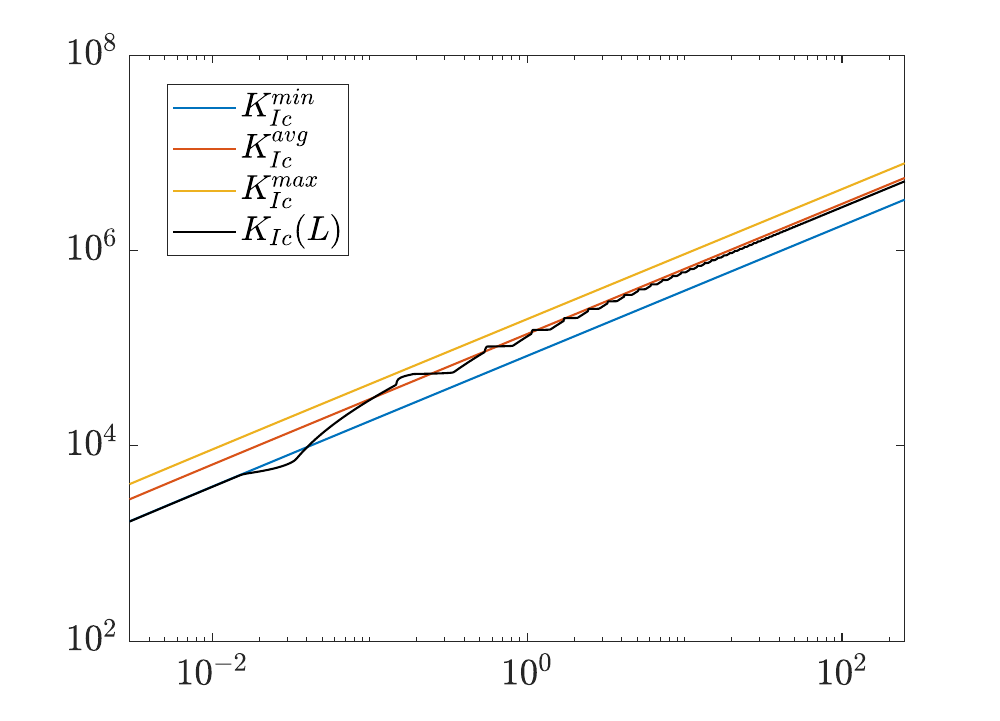}
	\put(-95,-5) {$l(t)$}
	\put(-185,120) {{\bf (f)}}
	
	\includegraphics[width=0.45\textwidth]{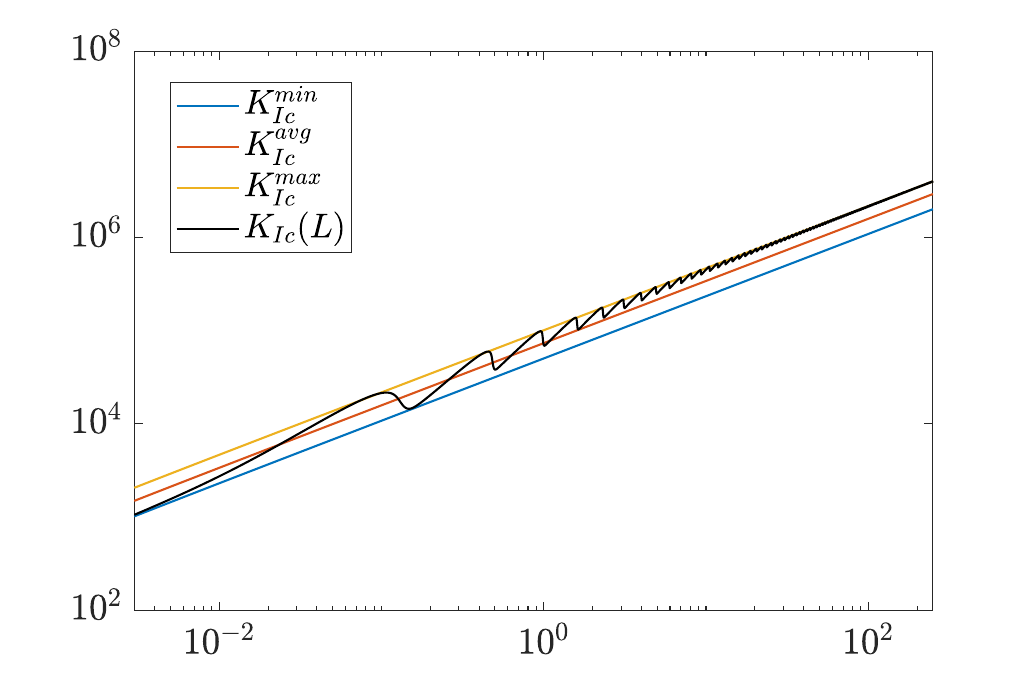}
	\put(-95,-5) {$l(t)$}
	\put(-220,60) {$\int_0^t U(\xi){\rm d}\xi$}
	\put(-185,120) {{\bf (g)}}
	\hspace{6mm}
	\includegraphics[width=0.45\textwidth]{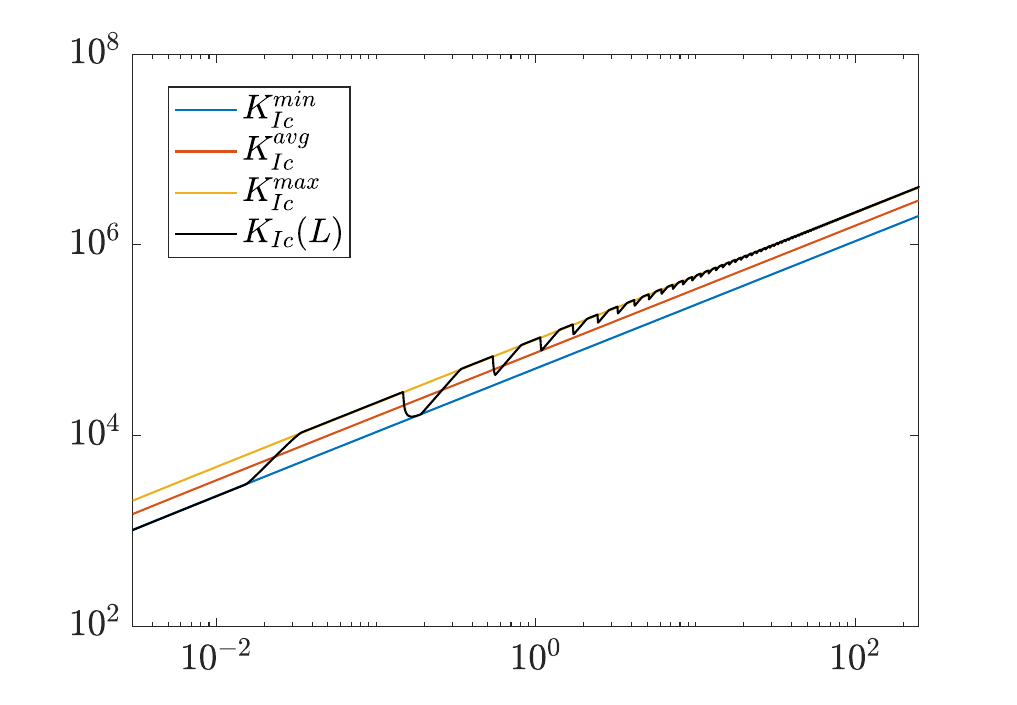}
	\put(-95,-5) {$l(t)$}
	\put(-185,120) {{\bf (h)}}
	\caption{The cumulative energy terms \eqref{energy} for the toughness-toughness case with (a), (c), (e), (g) sinusoidal, (b), (d), (f), (h) step-wise, distribution. These are shown alongside the behaviour for the case of homogeneous toughness with distributions: maximum toughness $K_{Ic}(l)\equiv K_{Ic}^{max}$, minimum toughness $K_{Ic}(l)\equiv K_{Ic}^{min}$, and the (arithmetic) average toughness $K_{Ic}(l)\equiv K_{Ic}^{avg}$. Here $P$ is the power injected into the fracture, $S$ the toughness (propagation energy), $D$ the viscous energy (stored in the fluid), and $U$ the elastic energy (stored in the solid).}
	\label{Energy_Limit_Main}
\end{figure}

With the instantaneous energy distribution considered, let us now investigate the {cumulative behaviour as the crack length becomes arbitrarily large. This case provides a means of analysing, in an averaged sense, the differing behaviour between fracture in homogeneous and heterogeneous material. It also allows the influence of the heterogeneous toughness distribution to be highlighted.} 

To {investigate the long-length crack behaviour}, we consider the cumulative energy for each term, and compare it against the behaviour expected for fracture in a homogeneous material with the toughness equal to: the maximum toughness {$K_{Ic}(l)\equiv K_{Ic}^{max}$}, the minimum toughness {$K_{Ic}(l)\equiv K_{Ic}^{min}$}, and the (arithmetic) average toughness {$K_{Ic}(l)\equiv K_{Ic}^{avg} = 0.5\left( K_{Ic}^{max} + K_{Ic}^{min}\right)$}. 

{In the viscosity-dominated case, the cumulative behaviour is largely independent of the toughness distribution. Therefore, the figures for the viscosity-dominated case has been relegated to Appendix.~B (Figs.~B.4-B.7). Meanwhile, figures for} the toughness-toughness case ($\delta_{max}=100$, $\delta_{min}=10$) are provided in Fig.~\ref{Energy_Limit_Main}. {As the cumulative behaviour of the energy strongly depends on whether toughness}\\

\noindent {\bf With both layers in the toughness-dominated regime (Fig.~\ref{Energy_Limit_Main})}\\
There is a clear difference in limiting behaviour {of the different energy terms for the heterogeneous toughness distributions in comparison for the homogeneous case(s). In the case where both layers are in the toughness-dominated regime (i.e.\ $\delta_{max}=100, \delta_{min}=10$, provided in Fig.~\ref{Energy_Limit_Main}).}
\begin{itemize}
	\item  For the power $P$ and elastic energy $U$ the oscillating (periodic) toughness tends to that of the maximum $K_{Ic}^{max}$. 
	\item The toughness term {$S$} meanwhile tends to a value just below that of the (arithmetic) average $K_{Ic}^{avg}$. 
	\item To balance this, the viscous energy $D$ {(used to overcome viscous friction on the crack faces)} {tends to a value significantly greater than those for any of the homogeneous cases. This would correspond to having a homogeneous toughness with $K_{Ic} \ll K_{Ic}^{min}$.} 
	\item {For all energy terms, there is minimal} dependence on the form of the toughness distribution, with both the sinusoidal and step-wise toughness distributions providing {very similar} results.
\end{itemize}

\noindent {\bf With {both layers in} the viscosity dominated/ intermediate regime}\\
{In the viscosity-dominated regime, the material toughness plays only a minor role in determining the fracture evolution. For this reason, all of the energy terms $P$, $U$, $S$, $D$ tend to a value close to that of the (arithmetic) mean toughness $K_{Ic}^{avg}$. There is only minor variation in results between the sinusoidal and step-wise toughness distribution. For this reason, figures for the intermediate-viscosity distribution are relegated to Appendix.~B (Figs.~B.4-B.7).}

\medskip

\noindent {\bf The intermediate case, with layers in differing regimes}\\
{The toughness-intermediate case is in between the two cases highlighted above. In this instance (see Figs.~B.4-B.7).
\begin{itemize}
	\item The power $P$ tend to a value in between that for the maximum $K_{Ic}^{max}$ and (arithmetic) average $K_{Ic}^{avg}$. 
	\item The elastic energy $U$  also tends to a value in between that for the maximum $K_{Ic}^{max}$ and (arithmetic) average $K_{Ic}^{avg}$, however in this case it is far closer to that for the maximum than $P$ is. 
	\item The toughness energy $S$ tends to that for the (arithmetic) average.
	\item The viscous energy $D$ tends to a value slightly below that for $K_{min}$. 
	\item These trends all hold for both the sinusoidal and step-wise distributions. However, $P$ and $U$ tend to the maximum faster (with increasing $\delta_{max}$) in the step-wise case than for the sinusoidal one.
\end{itemize}}
{The main difference compared to the case with both layers in the toughness-dominated regime is the lower value of viscous energy $D$, which now tends to a value below that for $K_{min}$. This is best explained by noting that the energy terms must be balanced \eqref{energy_balance_1}, and the fact that $U$ tends to that for the maximum faster than $P$.}

\section{Velocity of the fluid and the crack tip} \label{Sect:Velo}

With the energy distribution considered, let us now examine the velocity and acceleration. {This is first provided for the crack tip in Sect.~\ref{Sect:CrackTip}, to showcase the propagation behaviour. Next, in Sect.~\ref{Sect:Backflow} we highlight the movement of fluid within the fracture as it propagates through the heterogeneous material.}

\subsection{Crack tip propagation}\label{Sect:CrackTip}

As the fluid front and crack tip coincide (due to the no-lag assumption), we show graphically only the more general case of the fluid velocity throughout the fracture. These are provided in Fig.~\ref{Fluid_Velo_Overview}, for various normalised positions within the crack $\tilde x = x/l$. The case $\tilde x = 1$ therefore coincides with the velocity of the crack front. The associated fluid acceleration is meanwhile relegated to Appendix.~B (Fig.~B.8).

As expected, Fig.~\ref{Fluid_Velo_Overview} shows that the fracture rapidly propagates through the weaker material, while it slows as it permeates the tougher material, leading to rapid velocity oscillations. Comparing the various cases, it can be seen that the oscillations are far greater in the toughness-toughness case than the intermediate-viscosity one, with the formers crack tip experiencing both the fastest and slowest velocities of any case shown here. Comparisons by the authors (keeping $\delta_{max}=100$ and taking $\delta_{min}=10,1,0.1$) found that the extent of these oscillations primarily depends upon the maximum toughness, rather than the difference between the maximum and minimum toughness, although the latter does play a small role.

The toughness distribution meanwhile does have some impact, with the variations in the velocity being larger for the step-wise case than the sinusoidal one. The main impact of the toughness distribution however is on the acceleration of the crack front, as seen in Fig.~B.8 (with $\tilde x = 1$). It is apparent from those figures that the step-wise distribution experiences a peak acceleration that is consistently higher than that of the sinusoidal one. For example, in the toughness-toughness case ($\delta_{max}=100, \delta_{min}=10$) the peak acceleration of the step-wise distribution exceeds $10^5$ m/s$^2$, which is an order of magnitude greater than that of the sinusoidal one. This trend holds for all of the cases considered here, with the acceleration in the step-wise case always being an order or more greater than that for the sinusoidal distribution.

\subsection{Fluid velocity within the fracture: the backflow effect} \label{Sect:Backflow}

\begin{figure}[p!]
	\centering
	\includegraphics[scale=0.45]{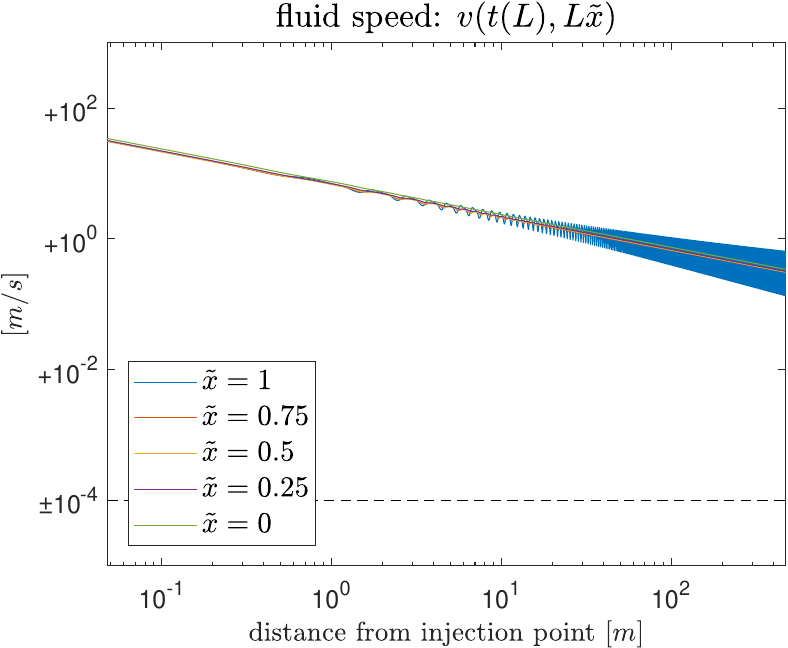}\hspace{5mm}
	\includegraphics[scale=0.45]{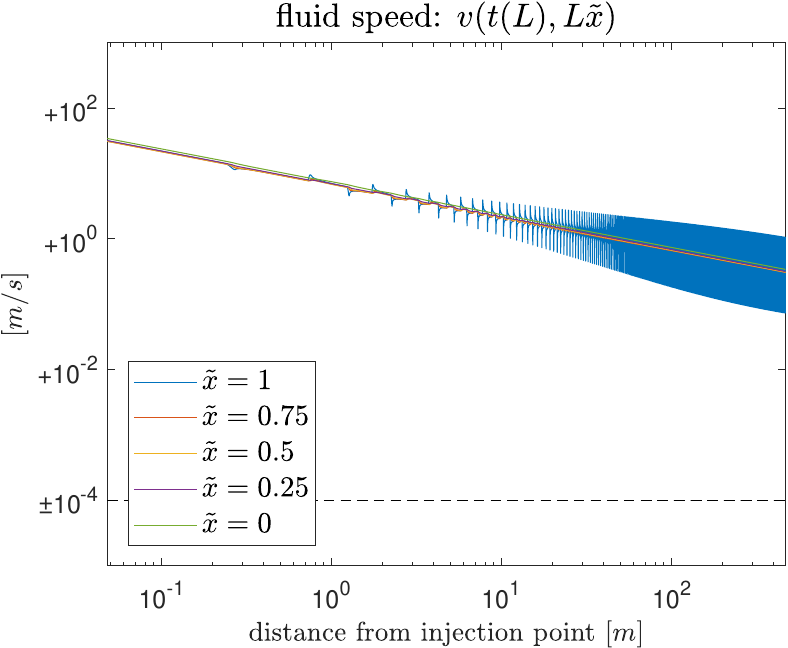}
	\includegraphics[scale=0.45]{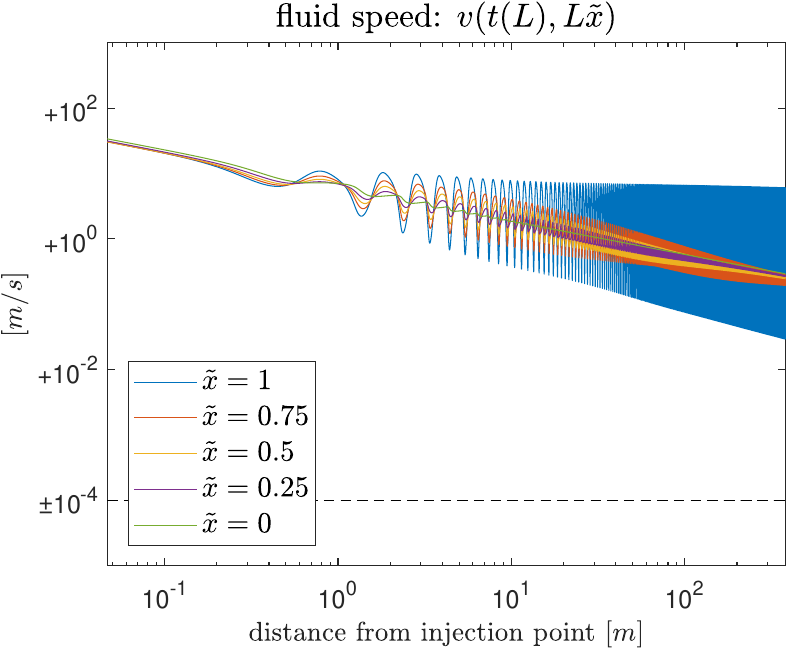}\hspace{5mm}
	\includegraphics[scale=0.45]{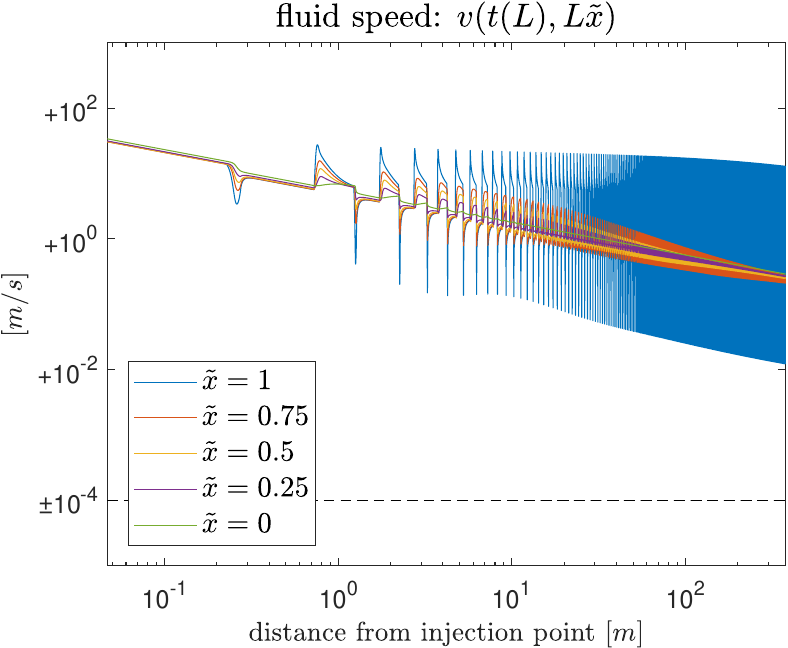}\\[4mm]
	\includegraphics[scale=0.45]{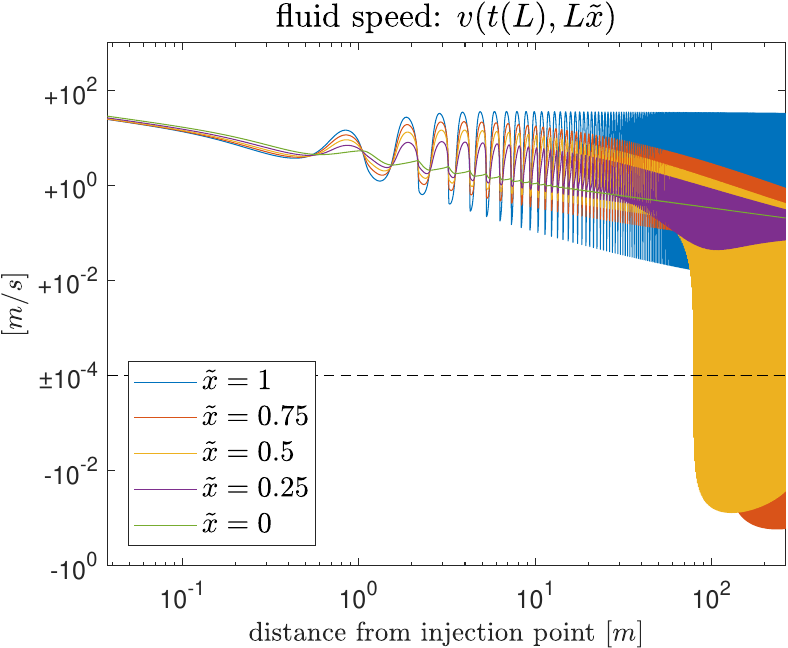}\hspace{5mm}
	\includegraphics[scale=0.45]{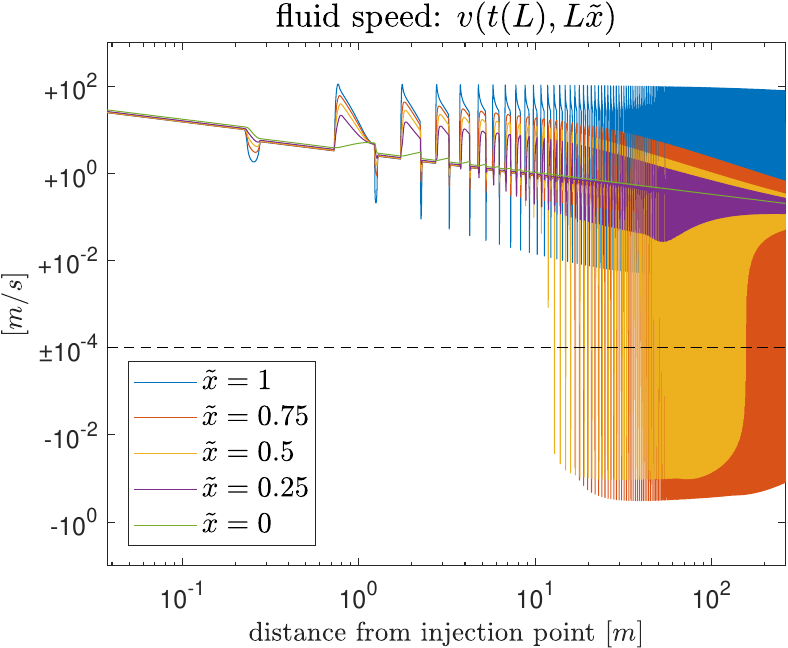}\\[4mm]
	\caption{Fluid velocity inside the fracture {as it propagates through position $l(t)$} for 5 chosen positions (beginning and the end of the fracture and at each quarter of the length). Note that the {normalised positions} move in time. (a), (b) $\delta_{max}=1$, $\delta_{min}=0.1$, (c), (d) $\delta_{max}=10$, $\delta_{min}=1$, (e), (f) $\delta_{max}=100$, $\delta_{min}=10$.}
	\label{Fluid_Velo_Overview}
\end{figure} 

{Recall that the elastic battery effect, discussed in Sect.~\ref{Sect:Battery}, means that the entire fracture geometry changes as the crack propagates through layers with differing toughness. Namely, the crack faces at a spacial position within the crack move further apart while propagating through tougher layers, before rapidly coming closer together when encountering weaker material layers. This shift in geometry has significant impacts on the fluid throughout the fracture, which we highlight in this section.}

{We therefore consider} the velocity of the fluid inside the fracture {throughout the whole crack length.} The fluid velocity within the fracture for fixed normalised spacial positions $\tilde x = x/l$ are again shown in Fig.~\ref{Fluid_Velo_Overview}, while the fluid acceleration is relegated to Appendix.~B (Fig.~B.8). 

We can see that in the intermediate-viscosity and toughness-intermediate cases there is a very similar trend for the fluid velocity as was seen for the crack tip velocity. Namely, the fluid velocity oscillates as the crack propagates between weaker and tougher layers. {Unsurprisingly, the trend is the inverse of that for the crack geometry (see Fig.~\ref{Energy_Pressure}c,d), with the rapid closing in of the crack faces associated with a high fluid velocity, while there is a significantly slower fluid velocity when the crack faces are gradually moving apart (and crack growth has slowed). This trends hold true throughout almost the whole crack length, although the size of this effect reduces away} from the crack tip and almost disappears at the fracture opening $x=0$. However, the extent to which this is the case varies depending on whether the crack is (locally) in the viscosity or toughness dominated regimes. 

Examining the toughness-toughness case however, we can see a significantly different behaviour occurring between the fracture front and midpoint. Namely, {although it behaves} similarly to the previous cases during initial propagation (until $l\approx 50$), after this point the fluid velocity begins to 'backflow' upon encountering weaker material layers. This can clearly be seen in Figs.~\ref{Fluid_Velo_Overview}e,f, where the fluid velocity at $\tilde x=0.5, 0.75$ is negative, and of order $10^0$, when transitioning from the tougher material layers to the weaker ones.

\begin{figure}[p!]
	\centering
	\includegraphics[width=0.48\textwidth]{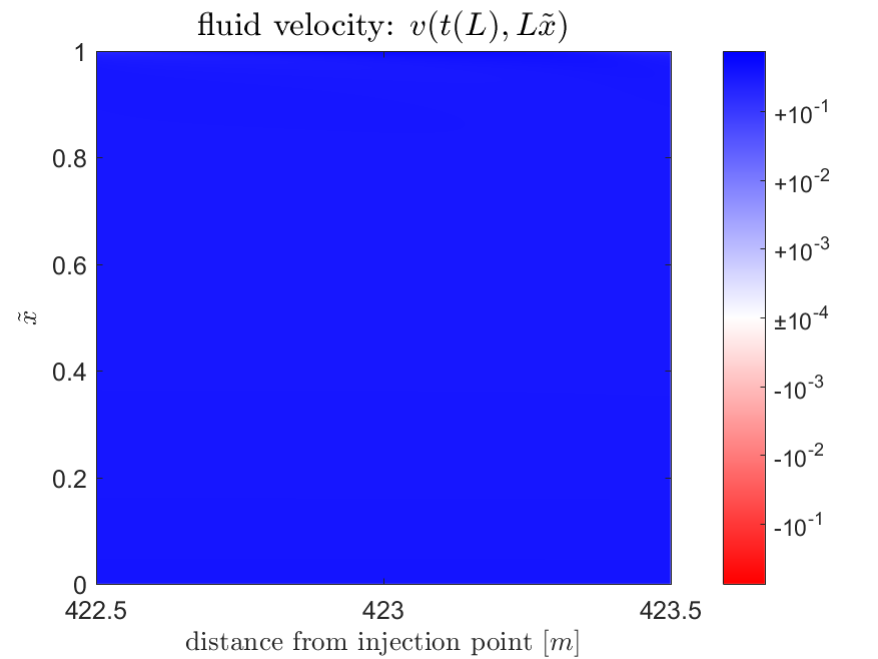}\hspace{3mm}
	\includegraphics[width=0.48\textwidth]{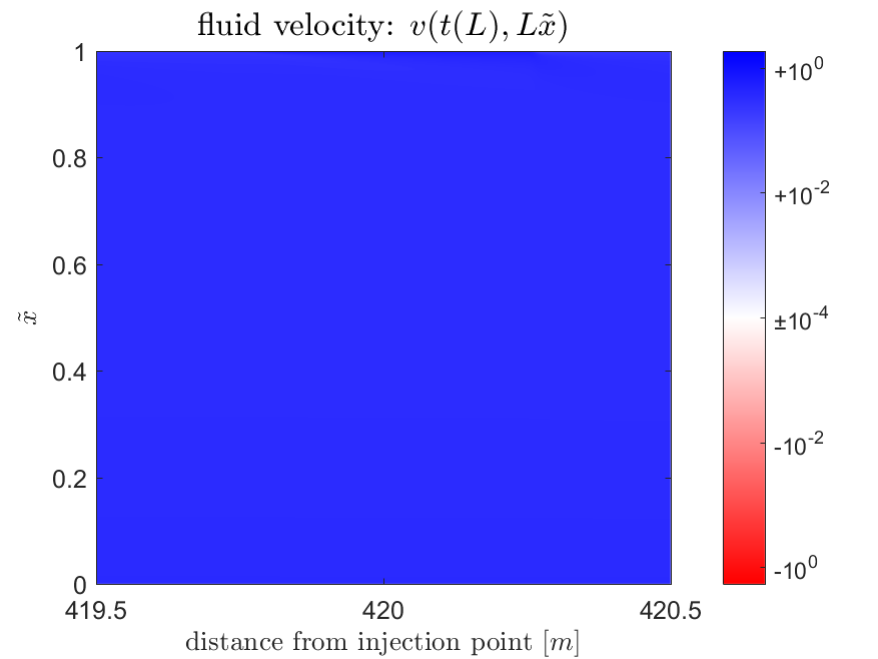}\\[3mm]
	\includegraphics[width=0.48\textwidth]{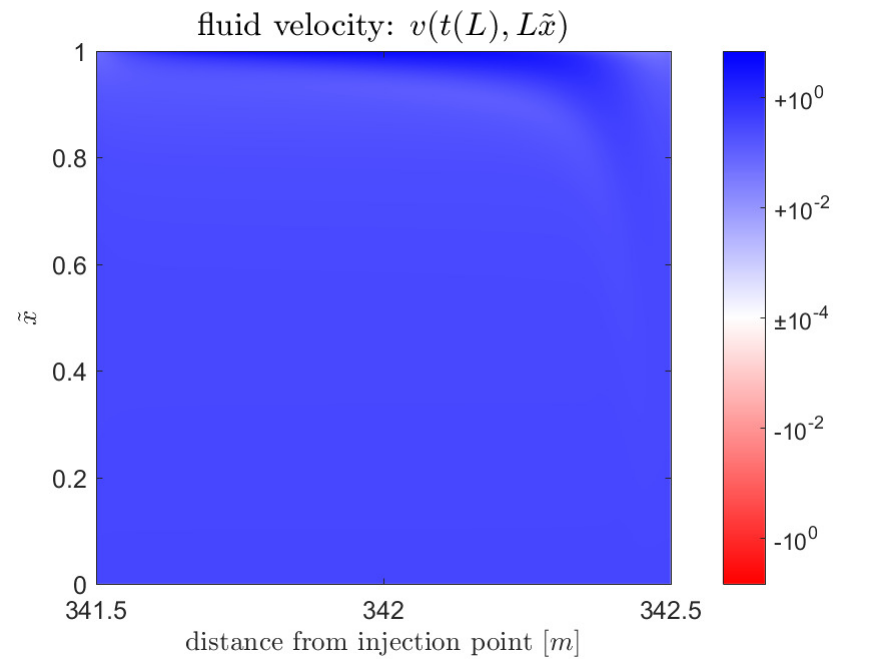}\hspace{3mm}
	\includegraphics[width=0.48\textwidth]{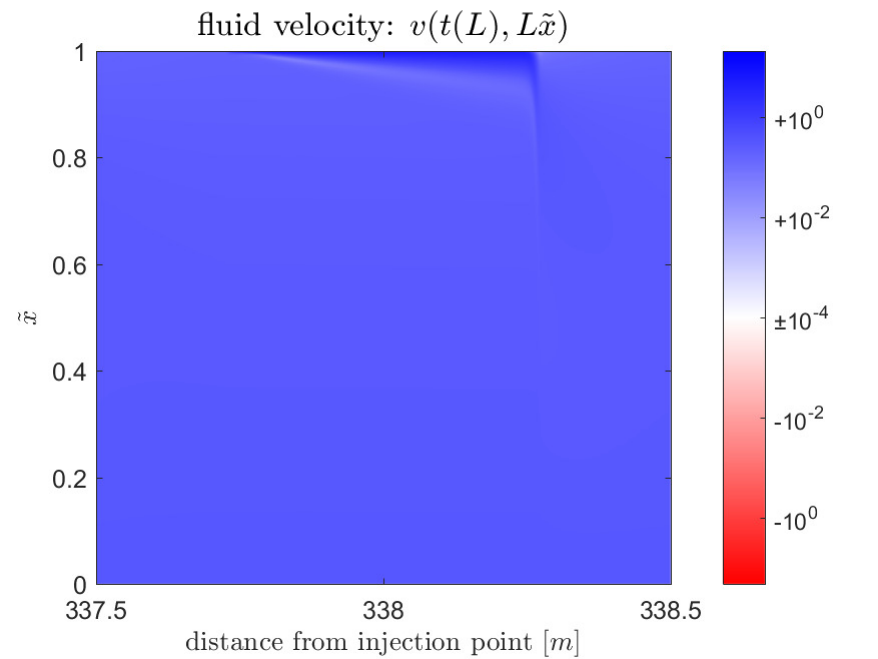}\\[3mm]
	\includegraphics[width=0.48\textwidth]{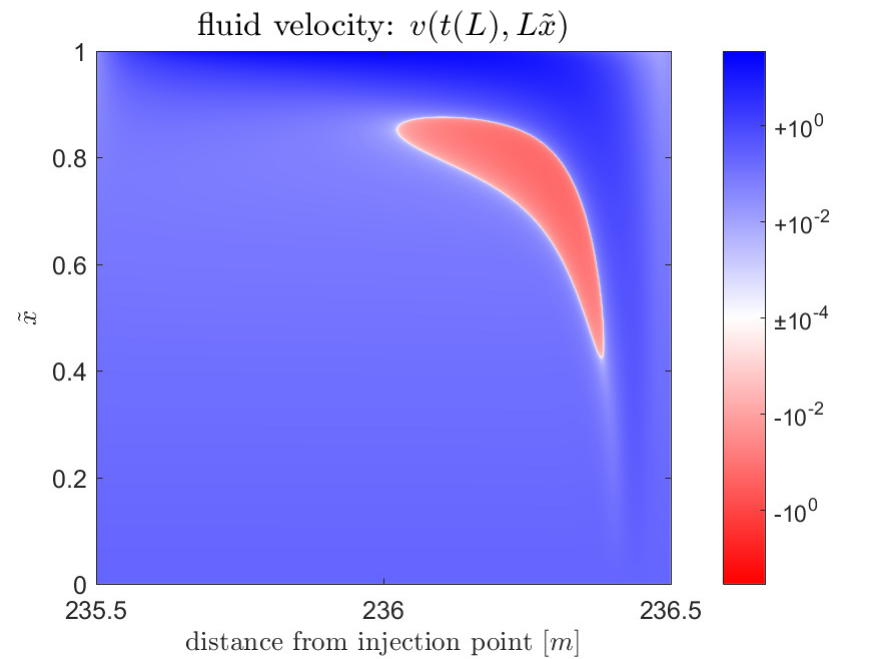}\hspace{3mm}
	\includegraphics[width=0.48\textwidth]{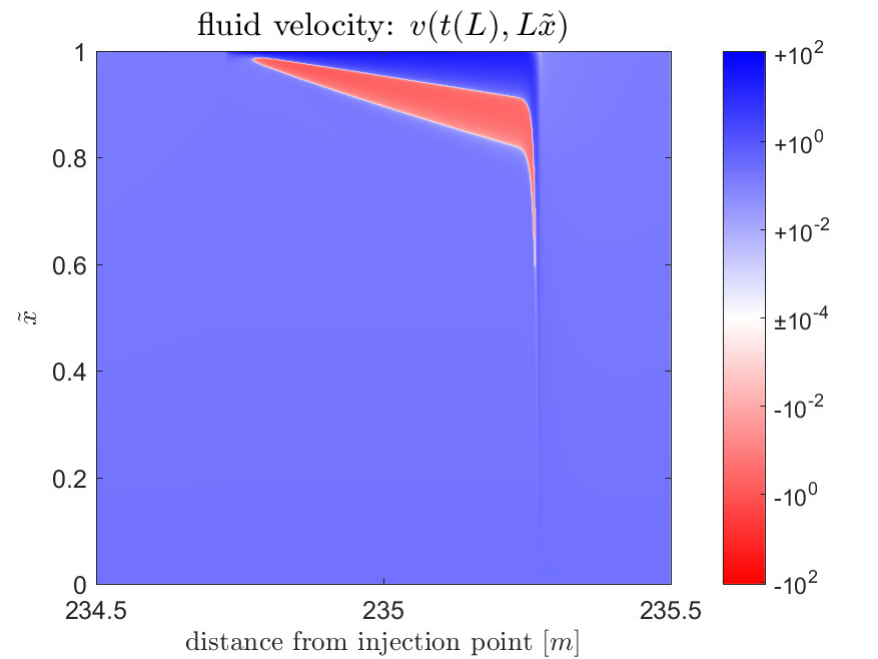}\\[3mm]
	\caption{Plots showing the fluid velocity {as the crack propagates through position $l(t)$} within a single period. (a), (b) $\delta_{max}=1$, $\delta_{min}=0.1$, (c), (d) $\delta_{max}=10$, $\delta_{min}=1$, (e), (f) $\delta_{max}=100$, $\delta_{min}=10$.}
	\label{Fluid_Velo_Zoom}
\end{figure}

{Let us} focus in more detail on the region of the fracture where this backflow occurs. In Fig.~\ref{Fluid_Velo_Zoom} we {show the} behaviour within a single toughness period. We can see that the extent of the backflow effect is strongly dependent on the maximum toughness, with there being a strong backflow effect in the toughness-toughness case, but in the toughness-intermediate case there are only regions of fluid stagnation (without backflow). These effects are entirely absent in the intermediate-viscosity case, for which fluid always travels towards the crack tip.  

\begin{figure}[t!]
	\centering
	\includegraphics[width=0.48\textwidth]{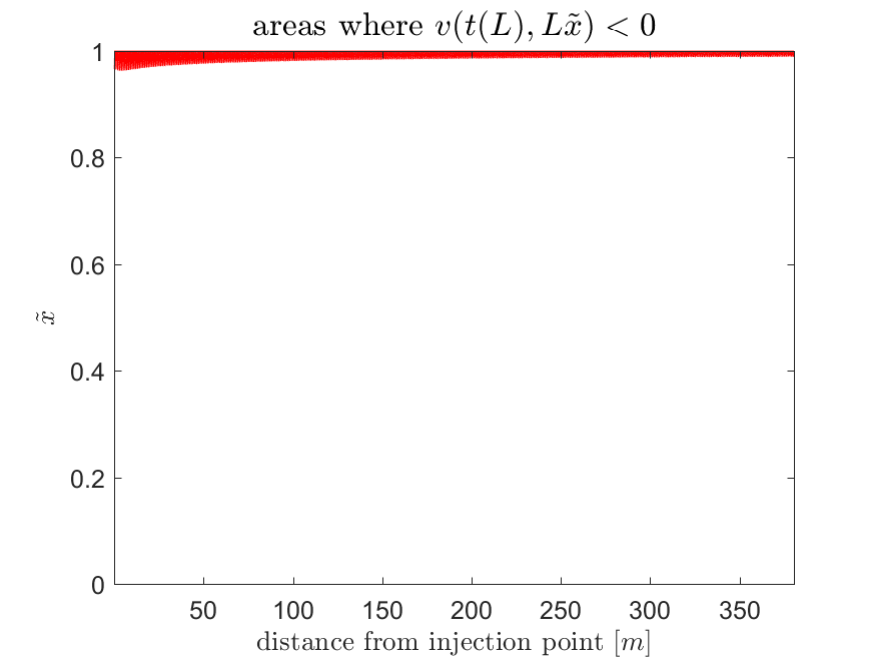}\hspace{3mm}
	\includegraphics[width=0.48\textwidth]{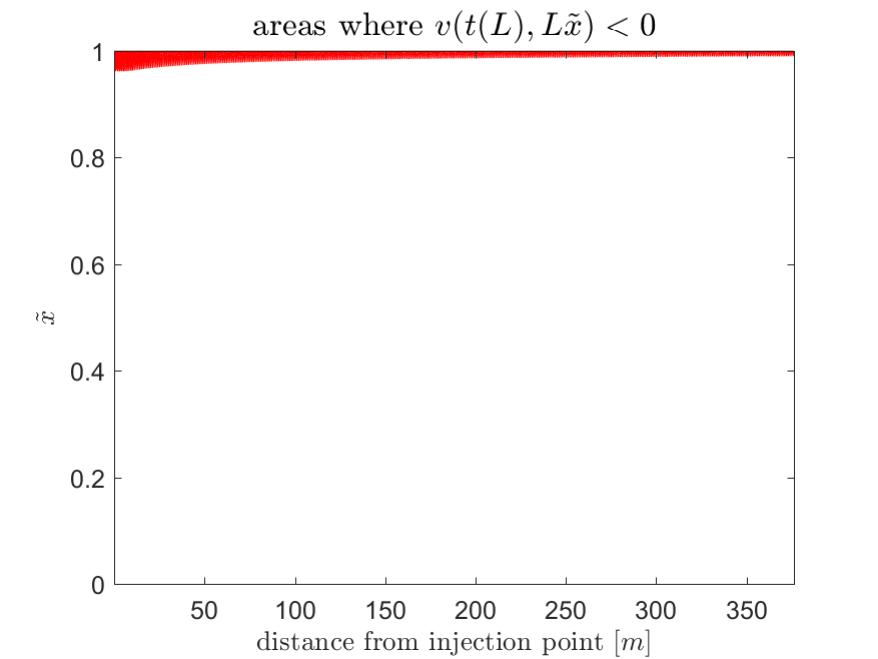}\\[3mm]
	\includegraphics[width=0.48\textwidth]{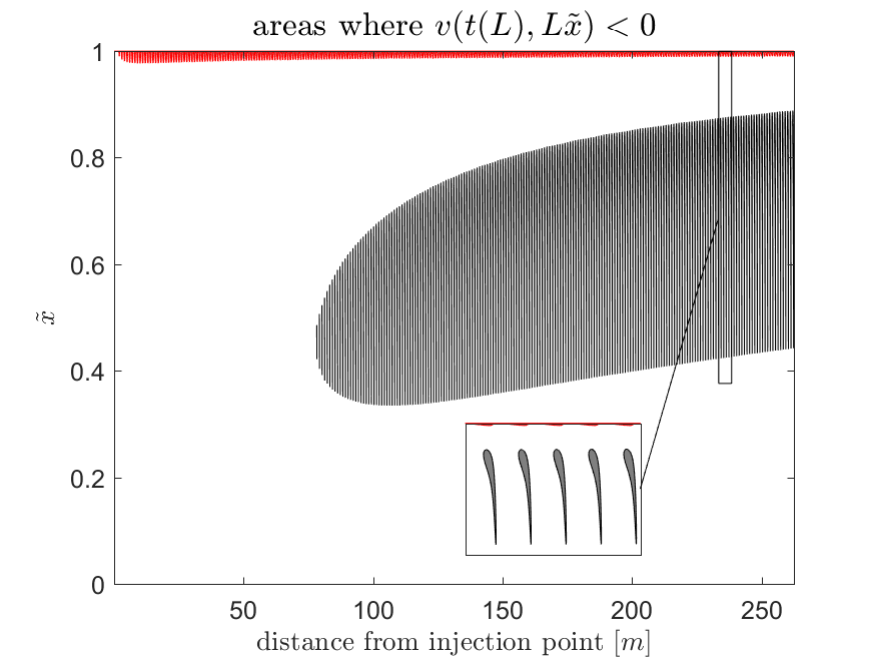}\hspace{3mm}
	\includegraphics[width=0.48\textwidth]{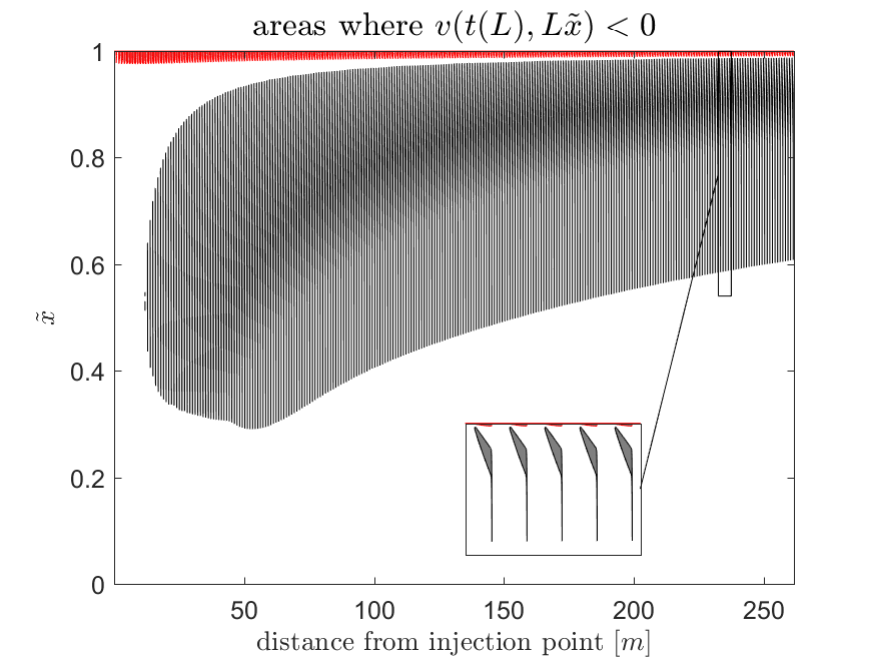}\\[3mm]
	\caption{Plots showing {which regions within the fracture are experiencing a negative velocity as the crack propagates through position $l(t)$ for:} 
		(a), (b) $\delta_{max}=10$, $\delta_{min}=1$, (c), (d) $\delta_{max}=100$, $\delta_{min}=10$. The regions coloured red denote the portion near to the crack tip where the pressure derivative is negative.}
	\label{Fluid_Velo_Neg}
\end{figure}

{Note that when backflow does occur, it is in part dependent on the toughness distribution. This is highlighted in Fig.~\ref{Fluid_Velo_Zoom}e,f, where it can be seen that the backflow occurs in different portion of the crack for the step-wise and sinusoidal distributions. Namely, for the step-wise toughness distribution the region with a negative fluid velocity begins immediately behind the crack tip, before moving backwards through a small portion of the crack. In the sinusoidal case, the backflow begins further away from the crack tip before moving backwards through the crack. Interestingly, even though the backflow has a significantly higher velocity in the step-wise case (of order $10^2$ for step-wise, $10^0$ for sinusoidal), the backflow propagates through more of the crack length for the sinusoidal distribution. This is due to the elastic battery `discharge' being an almost instantaneous process for the step-wise distribution, resulting in a higher velocity but more localised backflow. Meanwhile for the sinusoidal distribution the `discharge' is a more gradual process, producing a slower velocity backflow that is able to propagate further along the crack length. This difference in discharge behaviour also leads to the backflow initiating at differing start times, with it occurring for the step-wise distribution as soon as the weaker layer is encountered ($l=234.75$ in Fig.~\ref{Fluid_Velo_Zoom}f), but not until the crack has propagated half-way through the weaker layer for the sinusoidal distribution ($l=236$ in Fig.~\ref{Fluid_Velo_Zoom}e).}

{This behaviour within a single toughness period is mirrored through the later periods of fracture propagation. This can be seen in Fig.~\ref{Fluid_Velo_Neg}, which highlights the regions within the crack experiencing a negative fluid velocity over the whole HF process. It is clear that, while some critical initial crack length needs to be achieved before the fluid backflow initiates, once the crack length exceeds this value backflow will occur whenever a weaker material layer is encountered.} The toughness distribution plays a crucial role here, with the step-wise (layered) distribution exhibiting the backflow effect after a far shorter propagation distance (time), and over a far wider portion of the crack, than the sinusoidal toughness distribution.

This backflow effect is physical, and has been replicated by others whose solvers are based on significantly differing assumptions\footnote{Peruzzo et al replicated this effect in their solver for heterogeneous material (see \cite{PeruzzoThesis}, and upcoming papers). The present authors are grateful to them for these discussions.}. Crucially, this {phenomenon} can be explained in terms of the elastic battery outlined in Sect.~\ref{Sect:Battery}. The crack geometry changes as it transitions from propagating in a tougher material to a weaker one. Notably, the aperture decreases throughout the crack. The resulting redistribution of fluid is the cause of the backflow effect. 

\section{Discussions and Conclusions}\label{Sect:Conc}

Essential features of HF propagation in media with heterogeneous fracture toughness were investigated. A `universal solver'-type algorithm was implemented to provide high accuracy results for the fracture growth, profile and energy distribution. The key findings are as follows.
\begin{itemize}
	\item {\bf Limiting behaviour:} A new parameter $C_\infty$ was introduced to describe the limiting regime of a hydraulic fracture through heterogeneous media as crack length $l(t)\to\infty$. This value, provided it is constant (see analysis in Appendix.~A), can be used to quantify the impact of the heterogeneity on the crack behaviour. Crucially, the value of $C_\infty$ for oscillating toughness $K_{Ic}(l(t))$ does not converge to that for the maximum toughness $K_{Ic}^{max}$, and is dependent on the toughness distribution (i.e.\ sinusoidal or step-wise).
	\item {\bf Energy distribution:} The energy distribution within a hydraulic fracture travelling through heterogeneous rock was investigated. The four primary terms, the power, toughness, viscous and elastic energy were evaluated for the case of periodic toughness distribution with a step-wise and sinusoidal (infinitely layered) distribution.
	\item {\bf Elastic battery:} It was demonstrated that the solid media acts as an elastic battery. Elastic energy is stored during propagation through tougher layers, before being rapidly released to the fluid and to drive crack propagation once the crack tip reaches a weaker layer. This elastic battery {phenomenon} was most pronounced when the maximum material toughness promoted toughness-dominated behaviour, but was still present over long length-scales for viscosity-dominated fractures.
	\item {\bf Oscillation of the crack tip velocity:} As previously observed, there were rapid changes in fracture velocity during propagation through differing toughness layers. This follows from the elastic battery phenomena, which also describes more fundamental changes in crack geometry (e.g. the crack aperture).
	\item {\bf Fluid backflow:} A certain fraction of the fluid volume inside the crack moved in the opposite direction to the crack tip motion. {The extent (or existence) of backflow depends upon the maximum toughness of the material along the crack length.  When present, this fluid backflow initiates after the fracture reaches some critical crack length, which depends on the toughness distribution of the material, and occurs thereafter whenever the crack encounters a weaker material layer. The proportion of the crack length within which fluid backflow occurs also depends upon the toughness distribution.} 
\end{itemize}

{The presented work provides physical explanations for observed phenomena, as well as predicting new impacts of the toughness heterogeneity.} For example, the fluid backflow may lead to sediment redistribution during fracking operations, increasing the likelihood of blockages during flowback operations. The elastic battery phenomenon meanwhile indicates that the likelihood of crack arrest following injection shut-in (see e.g. \cite{MORI2021151}) should strongly depend on the toughness of the rock-layer at the time fluid injection ceases. Furthermore, as only Newtonian fluid was considered in this work, the impact of the fracturing fluid on the elastic battery is left unanswered; with it possible that it occurs even for dry cracks. {In the case with complex fluid rheology, such as Carreau fluids, it is possible that this may lead to a change in behaviour over an even larger portion of the crack length. The impact of leak-off on the fluid backflow is also unclear.} Targeted investigation of this phenomenon may also explain the accumulation of elastic energy before earthquakes in the lithosphere. More targeted investigations are required to quantify such effects. 

Meanwhile, the results for the crack tip velocity oscillation and the fluid backflow are physical effects arising from the change in crack geometry during fracture propagation. These results, and in particular the extreme acceleration of the fluid when propagating between material layers, raises immediate questions about the validity of neglecting the inertia term in the governing equations \eqref{continuity}-\eqref{fluid_balance}. Incorporation of this term may well dampen the oscillations seen for the acceleration of the crack tip in heterogeneous media, and lead to qualitative differences in the evolution of the fracture geometry over time. The energy balance may also be impacted by the incorporation of the kinetic energy term.

\section*{CRediT authorship contribution statement}

{\bf Daniel Peck} analysis of results, visualisation, produced the manuscript. {\bf Gaspare Da Fies} produced {the code} for KGD in heterogeneous media, performed initial analysis, initial visualisation. {\bf Ivan Virshylo} analysis of results, reviewed/edited the manuscript. {\bf Gennady Mishuris} oversaw all research, {contributed} ideas, analysis of results, reviewed/edited the manuscript.

\section*{Acknowledgements}
The research is supported by European project funded by Horizon 2020 Framework Programme for Research and Innovation
(2014–2020) (H2020-MSCA-RISE-2020) Grant Agreement number 101008140 EffectFact ``Effective Factorisation techniques for
matrix-functions: Developing theory, numerical methods and impactful applications’’.  The authors acknowledge support from the project within the Innovate Ukraine competition, funded by the UK International Development and hosted by the British Embassy Kyiv. {The authors are grateful to the reviewers for their helpful comments.}

\bibliography{Penny_Bib}
\bibliographystyle{apalike}

\setcounter{figure}{0}
\renewcommand{\thefigure}{\thesection.\arabic{figure}}
\setcounter{table}{0}
\renewcommand{\thetable}{\thesection.\arabic{figure}}
\setcounter{equation}{0}
\renewcommand{\theequation}{\thesection.\arabic{equation}}

\appendix



\section{The limiting value $C_\infty$}\label{Append_Cinfty}

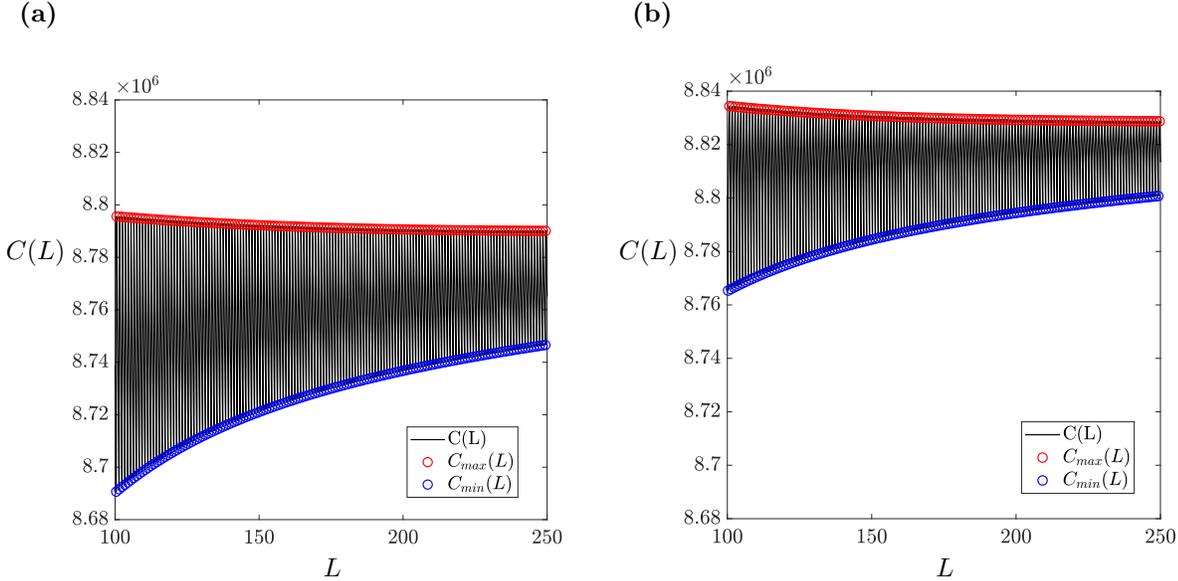
\begin{figure}[b!]
	\centering
	\includegraphics[width=0.45\textwidth]{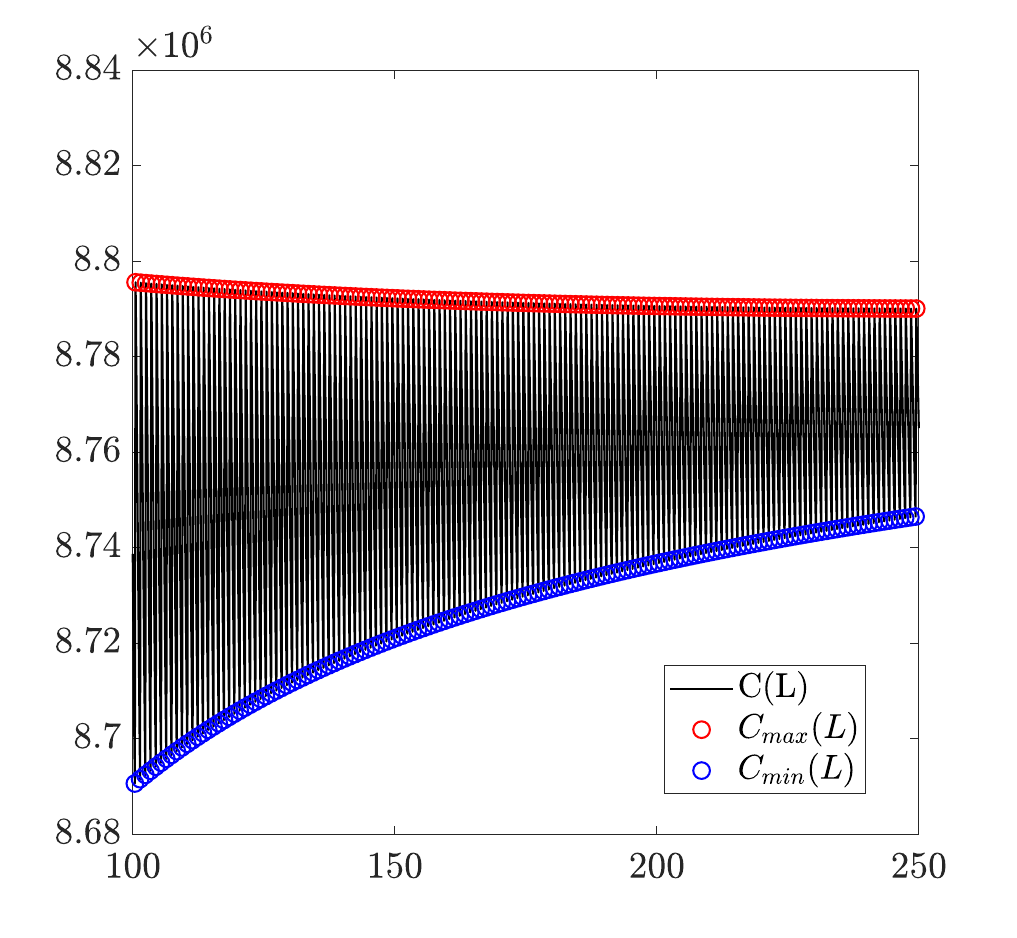}
	\put(-95,-5) {$l(t)$}
	\put(-195,85) {$C(l)$}
	\put(-195,155) {{\bf (a)}}
	\hspace{6mm}
	\includegraphics[width=0.45\textwidth]{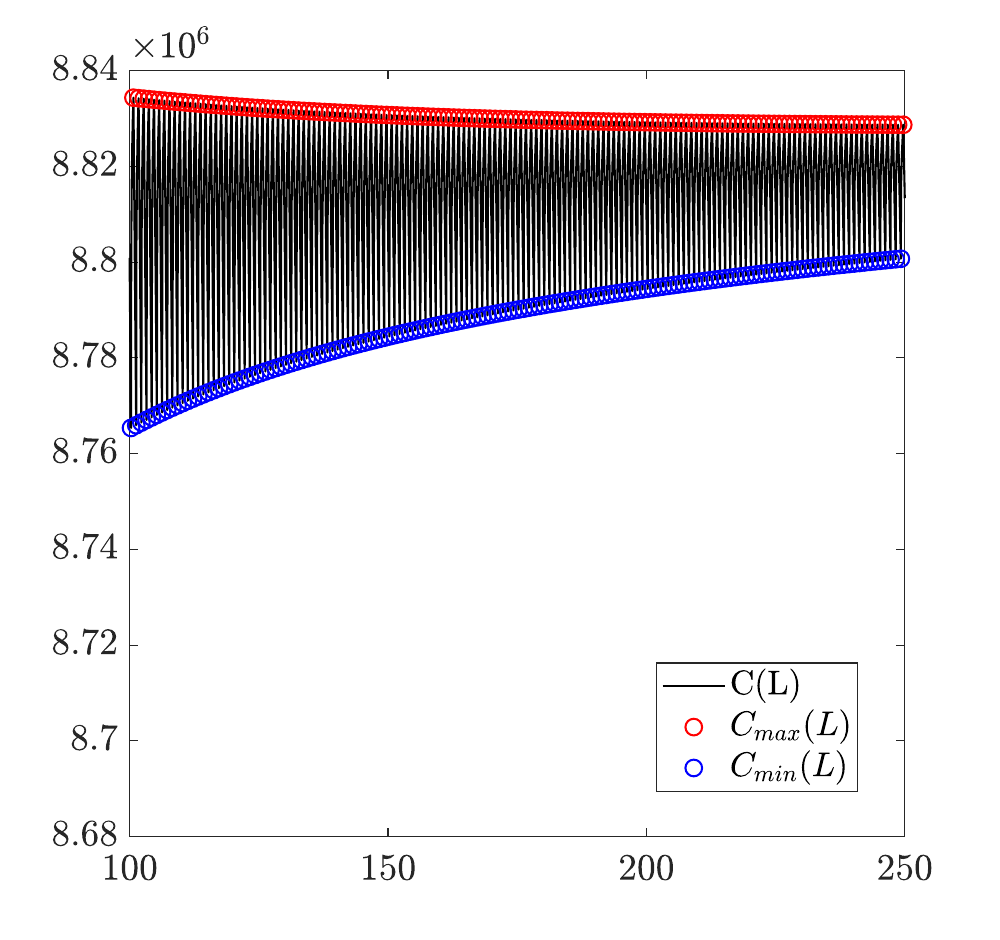}
	\put(-95,-5) {$l(t)$}
	\put(-195,85) {$C(l)$}
	\put(-195,155) {{\bf (b)}}
	\caption{The value of $C(L(l))$ (15), alongside the distribution of its maximum within each period $C^{max}(l(t))$ and minimum $C^{min}(l(t))$ when $\delta_{max}=100$, $\delta_{min}=10$ for the (a) sinusoidal, (b) step-wise, toughness distribution.}
	\label{Fig_Cmax_Cmin}
\end{figure}

To investigate the limiting value of the parameter $C(l(t))$ (15) as the crack length $l\to\infty$, we first extract the minimum and maximum values of the parameter within each period, denoted $C^{min}(l)$ and $C^{max}(l)$ respectively. Examples of these can be seen in Fig.~\ref{Fig_Cmax_Cmin}. Determining whether $C(l)\to C_\infty$, some constant, as $l\to\infty$ is now equivalent to demonstrating that $C^{min}(l)$ and $C^{max}(l)$ have the same limiting value as $l\to\infty$.

To investigate the limits of the two distributions $C^{max}$ and $C^{min}$, we utilize the least-squares method. We first seek to approximate each of these in the form
\begin{equation} \label{Cinf_LS1}
	C^{max/min} \left(l_k\right) = C_\infty^{max/min} + A l_k^n + B l_k^{2n},
\end{equation}
where the values of $C_\infty$, $A$, $B$, and $n$ are computed separately for $C^{max}(l)$ and $C^{min}(l)$ to obtain each approximation. The limiting values $C_\infty^{max}$ and $C_\infty^{min}$ are then compared, with their relative differences given in Table.~\ref{Table_RelDiff}. It can be seen that the relative differences are small in almost all cases, although it does exceed $10^{-1}$ in one instance (sinusoidal distribution with $\delta_{max}-\delta_{min}$ being $10-1$).

While it is not possible to conclude whether the values converge to the same constant $C_\infty^{max}, C_\infty^{min}\to C_\infty$ (and therefore $C(l)\to C_\infty$ as $l\to\infty$ in general), in part because differing values can be obtained depending on the assumed distribution, we can state from the above that this will be a reasonable approximation in the cases considered.

Finally, if we assume that $C(l)\to C_\infty$ as $l\to\infty$, then we compute $C_\infty$ using a combination of both distributions. Namely, noting that both distributions yielded $n\approx -0.5$ we seek a shared $C_\infty$ that minimises
$$
\left\{ \begin{array}{l}
	C^{max} (l_i) = C_\infty + A l_i^{-\frac{1}{2}} + B l_i^{-1} , \\
	C^{min} (l_j) = C_\infty + D l_j^{-\frac{1}{2}} + E l_j^{-1} .
\end{array}\right.
$$
This is obtained by seeking $C_\infty$, $A,B,C,D,E$ that minimise
$$
\sum_i \left( C^{max} (l_i) - C_\infty - A l_i^{-\frac{1}{2}} - B l_i^{-1} \right)^2 + \sum_j \left( C^{min} (l_j) - C_\infty - D l_j^{-\frac{1}{2}} - E l_j^{-1} \right)^2  .
$$
The values of $C_\infty$ provided in Table.~3 were obtained using this least-squares approximation applied to the final $50$ toughness periods (to avoid early-time behaviour impacting the long-time approximation). The parameter $\delta_\infty$ then follows immediately from (17).

\begin{table}[t]
	\centering
	\begin{tabular}{||l || c | c | c| c||}
		\hline \hline
		$\delta_{max} - \delta_{min}$ & $1-0.1$ & $10-1$ & $100-1$ & $100 - 10$ \\
		\hline \hline
		Sinusoidal & $6.11\times 10^{-3}$ & $1.44\times 10^{-1}$ & $1.07\times 10^{-3}$ & $6.46\times 10^{-2}$  \\
		\hline
		Square & $1.18\times 10^{-2}$  & $3.00\times 10^{-4}$ & $7.45\times 10^{-3}$ & $2.64\times 10^{-2}$   \\
		\hline\hline
	\end{tabular}
	\caption{Relative difference between $C_\infty^{max}$ and $C_\infty^{min}$ obtained using the least-squares approximation \eqref{Cinf_LS1} (to 3 significant figures).}
	\label{Table_RelDiff}
\end{table}




\setcounter{figure}{0}
\section{Figures}\label{Append_ExtraFigs}
For the sake of readability, not all figures were included in the main text. Here, we include the remaining figures, alongside some brief discussion. \\

\begin{figure}[hp!]
	\centering
	\includegraphics[width=0.45\textwidth]{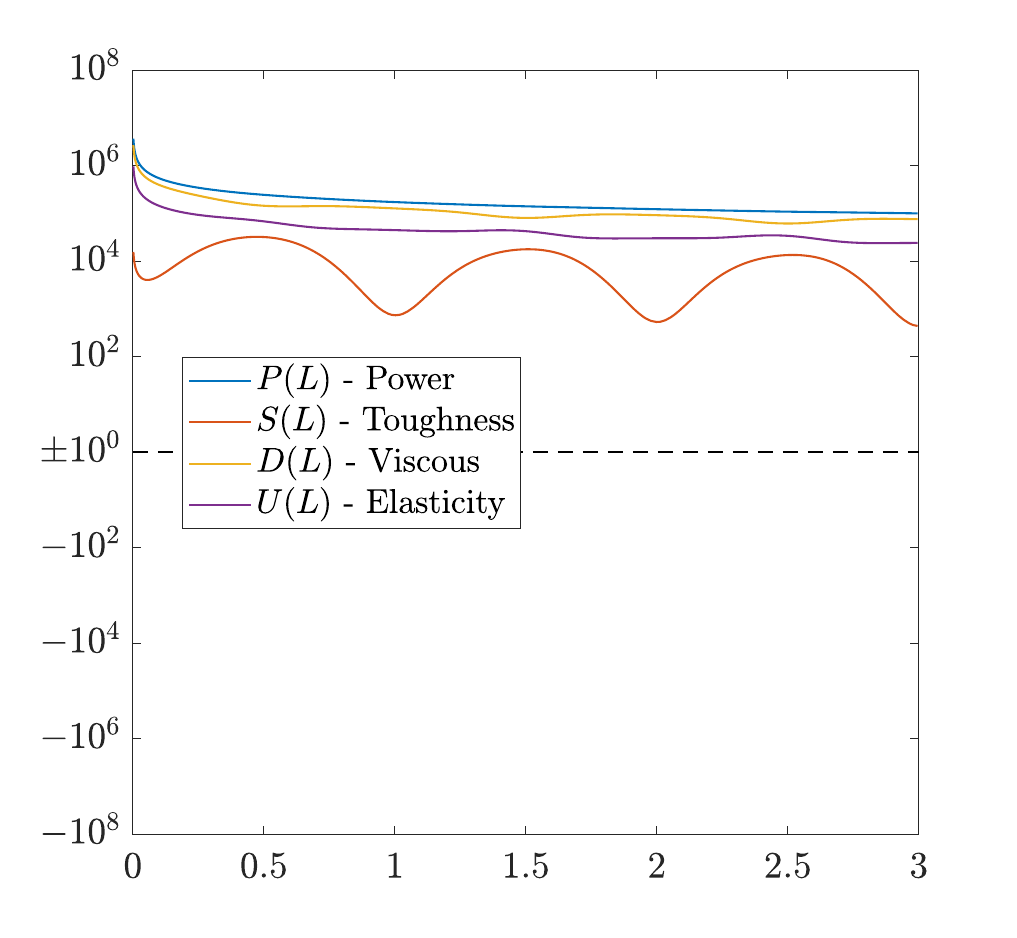}
	\put(-95,-3) {$l(t)$}
	\put(-205,85) {Power}
	\put(-195,155) {{\bf (a)}}
	\put(-105,155) {$k=1,2,3$}
	\hspace{6mm}
	\includegraphics[width=0.45\textwidth]{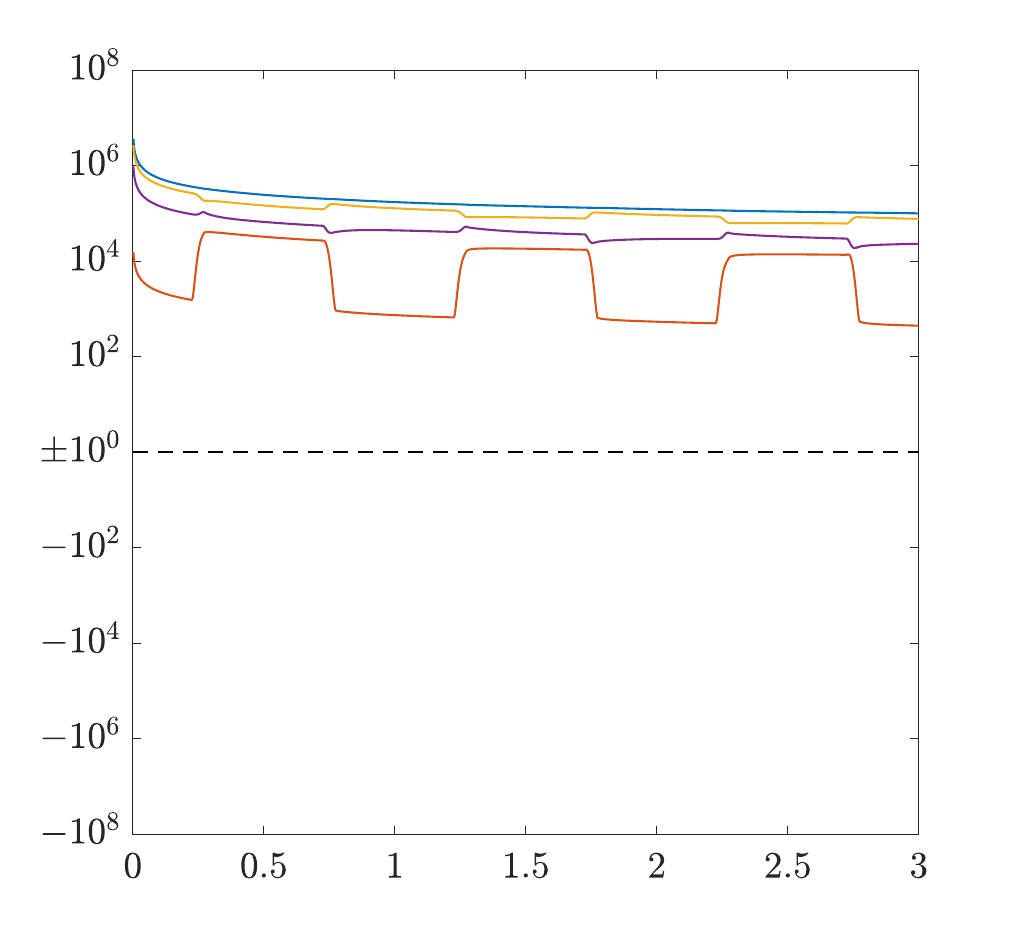}
	\put(-95,-3) {$l(t)$}
	\put(-195,155) {{\bf (b)}}
	\put(-105,155) {$k=1,2,3$}
	
	\vspace{2mm}
	
	\includegraphics[width=0.45\textwidth]{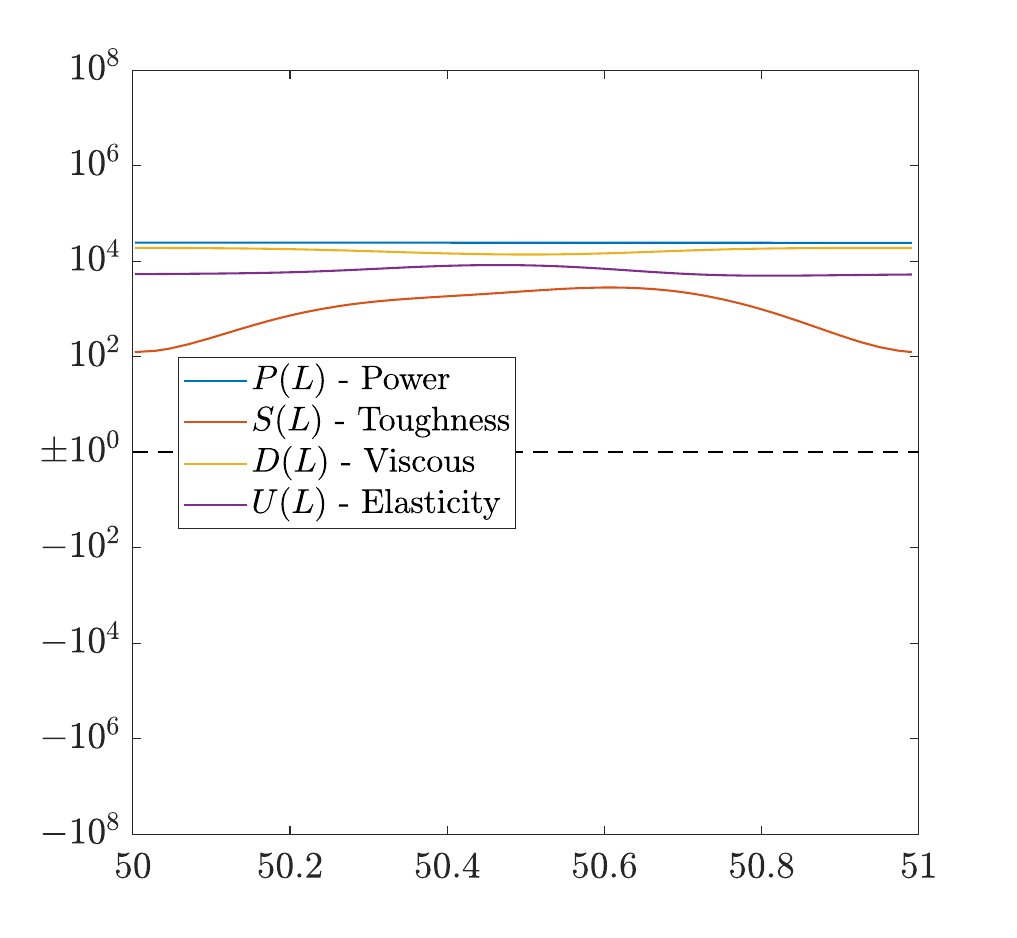}
	\put(-95,-3) {$l(t)$}
	\put(-205,85) {Power}
	\put(-195,155) {{\bf (c)}}
	\put(-105,155) {$k=50$}
	\hspace{6mm}
	\includegraphics[width=0.45\textwidth]{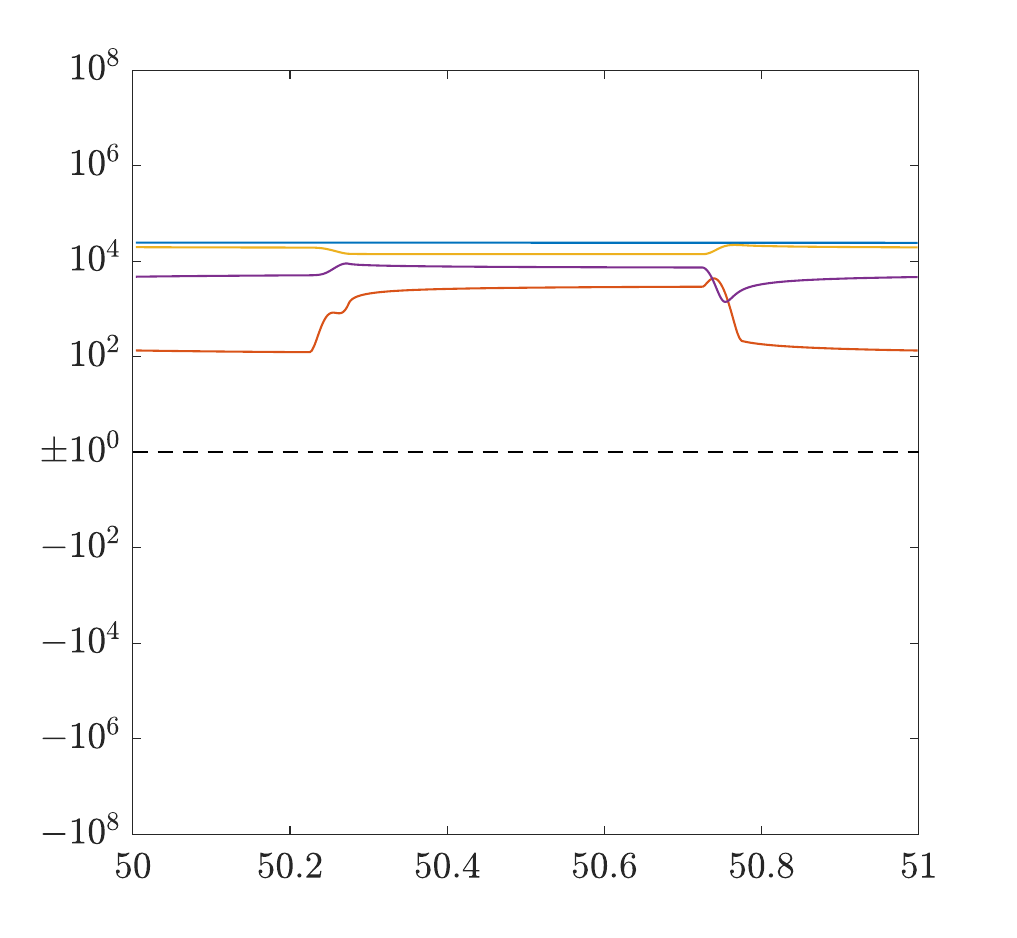}
	\put(-95,-3) {$l(t)$}
	\put(-195,155) {{\bf (d)}}
	\put(-105,155) {$k=50$}
	
	\vspace{2mm}
	
	\includegraphics[width=0.45\textwidth]{Power_Periods_250_Sin_1_01_250214-eps-converted-to.pdf}
	\put(-95,-3) {$l(t)$}
	\put(-205,85) {Power}
	\put(-195,155) {{\bf (e)}}
	\put(-105,155) {$k=250$}
	\hspace{6mm}
	\includegraphics[width=0.45\textwidth]{Power_Periods_250_Squ_1_01_250214-eps-converted-to.pdf}
	\put(-95,-3) {$l(t)$}
	\put(-195,155) {{\bf (f)}}
	\put(-105,155) {$k=250$}
	\caption{Distribution of the power [W] (19) within the hydraulic fracture as it propagates through (a), (b) the first three toughness periods, and the $k^{\text{th}}$ toughness period with (c), (d) $k=50$, (e), (f) $k=250$. This is for the intermediate-viscosity case ($\delta_{max}=1$, $\delta_{min}=0.1$) with the (a), (c), (e) sinusoidal, (b), (d), (f) step-wise, toughness distribution. Recall that for the stepwise distribution, $0.25<l-k<0.75$ corresponds to the maximum toughness layer, with the remainder of the domain corresponding to the minimum toughness layer.}
	\label{Energy_Periods_1_01}
\end{figure}
\begin{figure}[hp!]
	\centering
	\includegraphics[width=0.45\textwidth]{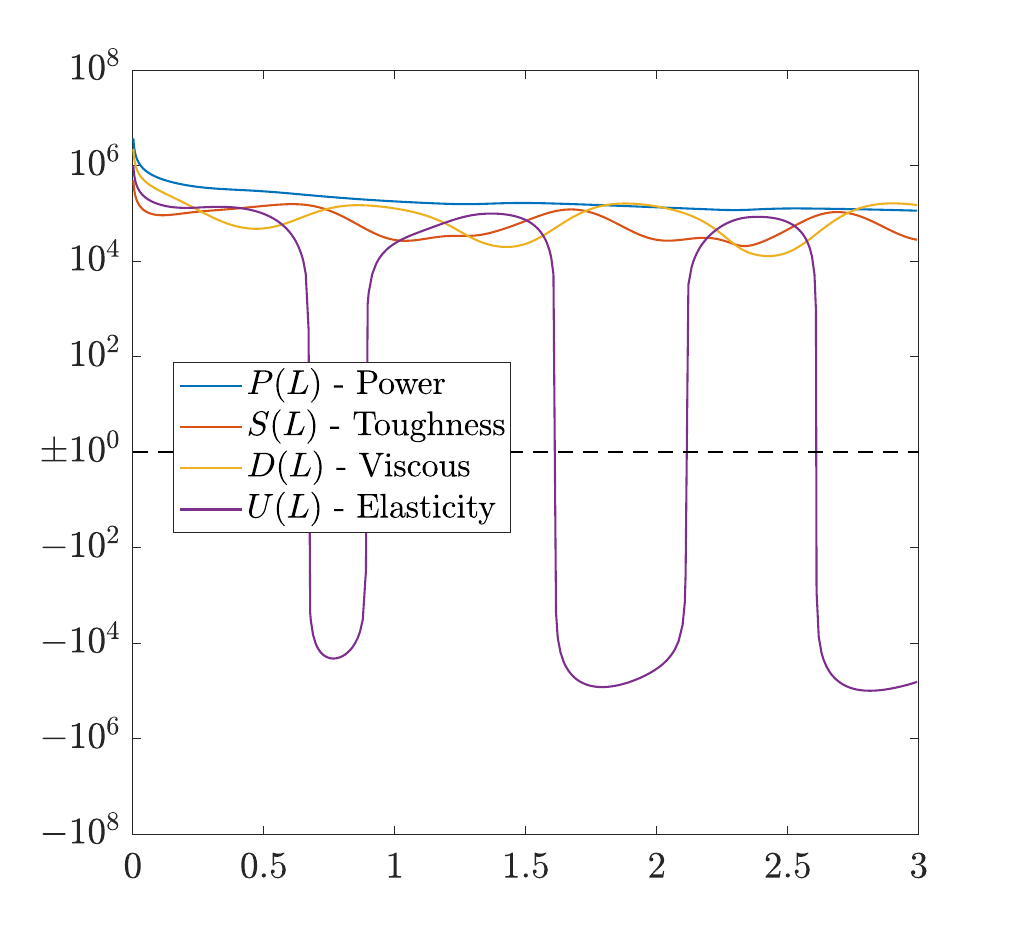}
	\put(-95,-3) {$l(t)$}
	\put(-205,85) {Power}
	\put(-195,155) {{\bf (a)}}
	\put(-105,155) {$k=1,2,3$}
	\hspace{6mm}
	\includegraphics[width=0.45\textwidth]{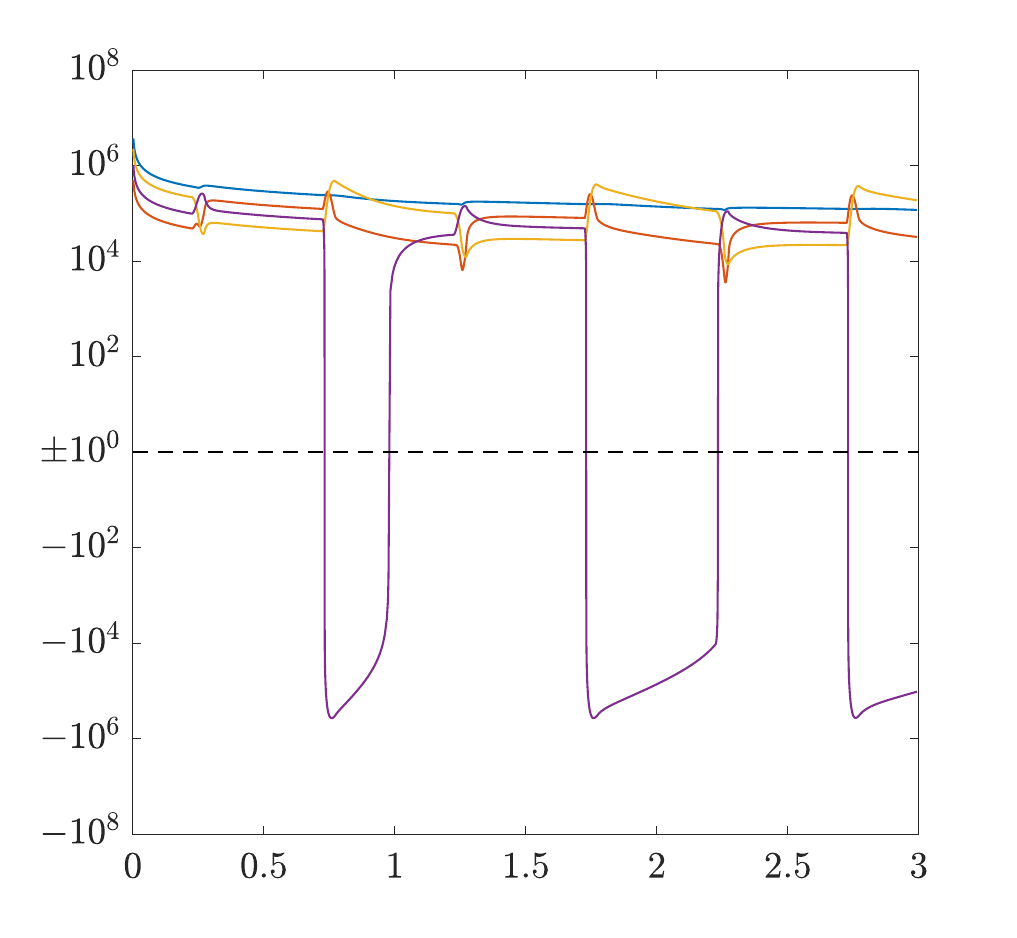}
	\put(-95,-3) {$l(t)$}
	\put(-195,155) {{\bf (b)}}
	\put(-105,155) {$k=1,2,3$}
	
	\vspace{2mm}
	
	\includegraphics[width=0.45\textwidth]{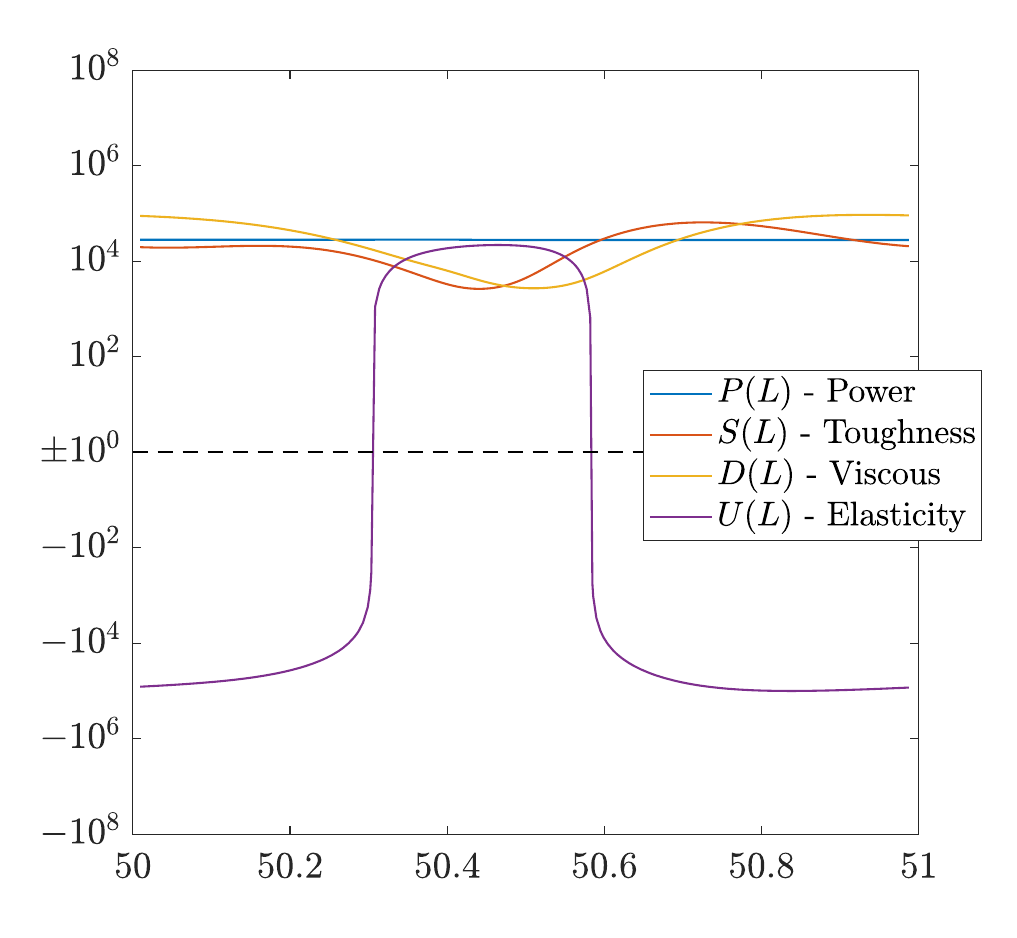}
	\put(-95,-3) {$l(t)$}
	\put(-205,85) {Power}
	\put(-195,155) {{\bf (c)}}
	\put(-105,155) {$k=50$}
	\hspace{6mm}
	\includegraphics[width=0.45\textwidth]{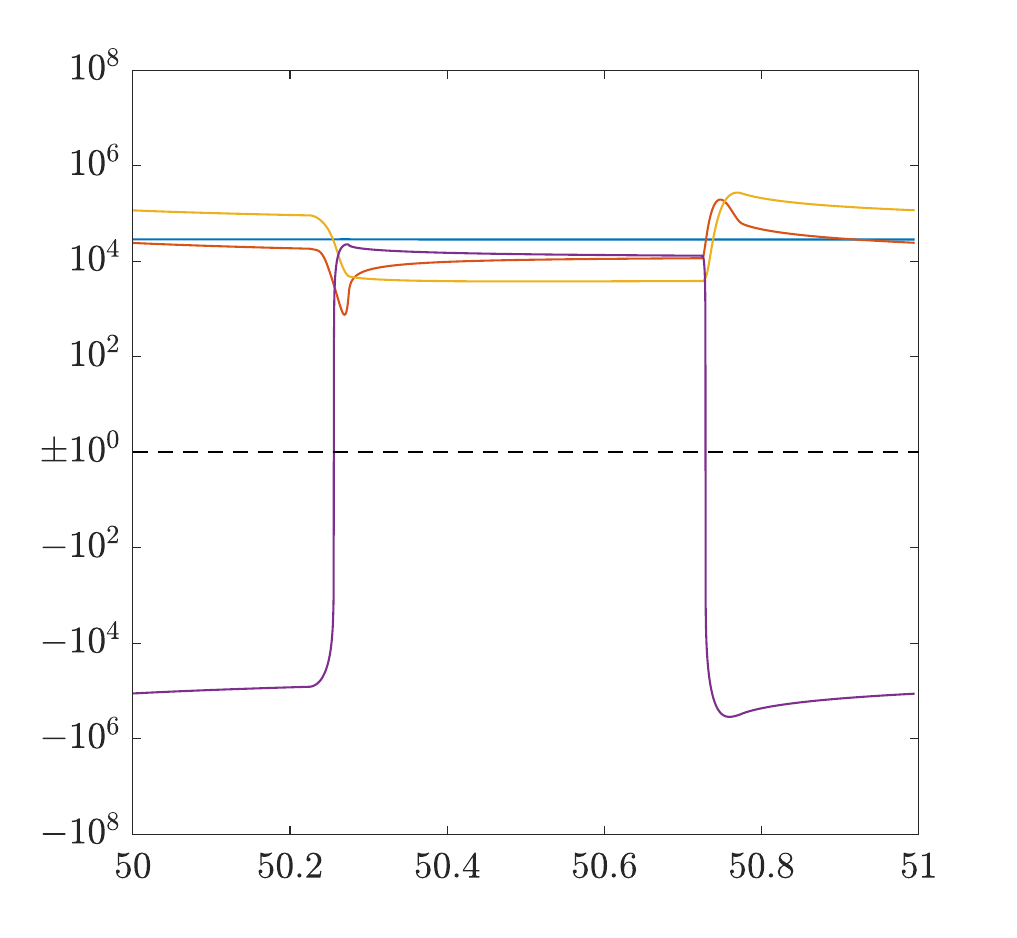}
	\put(-95,-3) {$l(t)$}
	\put(-195,155) {{\bf (d)}}
	\put(-105,155) {$k=50$}
	
	\vspace{2mm}
	
	\includegraphics[width=0.45\textwidth]{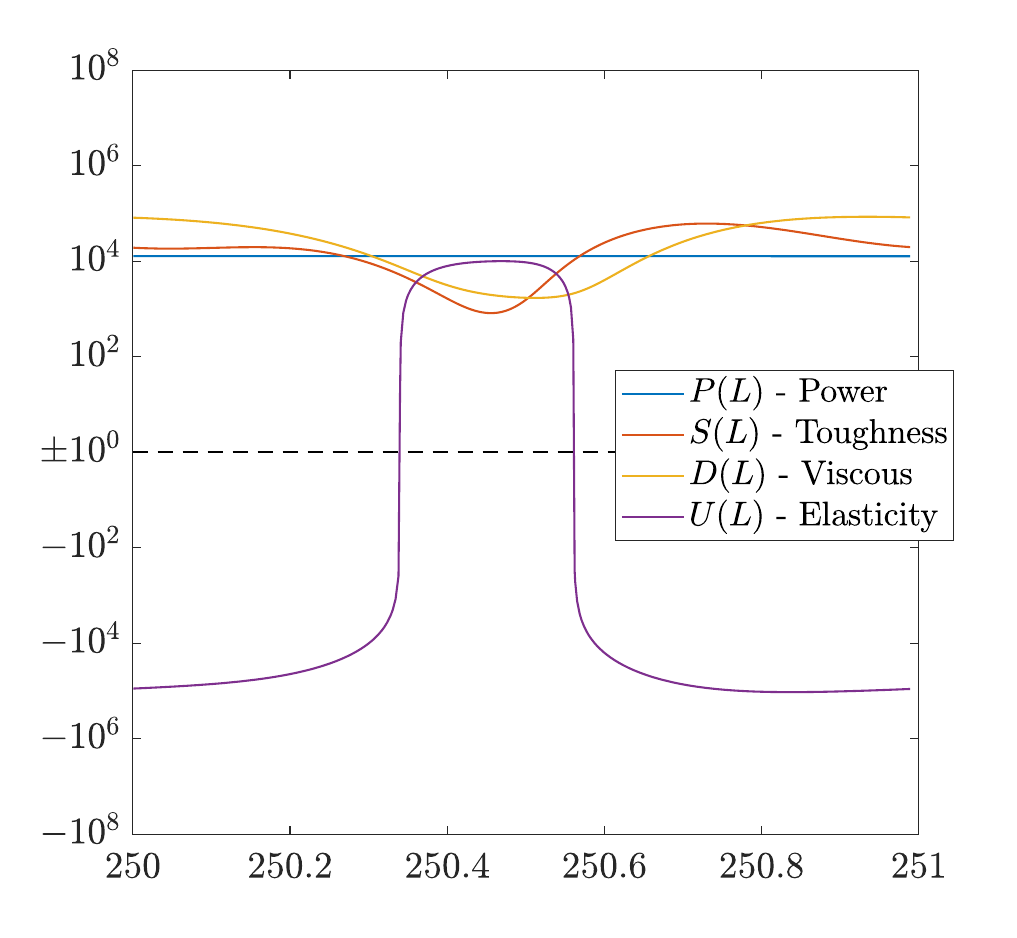}
	\put(-95,-3) {$l(t)$}
	\put(-205,85) {Power}
	\put(-195,155) {{\bf (e)}}
	\put(-105,155) {$k=250$}
	\hspace{6mm}
	\includegraphics[width=0.45\textwidth]{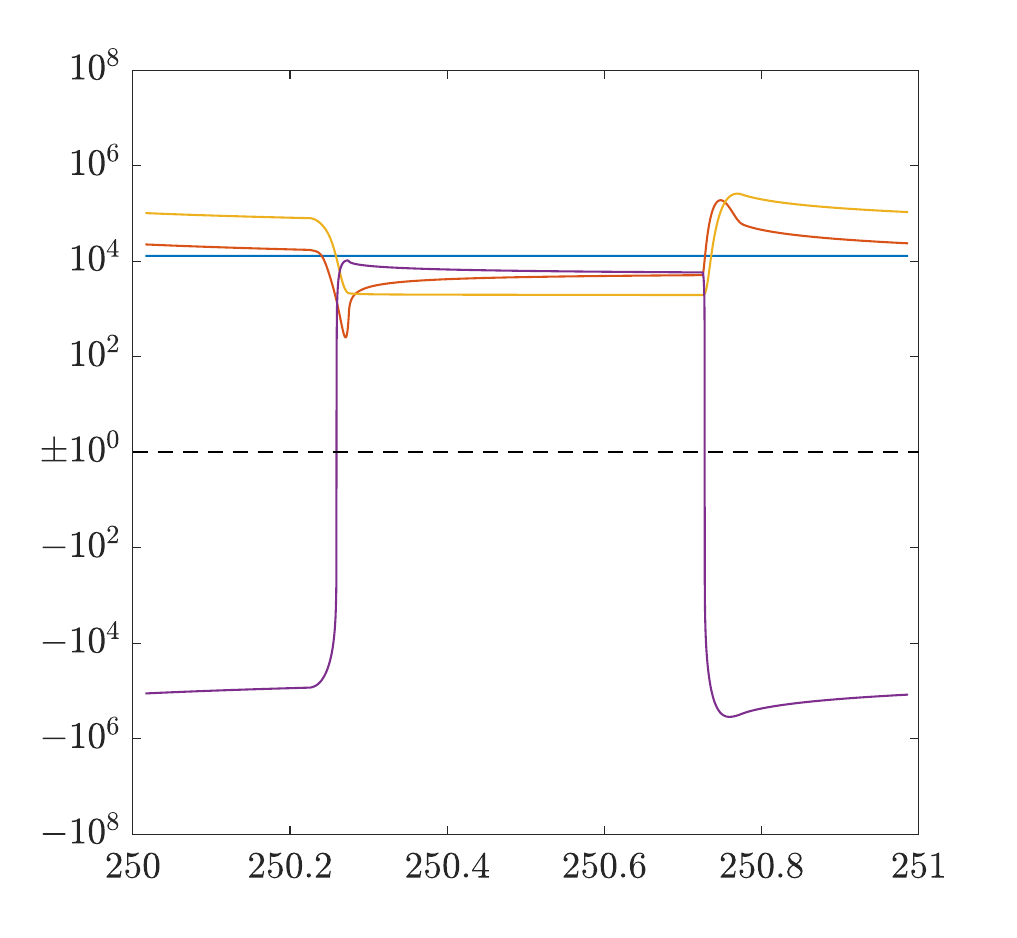}
	\put(-95,-3) {$l(t)$}
	\put(-195,155) {{\bf (f)}}
	\put(-105,155) {$k=250$}
	\caption{Distribution of the power [W] (19) within the hydraulic fracture as it propagates through (a), (b) the first three toughness periods, and the $k^{\text{th}}$ toughness period with (c), (d) $k=50$, (e), (f) $k=250$. This is for the toughness-intermediate case ($\delta_{max}=10$, $\delta_{min}=1$) with the (a), (c), (e) sinusoidal, (b), (d), (f) step-wise, toughness distribution. Recall that for the stepwise distribution, $0.25<l-k<0.75$ corresponds to the maximum toughness layer, with the remainder of the domain corresponding to the minimum toughness layer.}
	\label{Energy_Periods_10_1}
\end{figure}
\begin{figure}[hp!]
	\centering
	\includegraphics[width=0.45\textwidth]{Power_Periods_1_Sin_100_10_250214-eps-converted-to.pdf}
	\put(-95,-3) {$l(t)$}
	\put(-205,85) {Power}
	\put(-195,155) {{\bf (a)}}
	\put(-105,155) {$k=1,2,3$}
	\hspace{6mm}
	\includegraphics[width=0.45\textwidth]{Power_Periods_1_Squ_100_10_250214-eps-converted-to.pdf}
	\put(-95,-3) {$l(t)$}
	\put(-195,155) {{\bf (b)}}
	\put(-105,155) {$k=1,2,3$}
	
	\vspace{2mm}
	
	\includegraphics[width=0.45\textwidth]{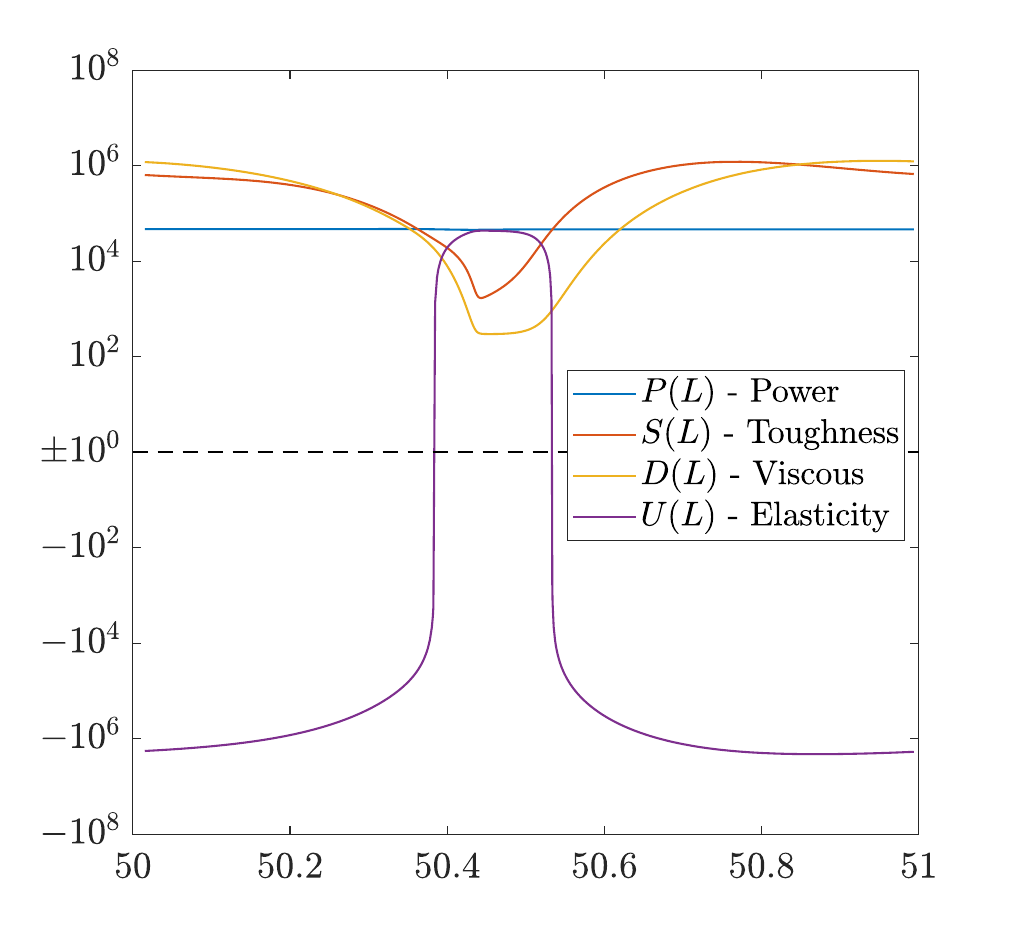}
	\put(-95,-3) {$l(t)$}
	\put(-205,85) {Power}
	\put(-195,155) {{\bf (c)}}
	\put(-105,155) {$k=50$}
	\hspace{6mm}
	\includegraphics[width=0.45\textwidth]{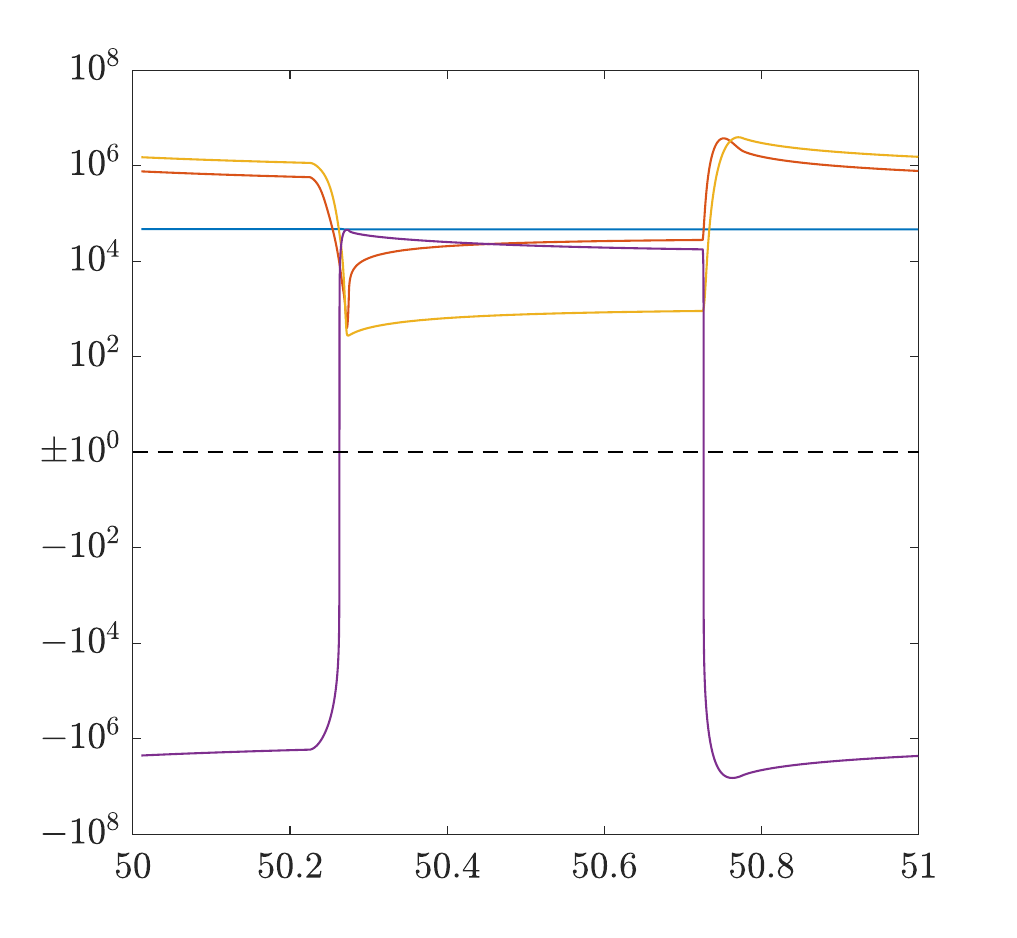}
	\put(-95,-3) {$l(t)$}
	\put(-195,155) {{\bf (d)}}
	\put(-105,155) {$k=50$}
	
	\vspace{2mm}
	
	\includegraphics[width=0.45\textwidth]{Power_Periods_250_Sin_100_10_250214-eps-converted-to.pdf}
	\put(-95,-3) {$l(t)$}
	\put(-205,85) {Power}
	\put(-195,155) {{\bf (e)}}
	\put(-105,155) {$k=250$}
	\hspace{6mm}
	\includegraphics[width=0.45\textwidth]{Power_Periods_250_Squ_100_10_250214-eps-converted-to.pdf}
	\put(-95,-3) {$l(t)$}
	\put(-195,155) {{\bf (f)}}
	\put(-105,155) {$k=250$}
	\caption{Distribution of the power [W] (19) within the hydraulic fracture as it propagates through (a), (b) the first three toughness periods, and the $k^{\text{th}}$ toughness period with (c), (d) $k=50$, (e), (f) $k=250$. This is for the toughness-toughness case ($\delta_{max}=100$, $\delta_{min}=10$) with the (a), (c), (e) sinusoidal, (b), (d), (f) step-wise, toughness distribution. Recall that for the stepwise distribution, $0.25<l-k<0.75$ corresponds to the maximum toughness layer, with the remainder of the domain corresponding to the minimum toughness layer.}
	\label{Energy_Periods_100_10}
\end{figure}

{\bf Power distribution as the hydraulic fracture propagates through the rock layers}, discussed in the main text in Sect.~3.2, are provided in Figs.~\ref{Energy_Periods_1_01}-\ref{Energy_Periods_100_10}. Note that some figures are repetitions of those given in the main text in Figs.~7,8, but are repeated here for ease of comparison. These figures clearly reinforce the conclusions stated in the main text, namely: the energy terms enter into a periodic oscillation to match the toughness after the first two periods; and the solid domain `stores' elastic energy during propagation through tougher layers, releasing it during propagation through weaker layers (the elastic battery). The latter effect is present in both the toughness-intermediate and toughness-toughness cases, but is only significant for the intermediate-viscosity toughness distribution over long length-scales.\\

\begin{figure}[t!]
	\centering
	\includegraphics[width=0.45\textwidth]{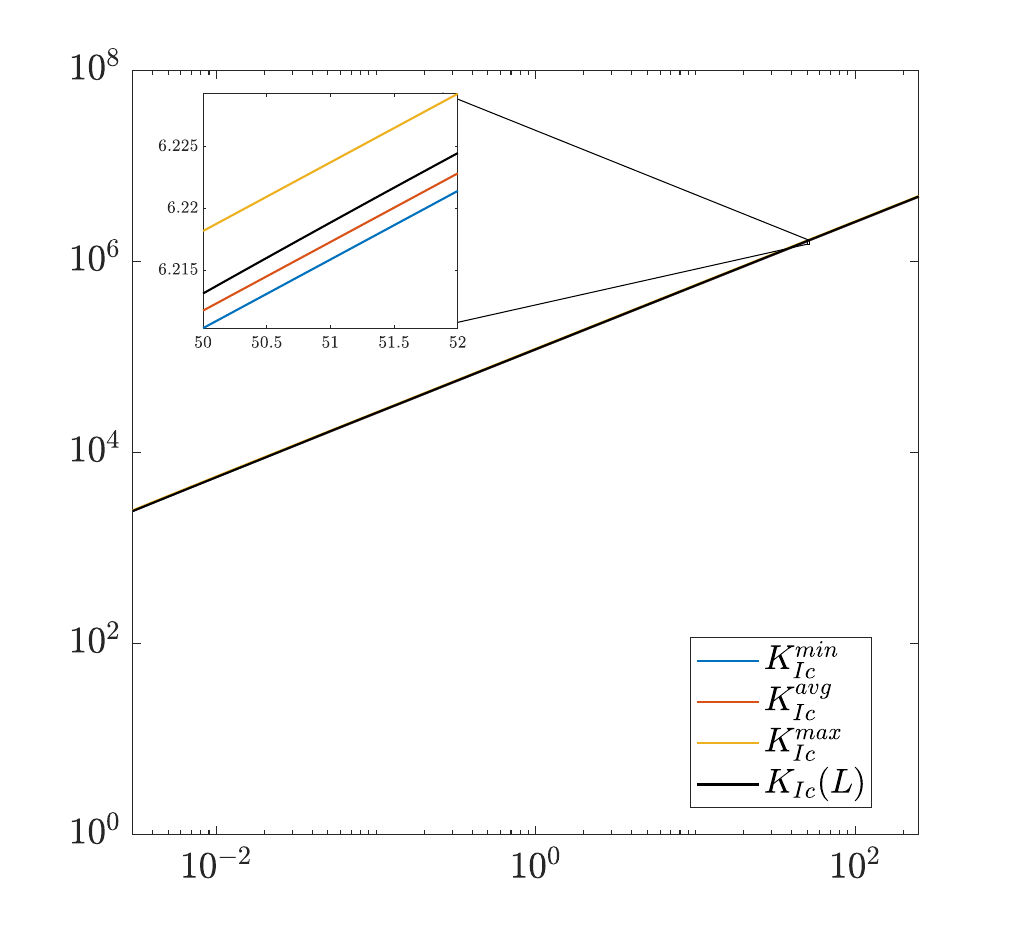}
	\put(-95,-3) {$l(t)$}
	\put(-235,85) {$\int_0^{l(t)} P(\eta)\,{\rm d}\eta$}
	\put(-195,155) {{\bf (a)}}
	\hspace{6mm}
	\includegraphics[width=0.45\textwidth]{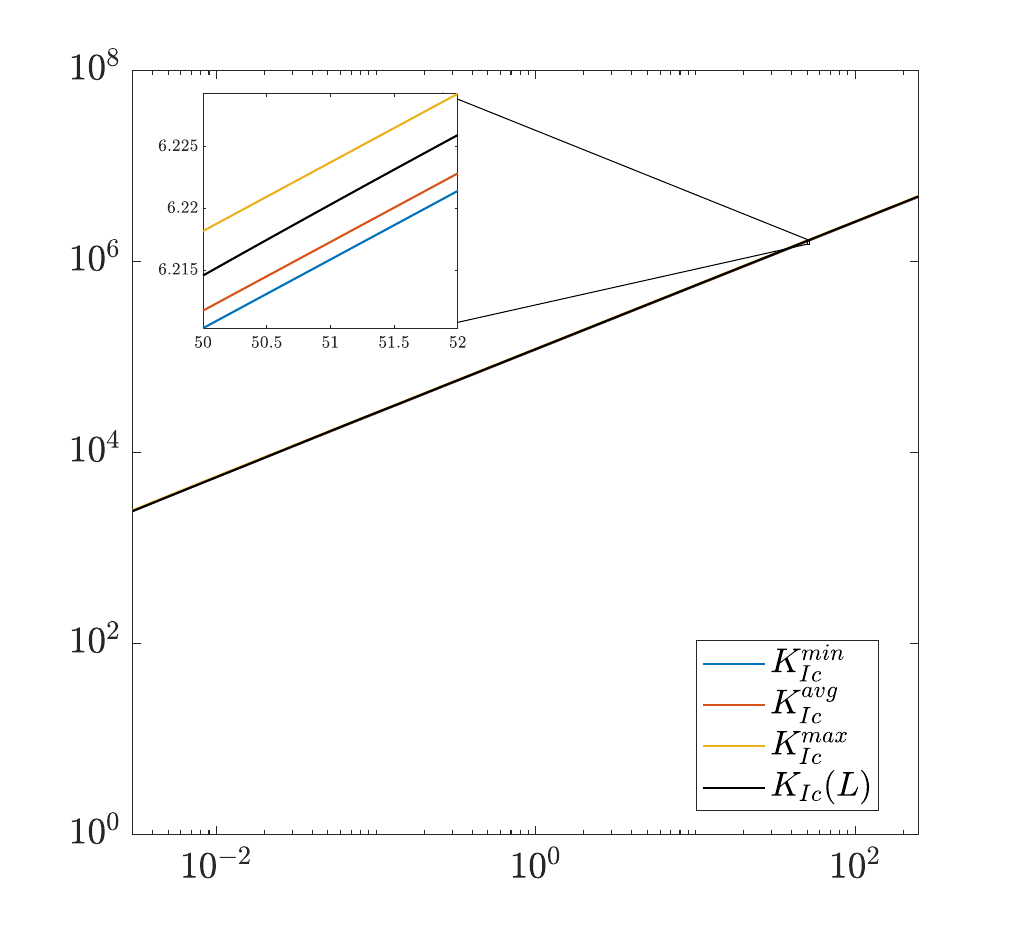}
	\put(-95,-3) {$l(t)$}
	\put(-195,155) {{\bf (b)}}
	
	\vspace{2mm}
	
	\includegraphics[width=0.45\textwidth]{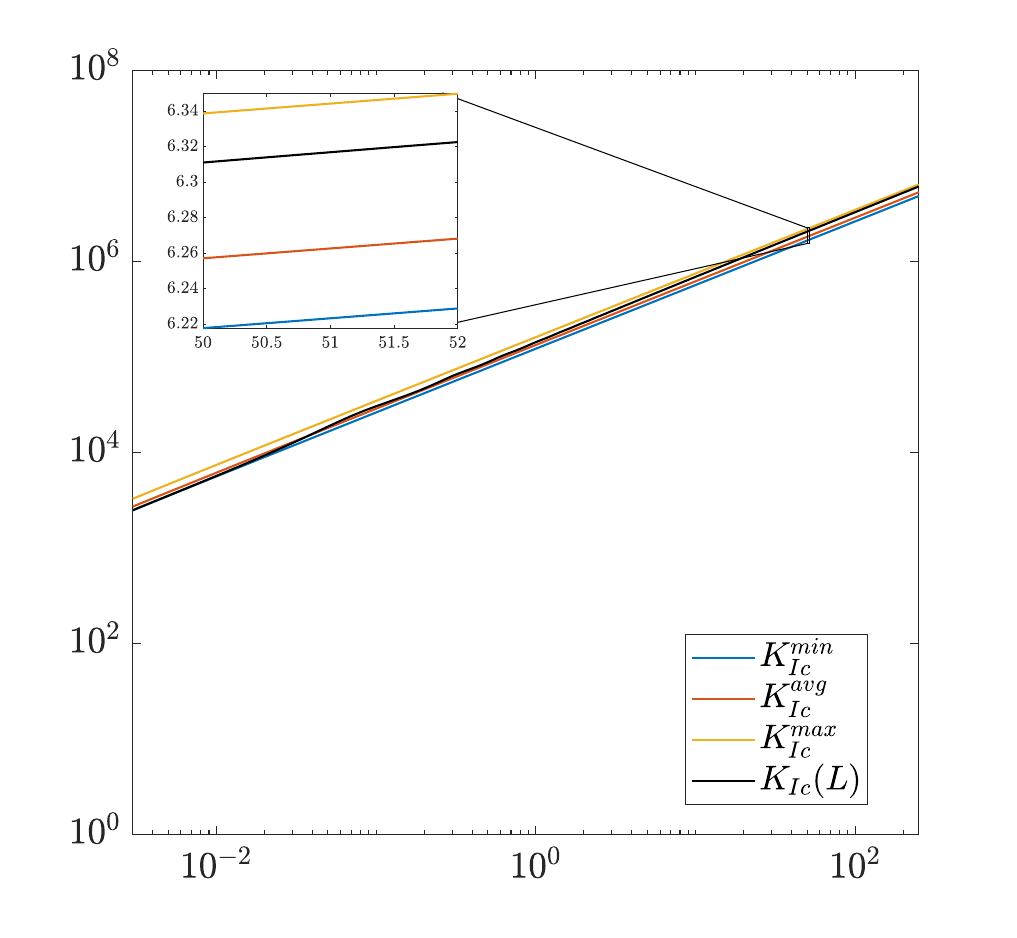}
	\put(-95,-3) {$l(t)$}
	\put(-235,85) {$\int_0^{l(t)} P(\eta)\,{\rm d}\eta$}
	\put(-195,155) {{\bf (c)}}
	\hspace{6mm}
	\includegraphics[width=0.45\textwidth]{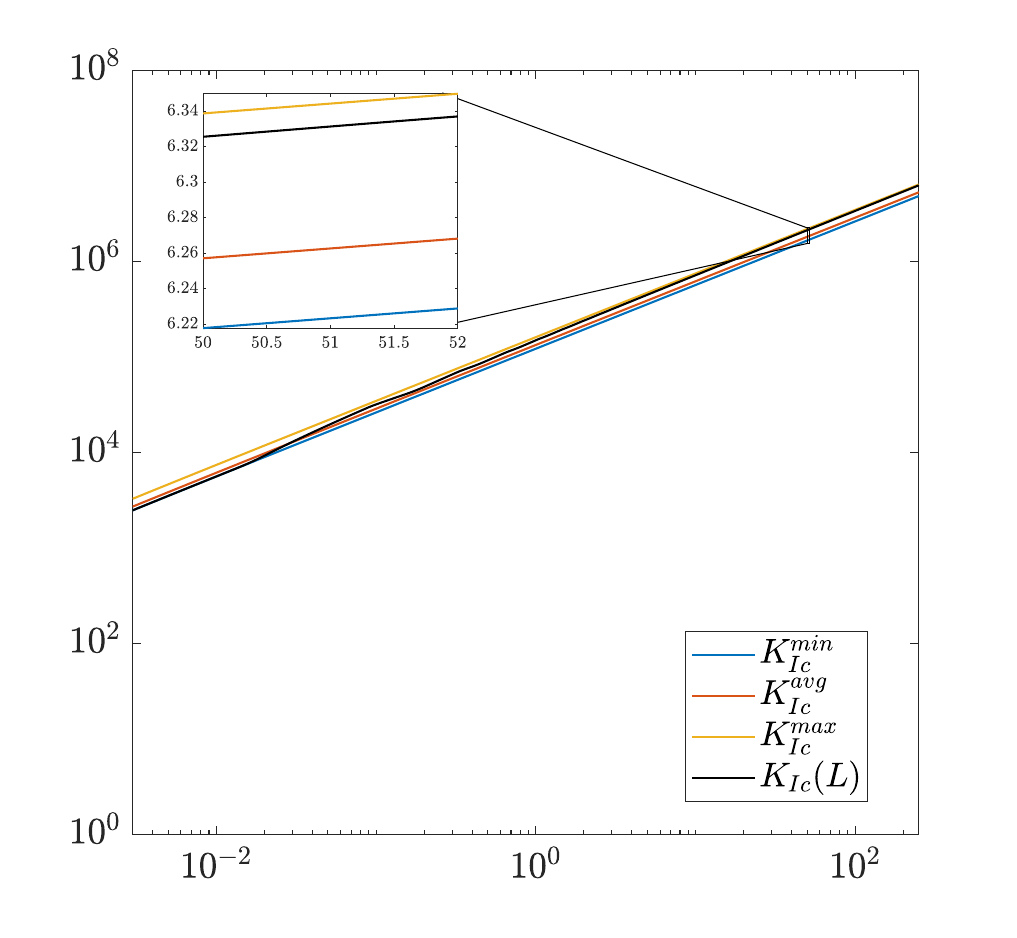}
	\put(-95,-3) {$l(t)$}
	\put(-195,155) {{\bf (d)}}
	
	\vspace{2mm}
	
	\includegraphics[width=0.45\textwidth]{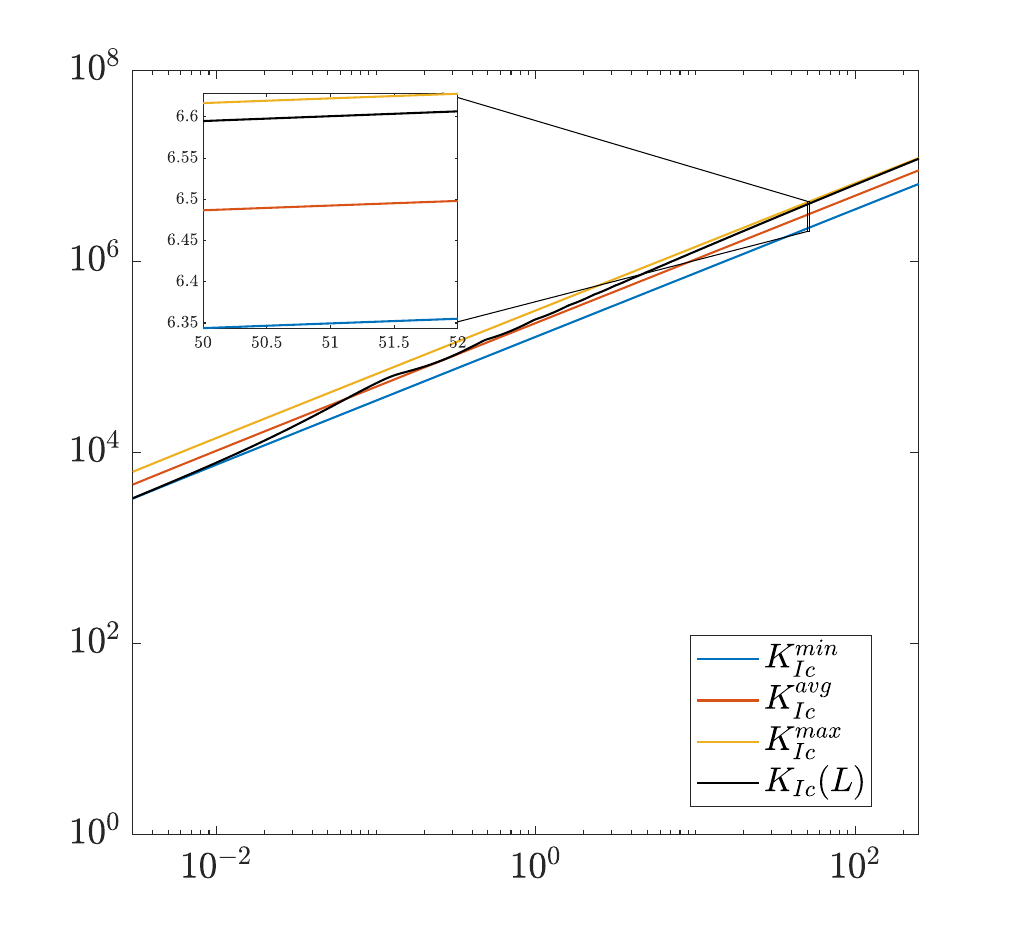}
	\put(-95,-3) {$l(t)$}
	\put(-235,85) {$\int_0^{l(t)} P(\eta)\,{\rm d}\eta$}
	\put(-195,155) {{\bf (e)}}
	\hspace{6mm}
	\includegraphics[width=0.45\textwidth]{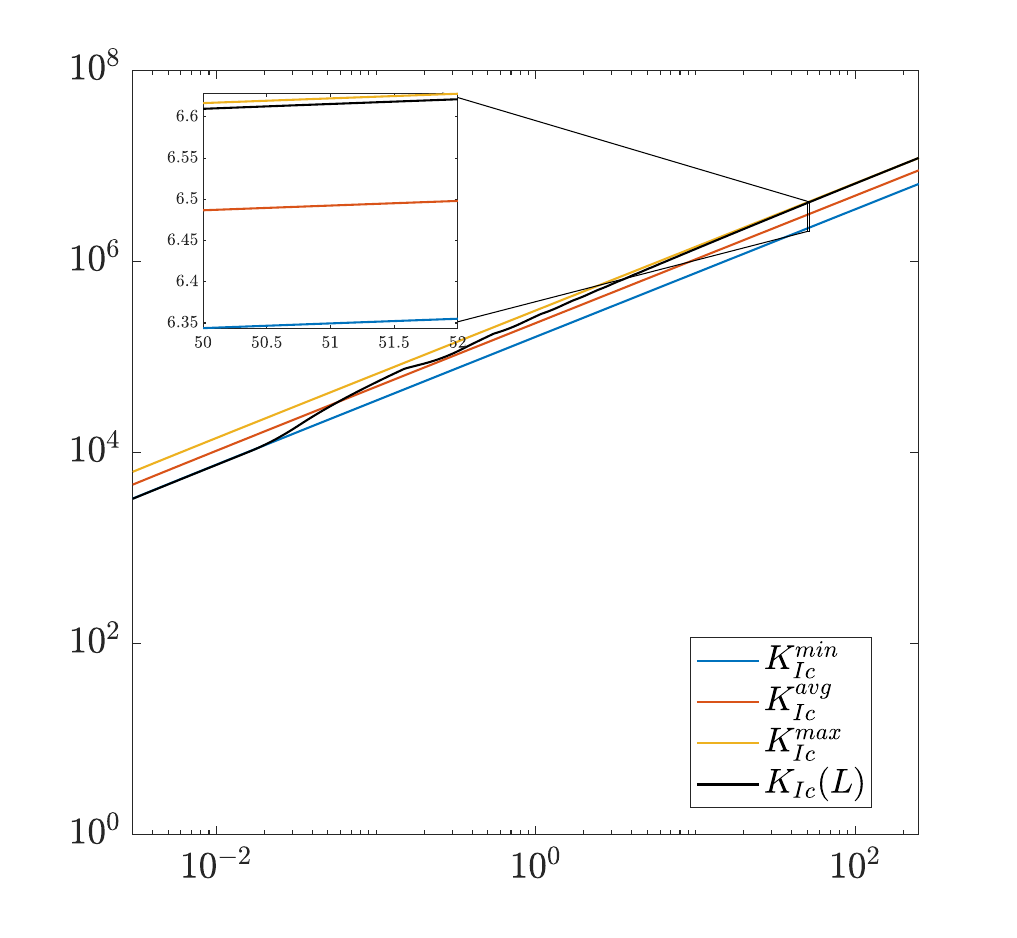}
	\put(-95,-3) {$l(t)$}
	\put(-195,155) {{\bf (f)}}
	\caption{The cumulative power injected into the fracture $P$, (19), alongside the behaviour for the case of homogeneous toughness with distributions: maximum toughness $K_{Ic}(l(t))\equiv K_{Ic}^{max}$, minimum toughness $K_{Ic}(l(t))\equiv K_{Ic}^{min}$, and the (arithmetic) average toughness $K_{Ic}(l(t))\equiv K_{Ic}^{avg}$. We consider toughness distributions (a), (b) $\delta_{max}=1$, $\delta_{min}=0.1$, (c), (d) $\delta_{max}=10$, $\delta_{min}=1$, (e), (f) $\delta_{max}=100$, $\delta_{min}=10$, in the (a), (c), (e) sinusoidal, (b), (d), (f) step-wise, cases. }
	\label{P_Cum_1}
\end{figure}

\begin{figure}[t!]
	\centering
	\includegraphics[width=0.45\textwidth]{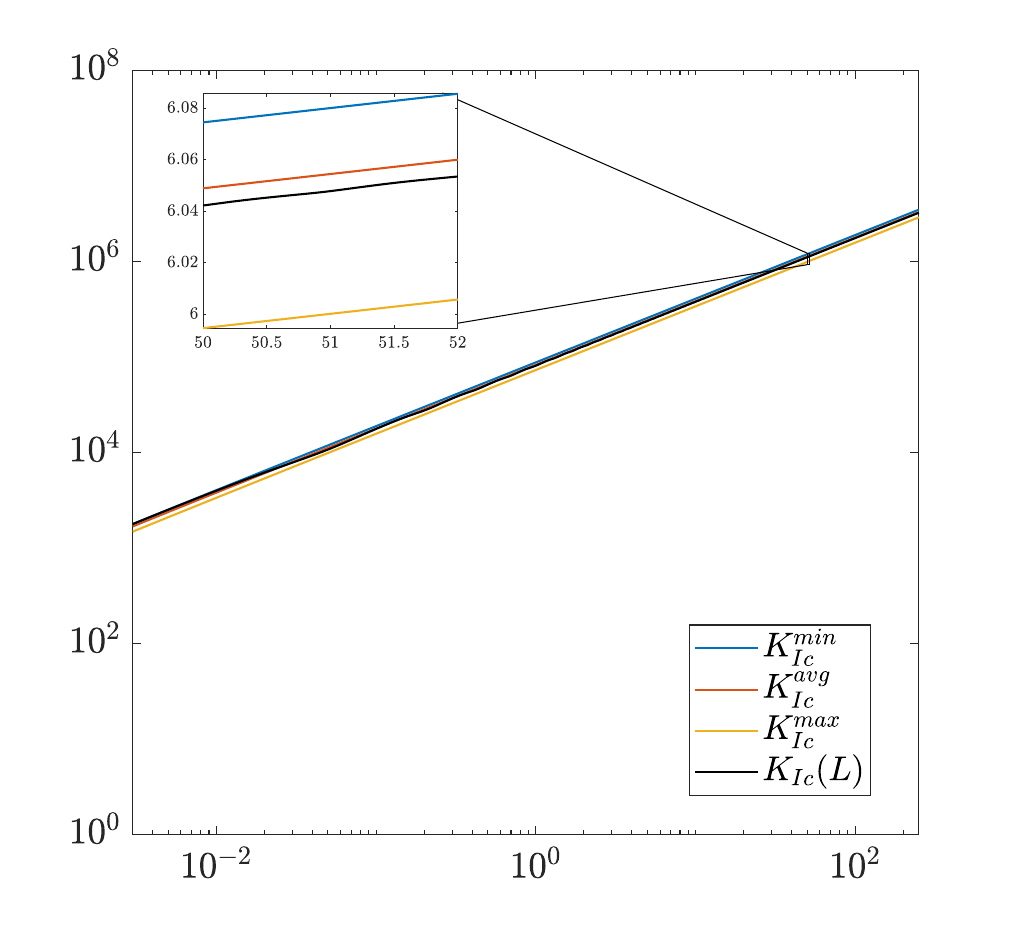}
	\put(-95,-3) {$l(t)$}
	\put(-235,85) {$\int_0^{l(t)} D(\eta)\,{\rm d}\eta$}
	\put(-195,155) {{\bf (a)}}
	\hspace{6mm}
	\includegraphics[width=0.45\textwidth]{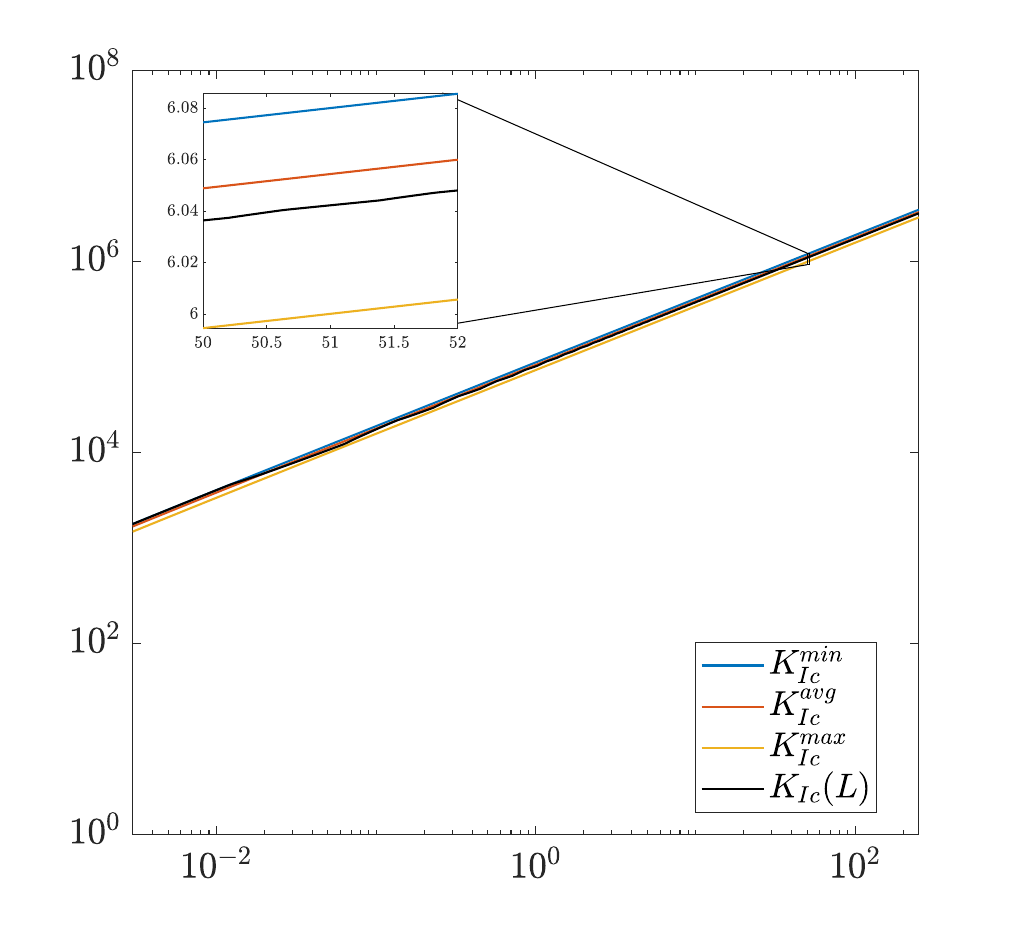}
	\put(-95,-3) {$l(t)$}
	\put(-195,155) {{\bf (b)}}
	
	\vspace{2mm}
	
	\includegraphics[width=0.45\textwidth]{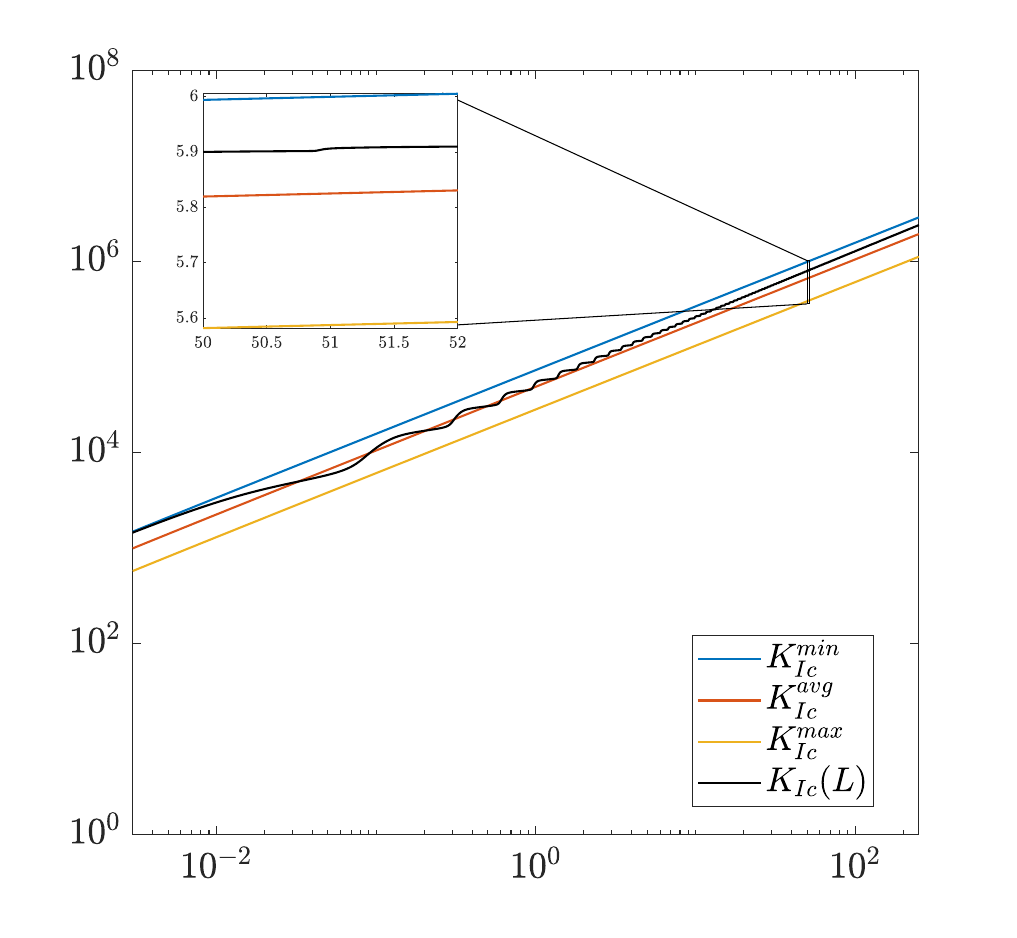}
	\put(-95,-3) {$l(t)$}
	\put(-235,85) {$\int_0^{l(t)} D(\eta)\,{\rm d}\eta$}
	\put(-195,155) {{\bf (c)}}
	\hspace{6mm}
	\includegraphics[width=0.45\textwidth]{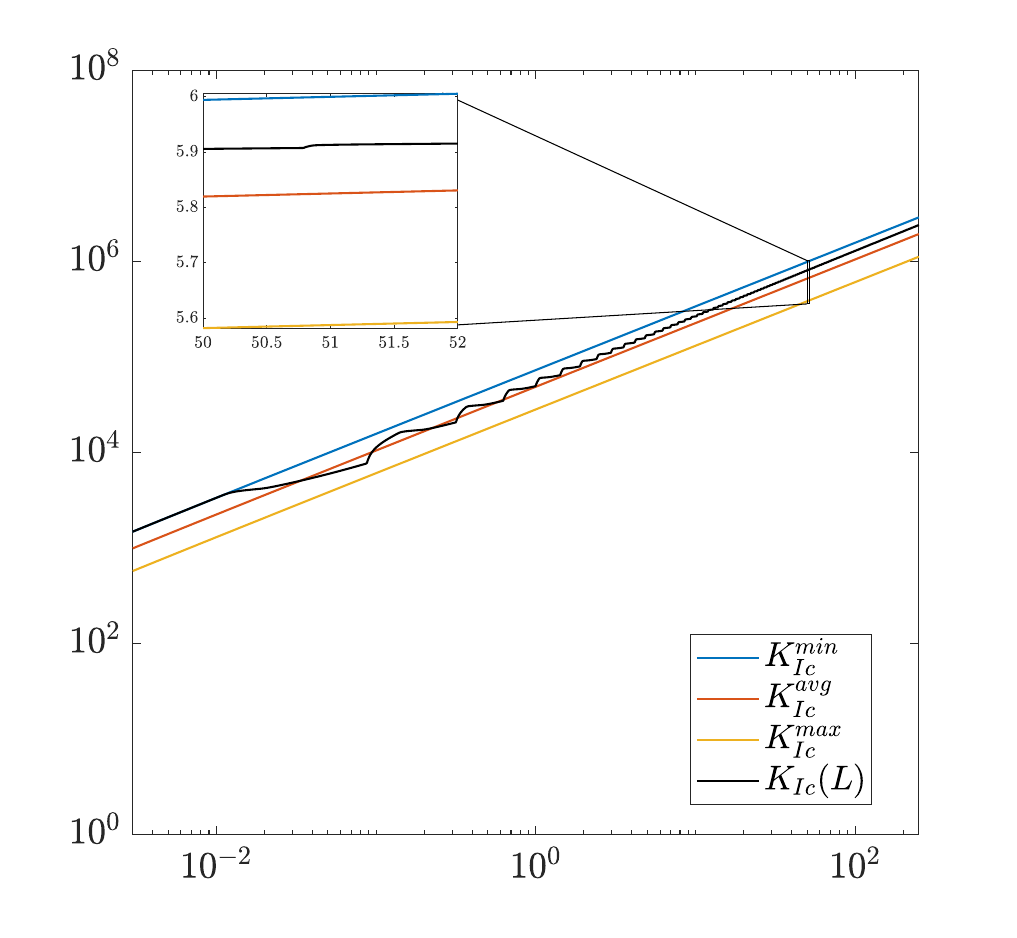}
	\put(-95,-3) {$l(t)$}
	\put(-195,155) {{\bf (d)}}
	
	\vspace{2mm}
	
	\includegraphics[width=0.45\textwidth]{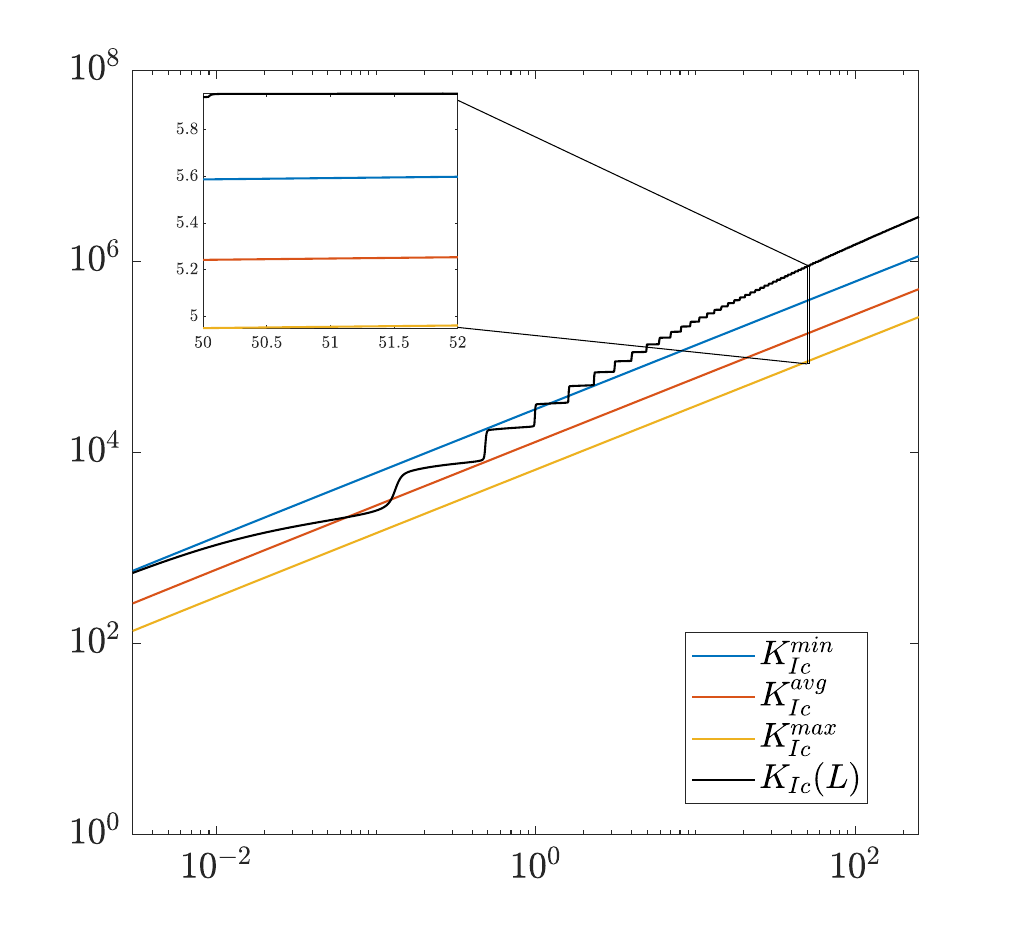}
	\put(-95,-3) {$l(t)$}
	\put(-235,85) {$\int_0^{l(t)} D(\eta)\,{\rm d}\eta$}
	\put(-195,155) {{\bf (e)}}
	\hspace{6mm}
	\includegraphics[width=0.45\textwidth]{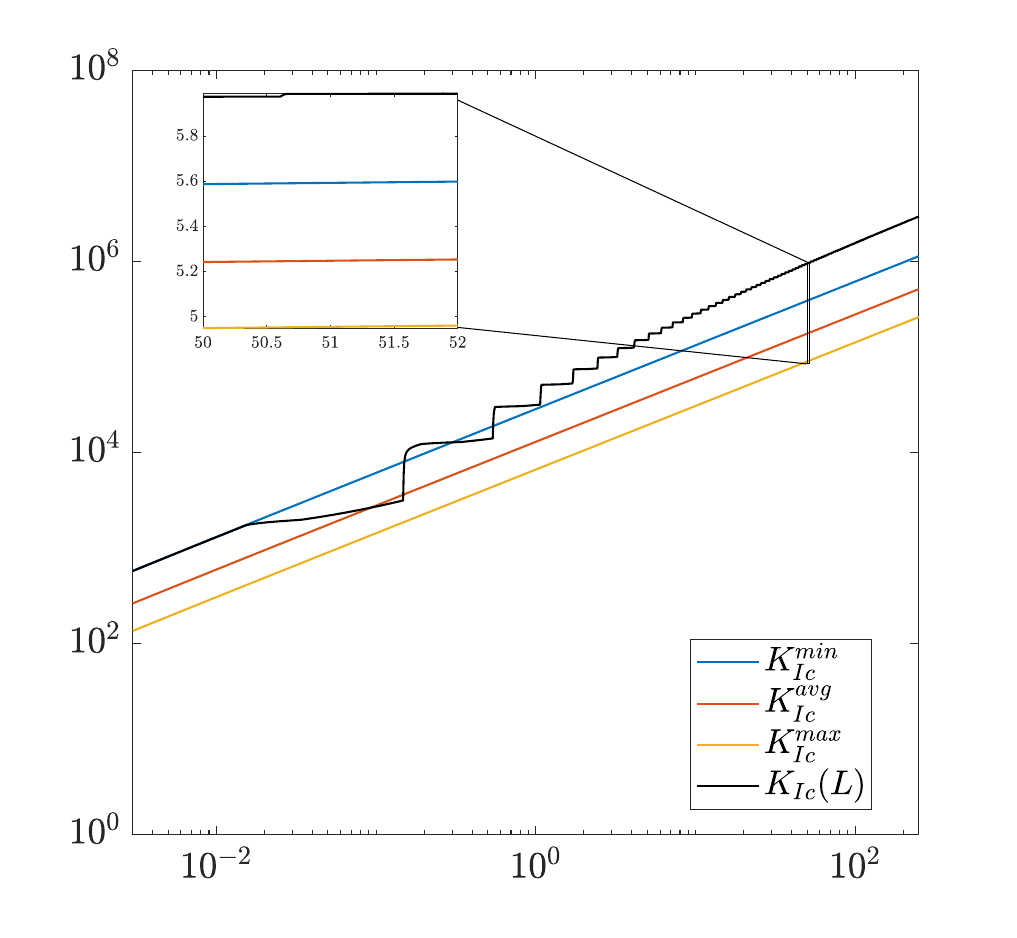}
	\put(-95,-3) {$l(t)$}
	\put(-195,155) {{\bf (f)}}
	\caption{The cumulative viscous energy $D$ expended overcoming viscous friction on the crack walls, (19), alongside the behaviour for the case of homogeneous toughness with distributions: maximum toughness $K_{Ic}(l(t))\equiv K_{Ic}^{max}$, minimum toughness $K_{Ic}(l(t))\equiv K_{Ic}^{min}$, and the (arithmetic) average toughness $K_{Ic}(l(t))\equiv K_{Ic}^{avg}$. We consider toughness distributions (a), (b) $\delta_{max}=1$, $\delta_{min}=0.1$, (c), (d) $\delta_{max}=10$, $\delta_{min}=1$, (e), (f) $\delta_{max}=100$, $\delta_{min}=10$, in the (a), (c), (e) sinusoidal, (b), (d), (f) step-wise, cases. }
	\label{D_Cum_1}
\end{figure}

\begin{figure}[t!]
	\centering
	\includegraphics[width=0.45\textwidth]{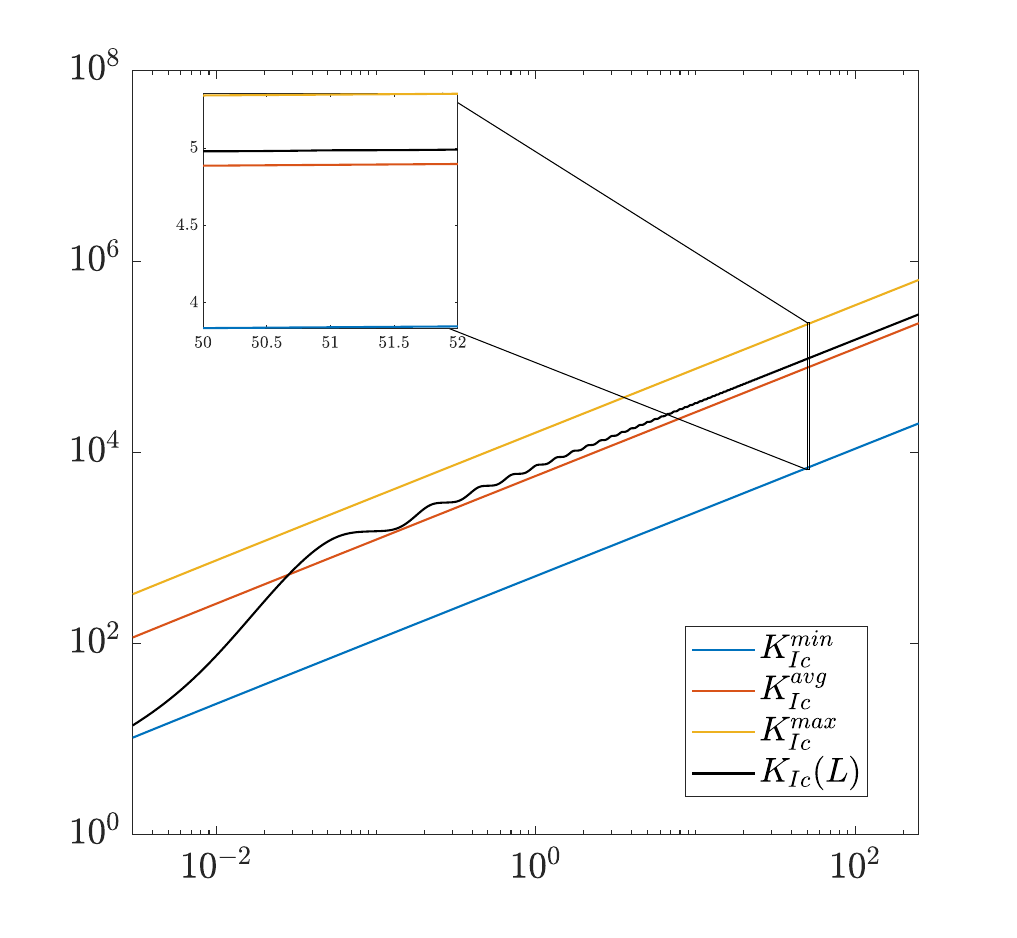}
	\put(-95,-3) {$l(t)$}
	\put(-235,85) {$\int_0^{l(t)} S(\eta)\,{\rm d}\eta$}
	\put(-195,155) {{\bf (a)}}
	\hspace{6mm}
	\includegraphics[width=0.45\textwidth]{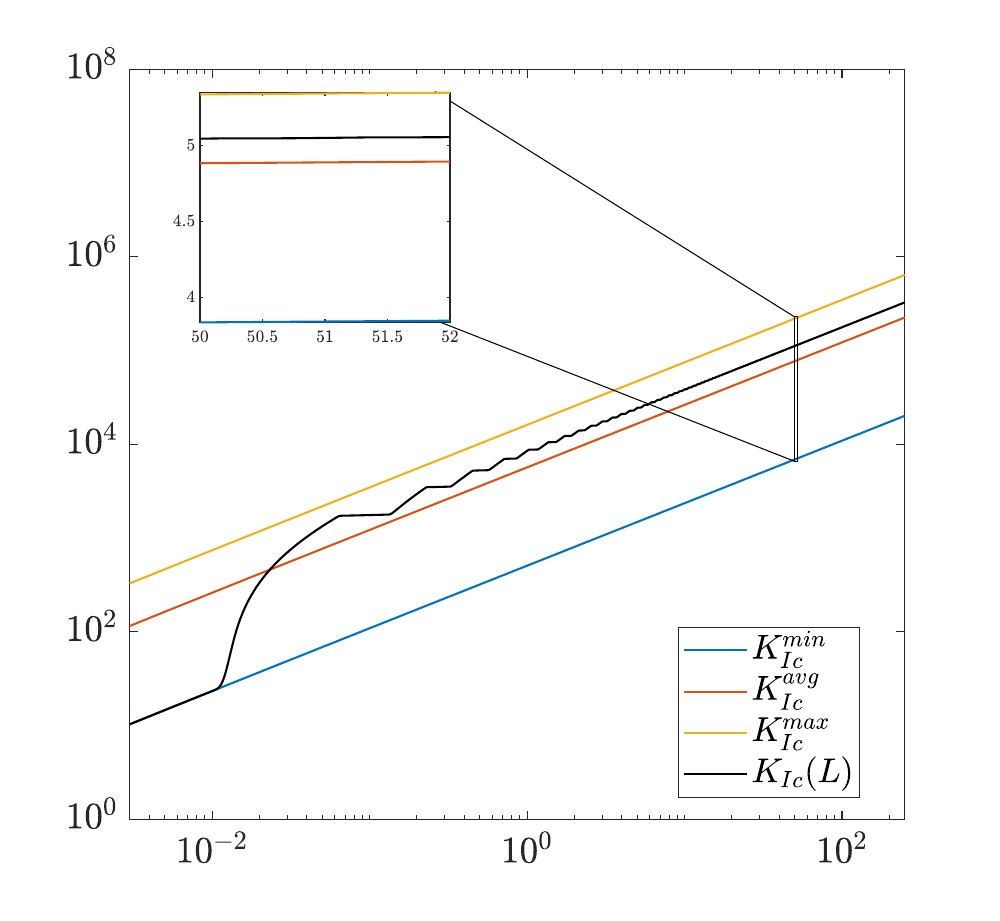}
	\put(-95,-3) {$l(t)$}
	\put(-195,155) {{\bf (b)}}
	
	\vspace{2mm}
	
	\includegraphics[width=0.45\textwidth]{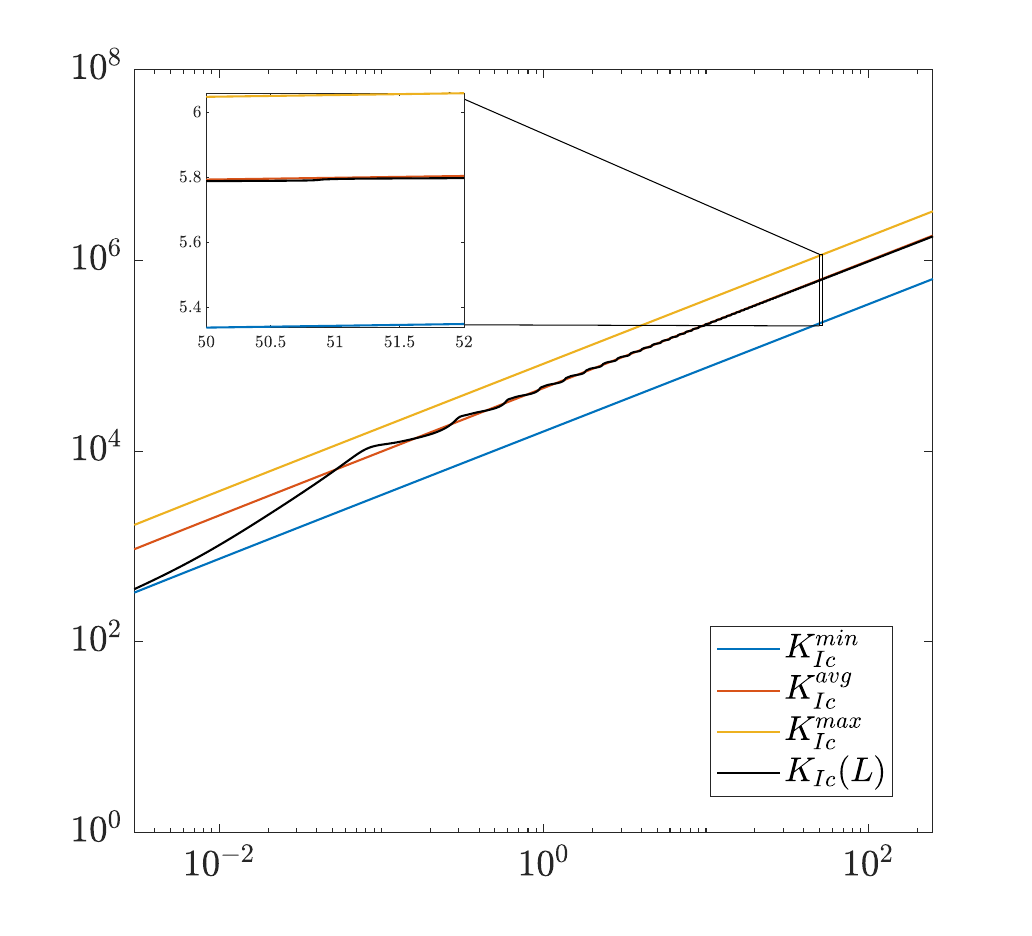}
	\put(-95,-3) {$l(t)$}
	\put(-235,85) {$\int_0^{l(t)} S(\eta)\,{\rm d}\eta$}
	\put(-195,155) {{\bf (c)}}
	\hspace{6mm}
	\includegraphics[width=0.45\textwidth]{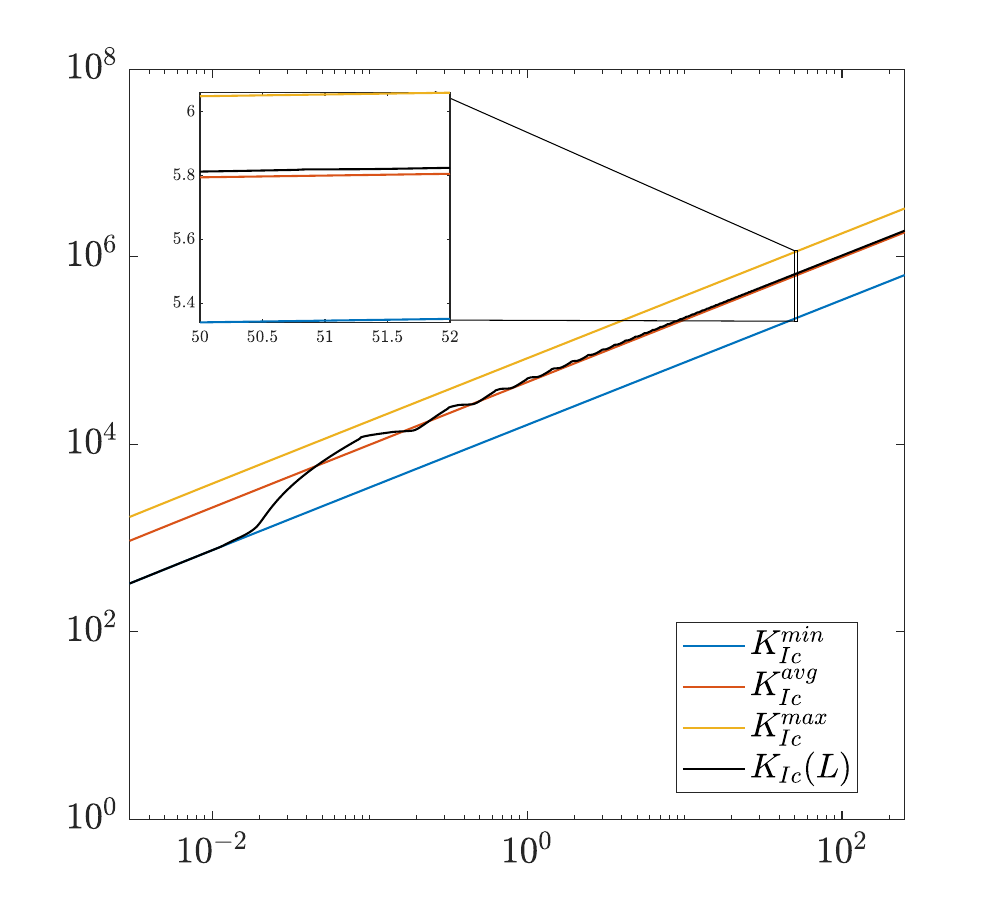}
	\put(-95,-3) {$l(t)$}
	\put(-195,155) {{\bf (d)}}
	
	\vspace{2mm}
	
	\includegraphics[width=0.45\textwidth]{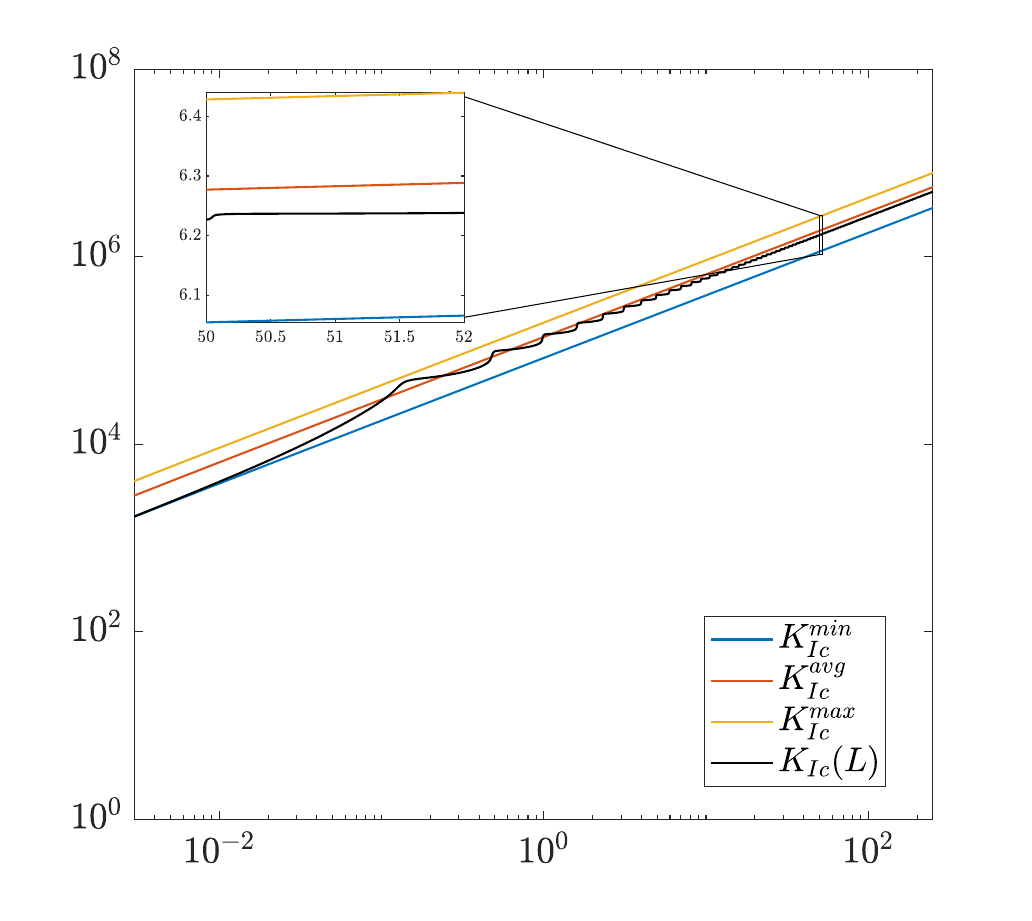}
	\put(-95,-3) {$l(t)$}
	\put(-235,85) {$\int_0^{l(t)} S(\eta)\,{\rm d}\eta$}
	\put(-195,155) {{\bf (e)}}
	\hspace{6mm}
	\includegraphics[width=0.45\textwidth]{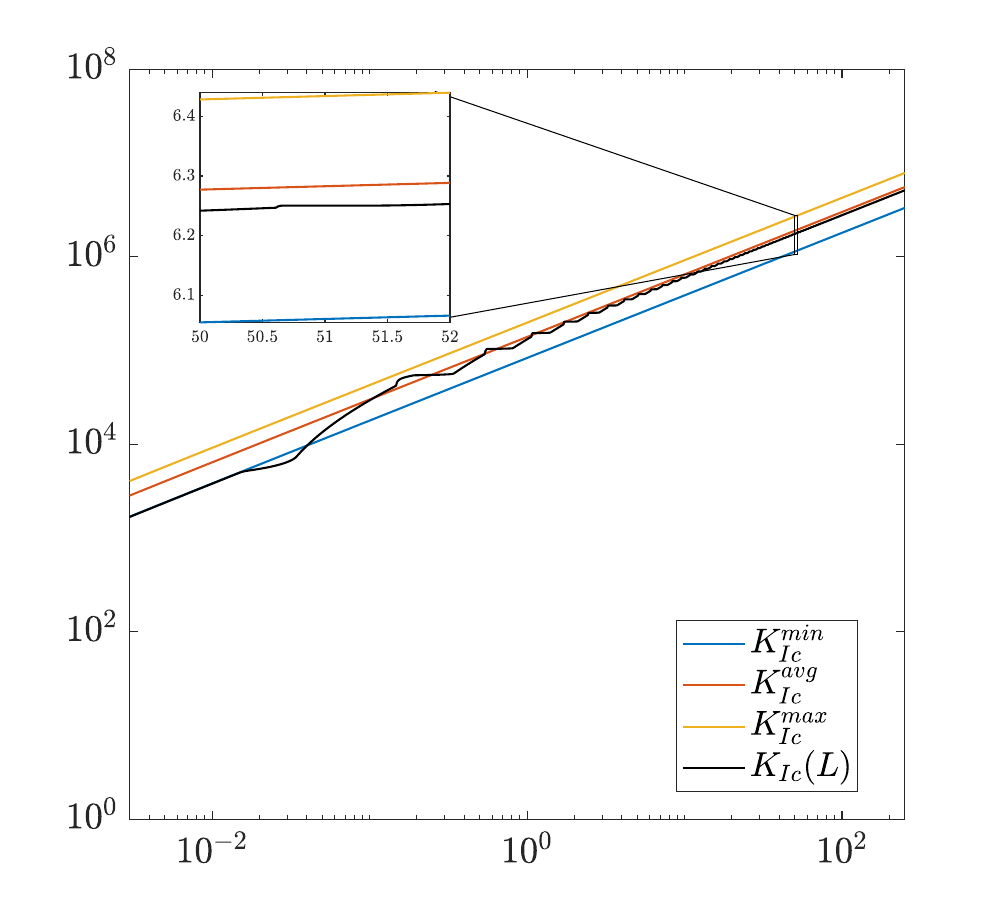}
	\put(-95,-3) {$l(t)$}
	\put(-195,155) {{\bf (f)}}
	\caption{The cumulative toughness energy expended fracturing the solid material $S$, (19), alongside the behaviour for the case of homogeneous toughness with distributions: maximum toughness $K_{Ic}(l(t))\equiv K_{Ic}^{max}$, minimum toughness $K_{Ic}(l(t))\equiv K_{Ic}^{min}$, and the (arithmetic) average toughness $K_{Ic}(l(t))\equiv K_{Ic}^{avg}$. We consider toughness distributions (a), (b) $\delta_{max}=1$, $\delta_{min}=0.1$, (c), (d) $\delta_{max}=10$, $\delta_{min}=1$, (e), (f) $\delta_{max}=100$, $\delta_{min}=10$, in the (a), (c), (e) sinusoidal, (b), (d), (f) step-wise, cases. }
	\label{S_Cum_1}
\end{figure}

\begin{figure}[t!]
	\centering
	\includegraphics[width=0.45\textwidth]{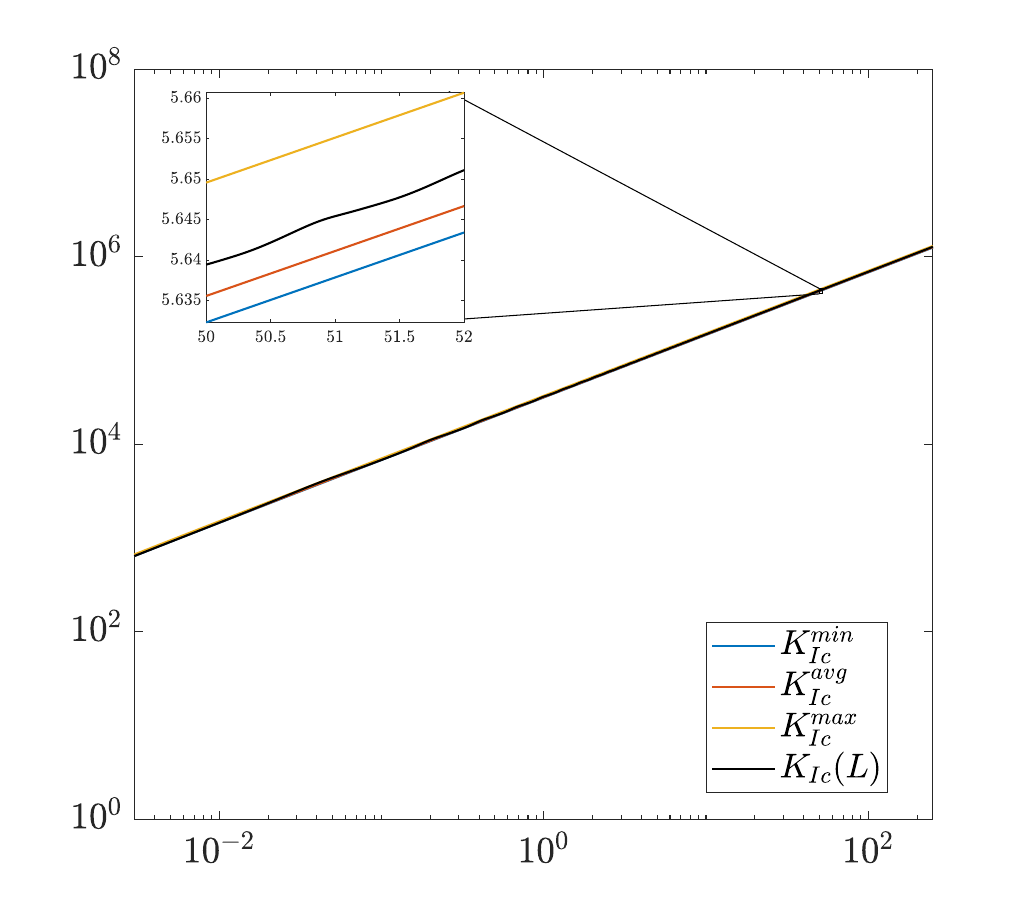}
	\put(-95,-3) {$l(t)$}
	\put(-235,85) {$\int_0^{l(t)} U(\eta)\,{\rm d}\eta$}
	\put(-195,155) {{\bf (a)}}
	\hspace{6mm}
	\includegraphics[width=0.45\textwidth]{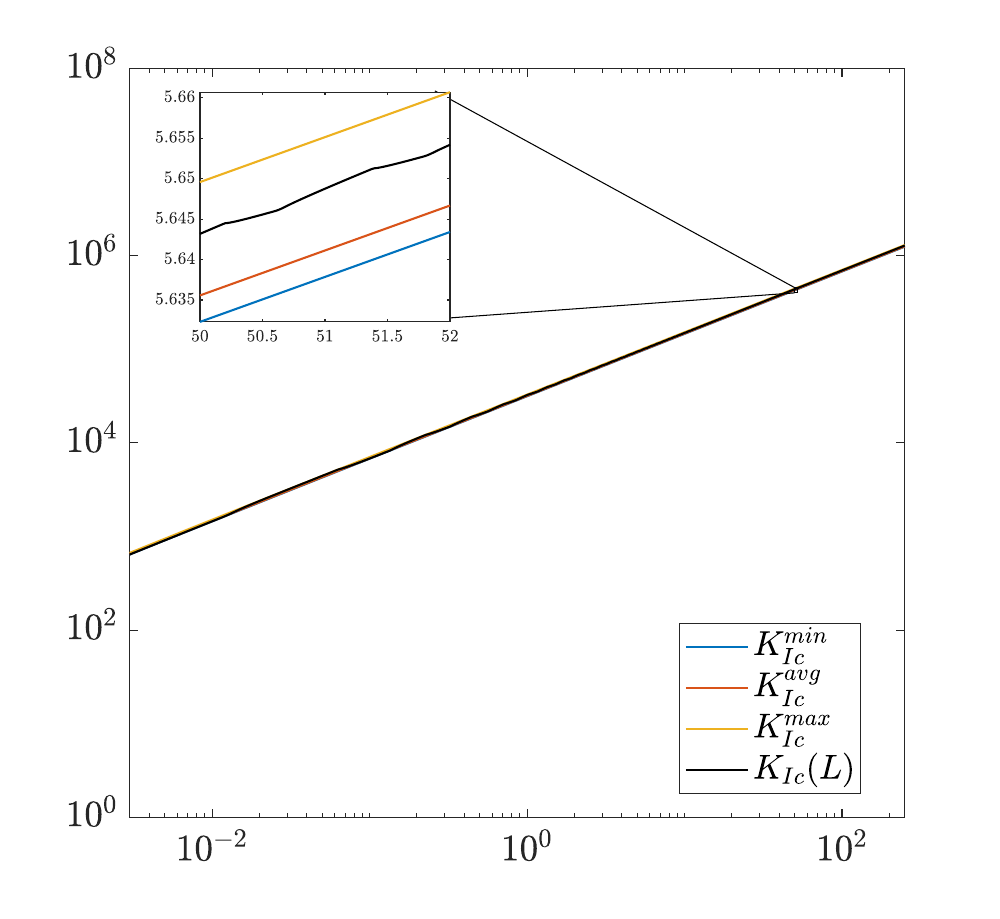}
	\put(-95,-3) {$l(t)$}
	\put(-195,155) {{\bf (b)}}
	
	\vspace{2mm}
	
	\includegraphics[width=0.45\textwidth]{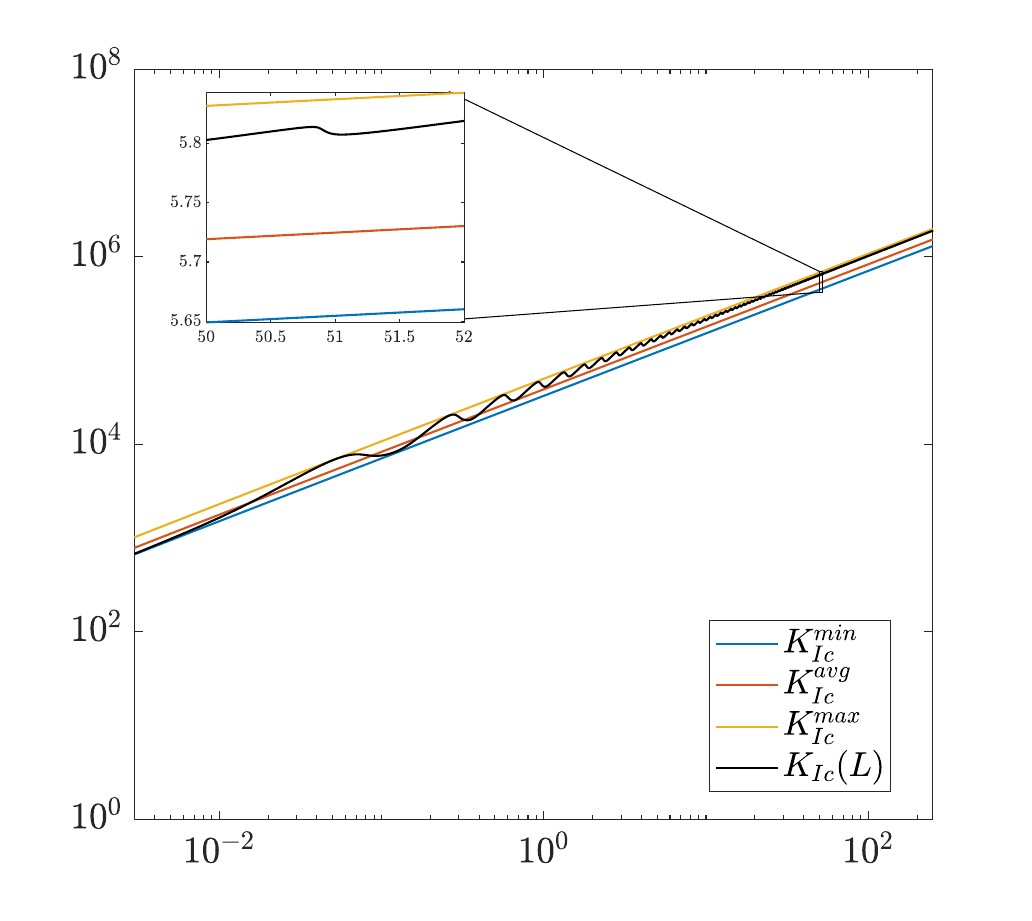}
	\put(-95,-3) {$l(t)$}
	\put(-235,85) {$\int_0^{l(t)} U(\eta)\,{\rm d}\eta$}
	\put(-195,155) {{\bf (c)}}
	\hspace{6mm}
	\includegraphics[width=0.45\textwidth]{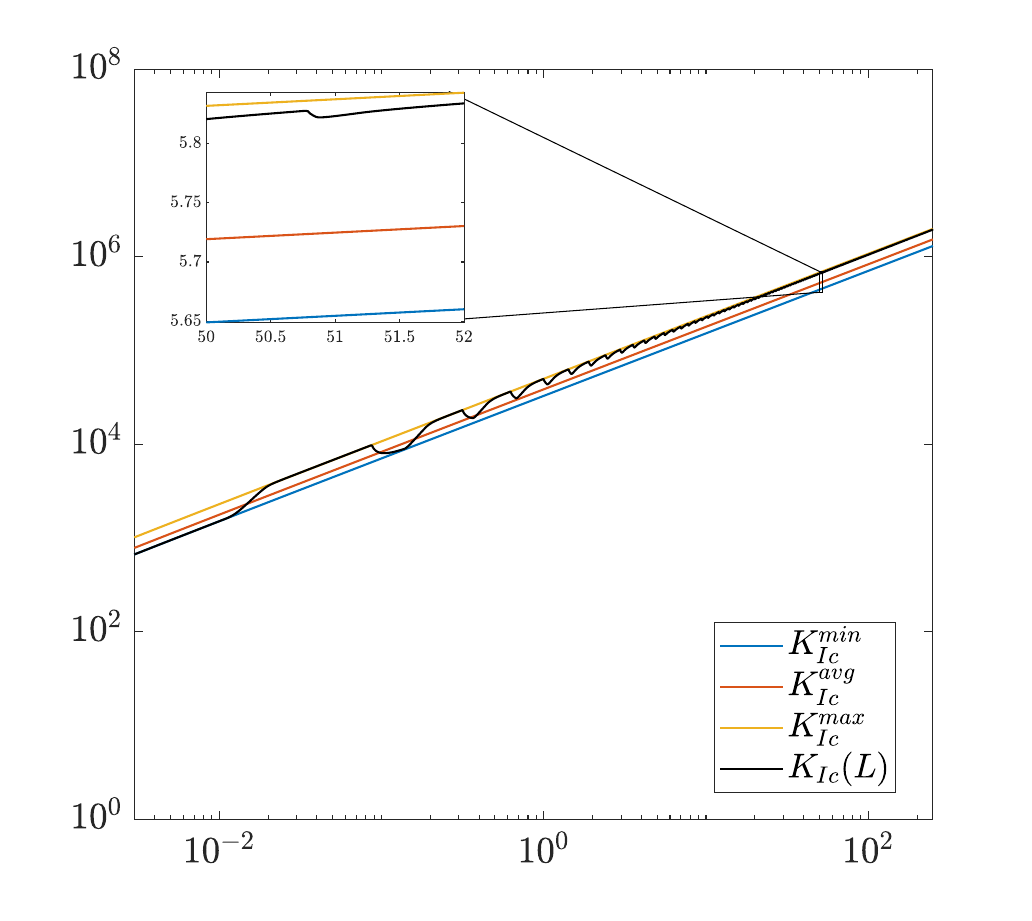}
	\put(-95,-3) {$l(t)$}
	\put(-195,155) {{\bf (d)}}
	
	\vspace{2mm}
	
	\includegraphics[width=0.45\textwidth]{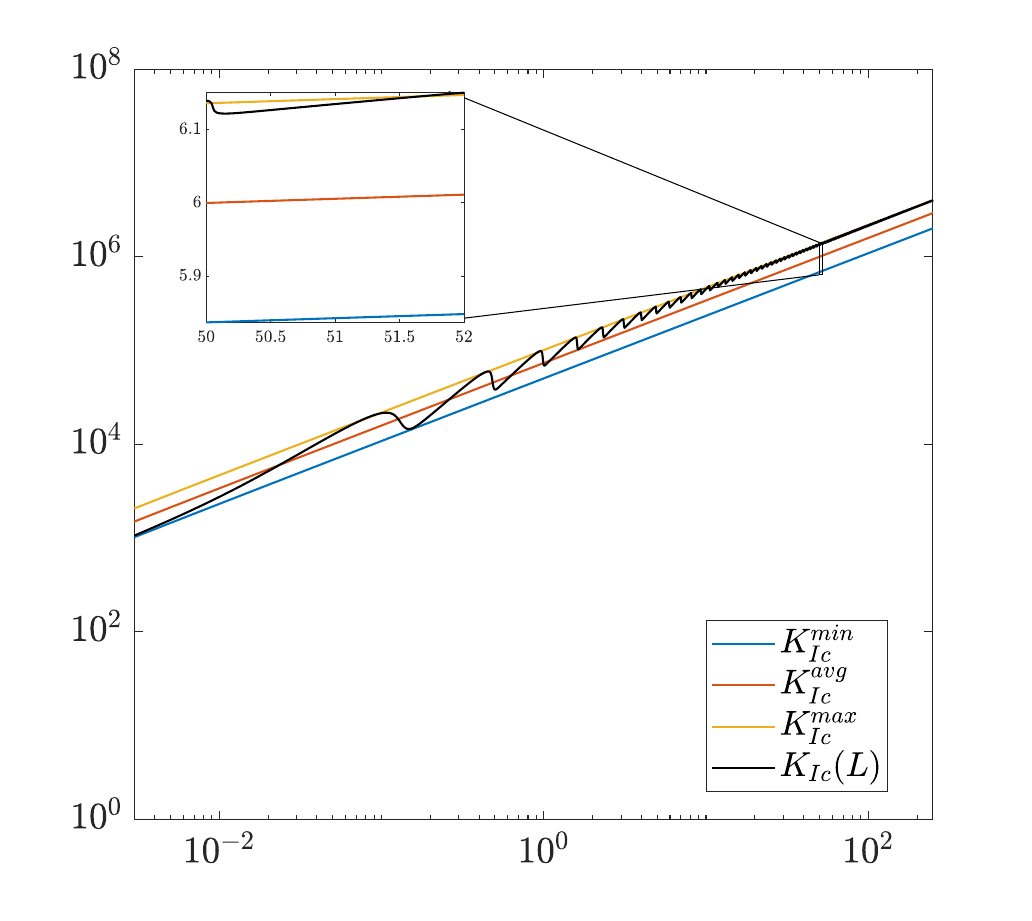}
	\put(-95,-3) {$l(t)$}
	\put(-235,85) {$\int_0^{l(t)} U(\eta)\,{\rm d}\eta$}
	\put(-195,155) {{\bf (e)}}
	\hspace{6mm}
	\includegraphics[width=0.45\textwidth]{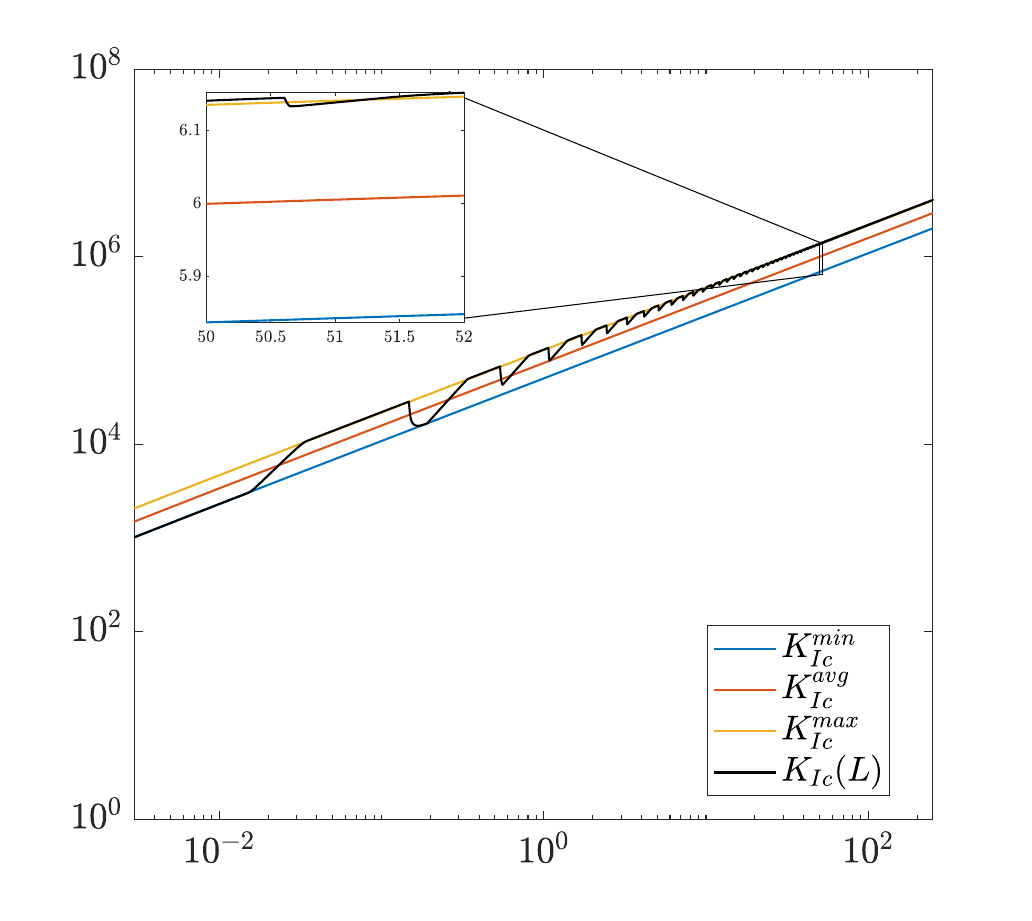}
	\put(-95,-3) {$l(t)$}
	\put(-195,155) {{\bf (f)}}
	\caption{The cumulative elastic energy $U$ stored within the solid material, (19), alongside the behaviour for the case of homogeneous toughness with distributions: maximum toughness $K_{Ic}(l(t))\equiv K_{Ic}^{max}$, minimum toughness $K_{Ic}(l(t))\equiv K_{Ic}^{min}$, and the (arithmetic) average toughness $K_{Ic}(l(t))\equiv K_{Ic}^{avg}$. We consider toughness distributions (a), (b) $\delta_{max}=1$, $\delta_{min}=0.1$, (c), (d) $\delta_{max}=10$, $\delta_{min}=1$, (e), (f) $\delta_{max}=100$, $\delta_{min}=10$, in the (a), (c), (e) sinusoidal, (b), (d), (f) step-wise, cases. }
	\label{U_Cum_1}
\end{figure}

{\bf The cumulative energy distribution}, discussed in the main text in Sect.~3.3, is provided in Figs.~\ref{P_Cum_1}-\ref{U_Cum_1}. The comparable figure in the main text is given in Fig.~10, providing all energy terms $P,D,S,U$ for the case $\delta_{max}=100$, $\delta_{min}=10$. Here, we provide each energy term separately and for all three toughness ratios considered in this paper. 

These figures include the cumulative energy distribution for oscillating toughness $K_{Ic}(l(t))$, alongside the cases with homogeneous toughness $K_{Ic}(l(t))\equiv K_{Ic}$ as:
$$
K_{Ic}^{max}= \max_{l\in [0,X]} \left(K_{Ic}(l)\right), \quad K_{Ic}^{min} = \min_{l\in [0,X]} \left(K_{Ic}(l)\right) \quad K_{Ic}^{avg} = \frac{1}{2}\left(K_{Ic}^{min} + K_{Ic}^{max}\right).
$$
To summarise the conclusions stated in the main text:
\begin{itemize}
	\item In the intermediate-viscosity case ($\delta_{max}=1$, $\delta_{min}=0.1$), the toughness distribution has little impact on the energy balance. Hence, all terms are close to that for the arithmetic average $K_{Ic}^{avg}$.
	\item In the toughness-toughness case ($\delta_{max}=100$, $\delta_{min}=10$), the power $P$ (injected into the crack) and elastic energy $U$ (stored in the solid) are close to that for the maximum toughness, while the toughness energy $S$ (spent extending the fracture) tends to a value just below that for the arithmetic average. To balance the other energy terms (19), the viscous term $D$ (spent overcoming viscous friction on the crack walls) tends to a value significantly above that for all of the homogeneous cases considered.
	\item The toughness-intermediate case ($\delta_{max}=10$, $\delta_{min}=1$), is in between these two cases.
	\item There is little dependence on the toughness distribution (sinusoidal or step-wise), with maximum value of the toughness distribution being the primary factor.
\end{itemize}

\begin{figure}[h]
	\centering
	\includegraphics[width=0.48\textwidth]{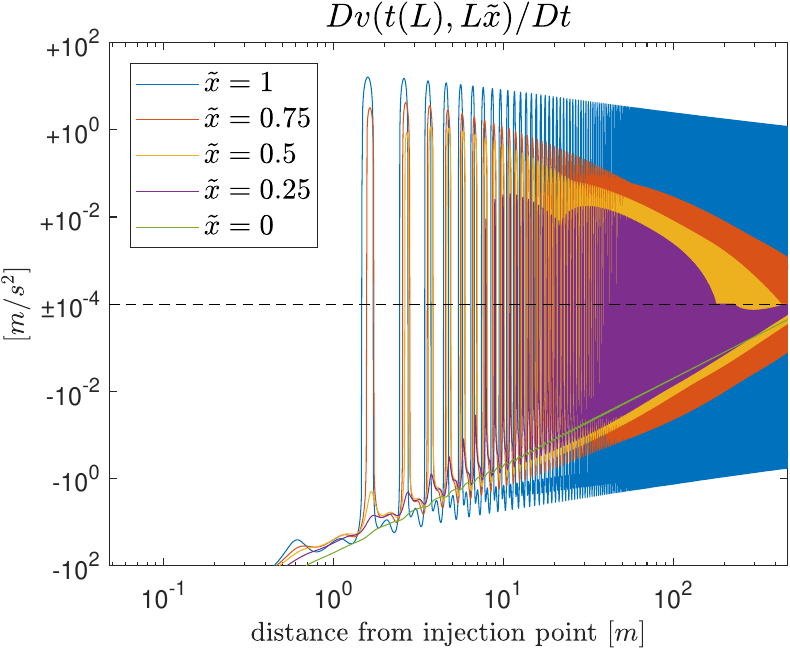}\hspace{3mm}
	\includegraphics[width=0.48\textwidth]{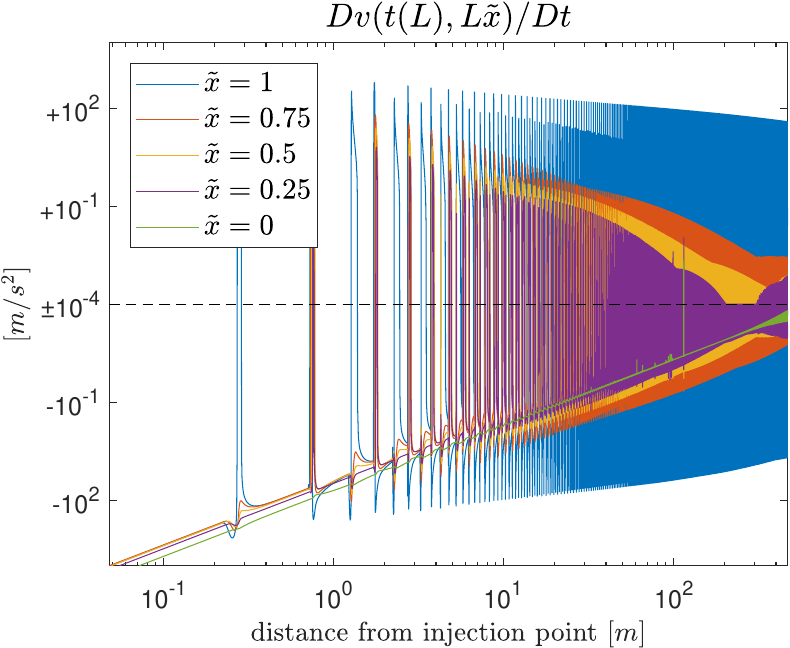}\\[3mm]
	\includegraphics[width=0.48\textwidth]{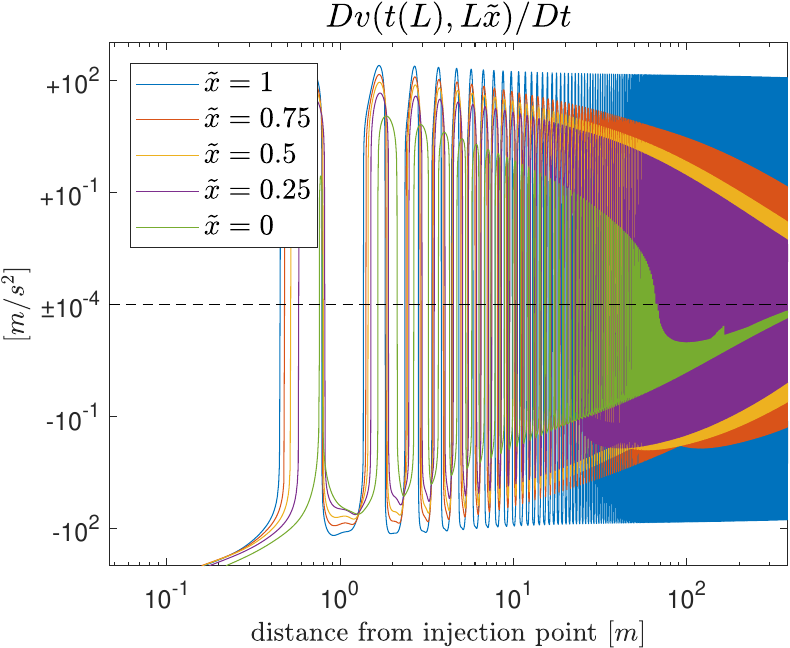}\hspace{3mm}
	\includegraphics[width=0.48\textwidth]{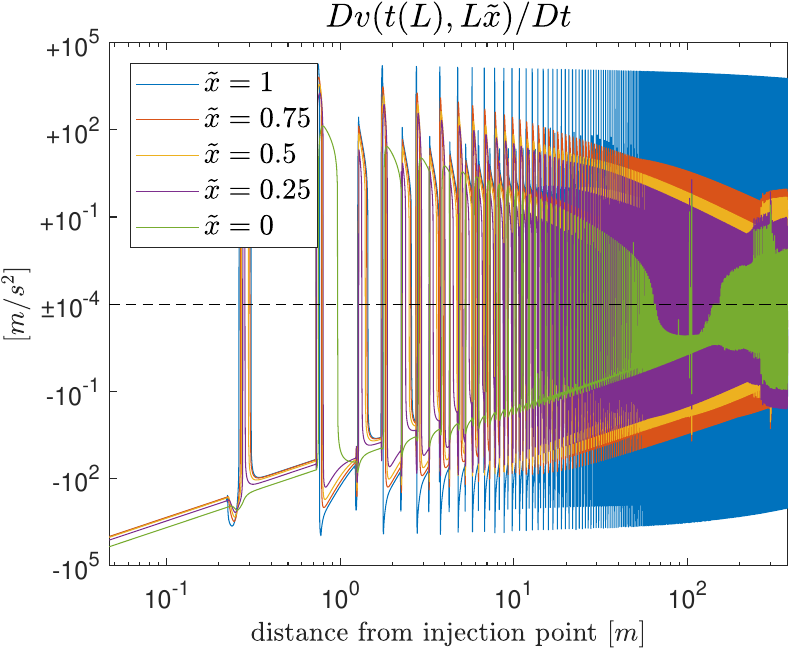}\\[3mm]
	\includegraphics[width=0.48\textwidth]{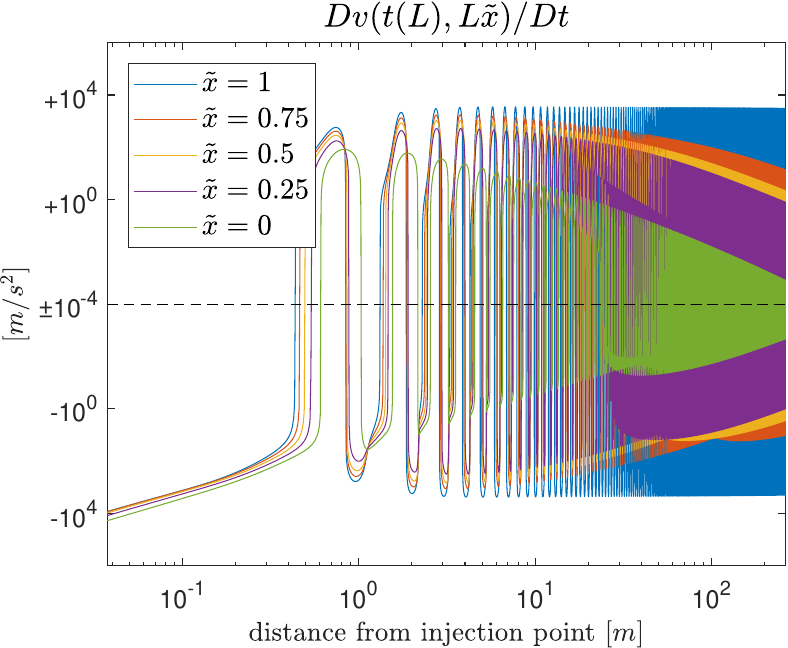}\hspace{3mm}
	\includegraphics[width=0.48\textwidth]{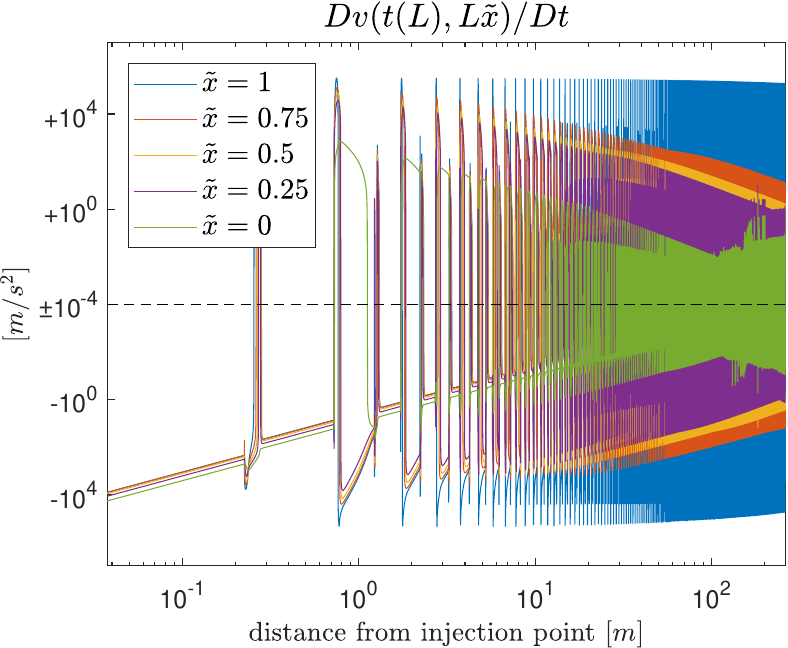}\\[3mm]
	\caption{Acceleration of the fluid particles inside the fracture for those 3 cases and 2 different toughness distributions in KGD model. (a), (b) $\delta_{max}=1$, $\delta_{min}=0.1$, (c), (d) $\delta_{max}=10$, $\delta_{min}=1$, (e), (f) $\delta_{max}=100$, $\delta_{min}=10$. }
	\label{Fluid_Accel_Overview}
\end{figure}

{\bf The acceleration of the fluid}, discussed in the main text in Sect.~4, is provided in Fig.~\ref{Fluid_Accel_Overview}. The corresponding figures for the fluid velocity are given in the main text as Fig.~11. Note that there is a high level of error for the fluid acceleration at late-time, due to the disproportionate impact of the local error for the fluid velocity $v$ on computations of this parameter. The general trends however are accurately displayed.

It is clear that there is an exceptionally high peak acceleration, which occurs when transitioning between different rock layers. The maximum acceleration is strongly dependent on the toughness distribution, being of order $10^5$m/s$^2$ for the step-wise distribution in the toughness-toughness case, but of order $10^1$m/s$^2$ for the sinusoidal distribution in the intermediate-viscosity case. Note also that the peak acceleration decreases as the crack propagates for the intermediate-viscosity regime, but remains almost constant for the toughness-toughness case (over the length-scale considered here).

\end{document}